\pdfoutput=1
\documentclass[twocolumn,fleqn,usenatbib]{mnras}

\usepackage{newtxtext,newtxmath}

\usepackage[T1]{fontenc}
\usepackage{ae,aecompl}

\usepackage{graphicx}	
\usepackage{amsmath}	
\usepackage{amssymb}	

\usepackage{flushend}

\title[{\rm ACS/WFC} {\it parallel fields}]{The {\it Hubble Space Telescope} UV Legacy Survey of Galactic Globular Clusters. XIII. {\rm ACS/WFC} Parallel-Field Catalogues\thanks{Based on observations with the NASA/ESA {\it Hubble Space Telescope}, obtained at the Space Telescope Science Institute, which is operated by AURA, Inc., under NASA contract NAS 5$-$26555.}}
\author[M.\,Simioni et al.]{
M.\,Simioni$^{1,2,3,4\thanks{email: msimioni@iac.es}}$,
L.\,R.\,Bedin$^4$,
A.\,Aparicio$^{2,1}$,
G.\,Piotto$^{3,4}$,
A.\,P.\,Milone$^{3,5}$,\newauthor
D.\,Nardiello$^{3,4}$,
J.\,Anderson$^{6}$,
A.\,Bellini$^{6}$,
T.\,M.\,Brown$^{6}$,
S.\,Cassisi$^{7,1}$,
A.\,Cunial$^{3}$,\newauthor
V.\,Granata$^{3,4}$,
S.\,Ortolani$^{3,4}$,
R.\,P.\,van\,der\,Marel$^{6,8}$,
E.\,Vesperini$^{9}$
\\
$^{1}$Instituto de Astrof{\'i}sica de Canarias, E-38200 La Laguna, Tenerife, Canary Islands, Spain\\
$^{2}$Department of Astrophysics, University of La Laguna, E-38200 La Laguna, Tenerife, Canary Islands, Spain\\
$^{3}$Dipartimento di Fisica e Astronomia ``Galileo Galilei'', Universit\`{a} degli Studi di Padova,  Vicolo dell'Osservatorio 3, Padova IT-35122\\
$^{4}$INAF - Osservatorio Astronomico di Padova, Vicolo dell'Osservatorio 5, I-35122 Padova, Italy\\
$^{5}$Research School of Astronomy and Astrophysics, Australian National University, Cotter Road, Weston, ACT 2611, Australia\\
$^{6}$Space Telescope Science Institute, 3700 San Martin Dr.,Baltimore, MD 21218, USA\\
$^{7}$INAF - Osservatorio Astronomico di Teramo, Via M. Maggini, I-64100 Teramo, Italy\\
$^{8}$Center for Astrophysical Sciences, Department of Physics \& Astronomy, Johns Hopkins University,\\
Baltimore, MD 21218, USA\\
$^{9}$Department of Astronomy, Indiana University, Bloomington, IN47401, USA\\
}
\date{Accepted 2018 January 17. Received 2018 January 17; in original form 2017 October 10}

\pubyear{2018}

\begin{document}
\label{firstpage}
\pagerange{\pageref{firstpage}--\pageref{lastpage}}
\maketitle

\begin{abstract}
  As part of the \textit{Hubble Space Telescope UV Legacy Survey of Galactic Globular Clusters}, 110 parallel fields were observed with the Wide Field Channel of the Advanced Camera for Surveys, in the outskirts of 48 globular clusters, plus the open cluster NGC\,6791. Totalling about $0.3$ square degrees of observed sky, this is the largest homogeneous {\it Hubble Space Telescope} photometric survey of Galalctic globular clusters outskirts to date. In particular, two distinct pointings have been obtained for each target on average, all centred at about $6.5$ arcmin from the cluster centre, thus covering a mean area of about $23\,{\rm arcmin^{2}}$ for each globular cluster.
For each field, at least one exposure in both F475W and F814W filters was collected. 
In this work, we publicly release the astrometric and photometric catalogues and the astrometrised atlases for each of these fields.
\end{abstract}


\defcitealias{2015AJ....149...91P}{Paper\,I}
\defcitealias{2017MNRAS.464.3636M}{Paper\,IX}

\section{Introduction}\label{sec:intro}

For almost three decades, the Milky Way Globular Clusters (GCs) have been the target of large CCD photometric surveys aimed at sampling their stellar populations in a homogeneous way (\citealt{2000A&AS..144....5R}, \citealt{2000A&AS..145..451R}, \citealt{2002A&A...391..945P}, \citealt{2007AJ....133.1658S}) using both space- and ground-based instruments.
The growing sample of data, and the advent of increasingly sophisticated data-analysis techniques, have clearly demonstrated that GCs host distinct stellar populations with different chemical abundances. High-precision photometric measurements have revealed that the colour-magnitude diagrams (CMDs) show distinct sequences in various evolutionary stages (see e.g. \citealt{1997PhDT.........8A}, \citealt{1999Natur.402...55L}, \citealt{2000ApJ...534L..83P}, \citealt{2004ApJ...605L.125B}, \citealt{2007ApJ...661L..53P}, \citealt{2008ApJ...673..241M}, \citealt{2010AJ....140..631B}). These findings are also supported by spectroscopical evidence that the stellar populations of these systems are not as simple as thought (see e.g. \citealt{2008A&A...490..625M}, \citealt{2008ApJ...672L..29Y}, \citealt{1994PASP..106..553K}, \citealt{2009A&A...505..117C}, \citealt{2009A&A...505..139C}, \citealt{2004ARA&A..42..385G}, \citealt{2012A&ARv..20...50G}).

The {\it Hubble Space Telescope UV Legacy Survey of Galactic Globular Clusters} (GO-13297; PI:Piotto) has been specifically designed to further investigate this phenomenon and it now appears likely that all Galactic GCs host multiple stellar populations (\citealt{2015AJ....149...91P} -- hereafter \citetalias{2015AJ....149...91P}, \citealt{2017MNRAS.464.3636M} -- hereafter \citetalias{2017MNRAS.464.3636M}).
In the context of this survey, parallel Advanced Camera for Surveys (ACS) observations have been obtained. While the main observations were taken using a combination of UV and optical filters of the Wide Field Camera 3 ({\rm WFC3}), the lack of filters bluer than F435W dictated the use of the F475W and F814W filters of the Wide Field Channel of the ACS ({\rm ACS/WFC}) in the parallel observations. The large colour baseline provided by this filter combination guarantees sensitivity to helium abundance differences, while being largely insensitive to star-to-star variations in light-element abundances (\citealt{2011A&A...534A...9S}, \citealt{2017MNRAS.464.2341C}).

One of the main objectives for which these observations were planned is to investigate how different stellar populations formed in GCs.
Strong observational constraints come from the analysis of the radial distribution of each stellar population (\citealt{2008MNRAS.391..825D}; \citealt{2009A&A...507.1393B}, \citealt{2013MNRAS.429.1913V}).
As an example, \citet{2016MNRAS.tmp.1125S} complemented {\rm WFC3} data of the central regions of NGC\,2808 with {\rm ACS} parallel observations and found evidence of different radial trends associated with distinct stellar populations hosted by the cluster. 
Thus, clusters with large helium variations among their stellar populations are the preferred target of investigation with the current data-sample. Other interesting targets, albeit extensively studied, are those defined as Type-II clusters in \citetalias{2017MNRAS.464.3636M}, which displays multiple sub-giant branches in optical CMDs.

We stress the fact that this is the first homogeneous {\it HST} photometric survey of the outskirts of Galactic GCs.
The observations presented here represent a first epoch for future studies aimed at systematical measurements of absolute, relative and internal proper motions of stars in these regions. Archival {\it HST} observations matching a sub-sample of the observed fields exist, and proper motions will be published separately.
In the imaged stellar fields, the stellar density is not as high as in the central regions. As a consequence, crowding is not a serious issue for these data. That makes them particularly suitable to be used as input catalogue for future spectroscopic surveys.

The present catalogues can be used to perform several interesting analyses. For example, dynamical interactions between stars in GCs is at the origin of the mass segregation phenomenon. A precise estimate of its effects is fundamental for the derivation of a global mass function for a GC (\citealt{1997MNRAS.289..898V}, \citealt{2010AJ....139..476P}, \citealt{2017MNRAS.471.3668S}).
The measurement of the fraction of binaries is also fundamental for this kind of analysis and could provide useful constraints for dynamical models \citep{2012A&A...540A..16M}. We note, also, that in some cases white dwarf cooling sequences are visible in the obtained CMDs.
Finally, it is interesting to note that due to the presence of many extra-galactic objects in the observed field, other studies could benefit from these observations.

In this work, we present the first photometric catalogues from the {\rm ACS/WFC} parallel observations of the GO-13297 program. All data have been reduced in a homogeneous manner, making these catalogues particularly suitable for inter-comparison.
The article is organised as follows: in Section \ref{sec:data} the data are presented along with some information about the observing strategy, together with a detailed description of the data reduction. The extracted CMDs are presented in Section \ref{sec:cmd}. Details on the selection of well measured stars are given in Section \ref{sec:ex}. In Section \ref{sec:cat} the catalogues and the released electronic material are described in detail. Finally, in Section \ref{sec:sum}, after a summary, we briefly discuss some of the main scientific questions we will address with these catalogues in subsequent papers.

\section{Observations and Data Reduction}\label{sec:data}

\begin{table*}
  \footnotesize
  \caption{\normalsize Observation log. For each GC in the survey, and the open cluster NGC\,6791,} we show right ascension and declination of each distinct parallel field, referred to the centre of {\rm ACS/WFC}. We also report the number of orbits, Telescope orientation (V3 PA) for each orbit and exposure time in each filter.
  \label{tab:log}
  \begin{tabular}{c c c l c c c l l}
     \hline
     \#   & CLUSTER   & ORBITS & FIELD      & RA (J2000)           & DEC (J2000)                          & EPOCH    & EXP. TIME             & EXP. TIME    \\
          &           &        & [PA (deg)] & $(^{h}\, ^{m}\, ^{s})$ & $(^{\circ}\, ^{\prime}\, ^{\prime\prime})$ &          & F475W (s)             & F814W (s)    \\
      \hline
%
01   & NGC1261   & 5      & F1 [92]    & 03:12:49.68            & -55:17:25.2         		             & 31/08/13 & 770                             & 694                \\
     &           &        & F2 [138]   & 03:12:16.96            & -55:19:29.6                                & 11/09/13 & 745                             & 669                \\
     &           &        & F3 [182]   & 03:11:44.35            & -55:17:39.3                                & 08/11/13 & 766                             & 690                \\
     &           &        & F4 [225]   & 03:11:30.53            & -55:13:17.9                                & 07/12/13 & 745                             & 669                \\
     &           &        & F5 [48]    & 03:13:01.92            & -55:12:51.4                                & 29/06/14 & 829                             & 753                \\
02   & NGC1851   & 7      & F1 [195]   & 05:13:37.57            & -40:06:44.0                                & 27/12/10 & 2x40;2x1277;1237                & 6x488;1x40         \\
     &           &        & F2 [164]   & 05:13:51.57            & -40:08:55.8                                & 11/11/10 & 2x40;2x1277;2x1237              & 8x488;2x40         \\
03   & NGC2298   & 4      & F1 [185]   & 06:48:36.04            & -36:05:08.7                                & 18/12/13 & 2x785                           & 2x683              \\
     &           &        & F2 [273]   & 06:48:35.14            & -35:55:40.7                                & 07/03/14 & 885                             & 816                \\
     &           &        &            &                        &                                            & 15/03/15 & 887                             & 815                \\
04   & NGC3201   & 2      & F1 [25]    & 10:18:12.19            & -46:21:47.1                                & 13/09/13 & 685                             & 612                \\
     &           &        & F2 [115]   & 10:17:54.00            & -46:30:51.0                                & 01/01/14 & 689                             & 616                \\
05   & NGC4590   & 2      & F1 [112]   & 12:39:42.64            & -26:50:34.9                                & 21/12/13 & 627                             & 554                \\
     &           &        & F2 [202]   & 12:39:01.10            & -26:47:48.5                                & 30/03/14 & 627                             & 554                \\
06   & NGC4833   & 4	  & F1 [113]   & 13:00:10.12            & -70:58:29.4                                & 17/01/14 & 2x840                           & 2x771              \\
     &           &        & F2 [202]   & 12:58:20.68            & -70:55:42.6                                & 09/04/14 & 2x806                           & 2x730              \\
07   & NGC5024   & 6      & F1 [31]    & 13:13:21.59            & +18:12:01.1                                & 24/03/14 & 4x725;2x723                     & 3x370              \\
     &           &        & F2 [120]   & 13:13:03.82            & 18:03:52.0                                 & 08/12/13 & 4x775;2x774                     & 3x375              \\
08   & NGC5053   & 5      & F1 [352]   & 13:16:42.03            & +17:47:28.7                                & 01/04/14 & 740                             & 664                \\
     &           &        & F2 [37]    & 13:16;53.97            & +17:43:15.0                                & 16/03/14 & 740                             & 664                \\
     &           &        & F3 [80]    & 13:16;50.28            & +17:38:33.1                                & 23/01/14 & 790                             & 714                \\
     &           &        & F4 [125]   & 13:16;33.21            & +17:35:39.5                                & 05/12/13 & 790                             & 714                \\
     &           &        & F5 [308]   & 13:16;21.88            & +17:48:25.0                                & 16/05/14 & 765                             & 689                \\
09   & NGC5286   & 2      & F1 [73]    & 13:47:07.09            & -51:25:00.2                                & 14/12/13 & 728                             & 655                \\
     &           &        & F2 [162]   & 13:46:11.09            & -51:28:46.7                                & 15/03/14 & 603                             & 559                \\
10   & NGC5466   & 4      & F1 [112]   & 14:05:41.08            & +28:26:12.1                                & 05/01/14 & 834;835                         & 763;765          \\
     &           &        & F2 [21]    & 14:05:54.29            & +28:35:18.1                                & 29/03/14 & 2x776                           & 2x700            \\
11   & NGC5897   & 4      & F1 [112]   & 15:17:37.48            & -21:06:27.5                                & 12/02/14 & 830;833                         & 2x761            \\
     &           &        & F2 [202]   & 15:16:58.75            & -21:03:44.3                                & 13/05/14 & 779;781                         & 710;709            \\
12   & NGC5904   & 2      & F1 [323]   & 15:18:34.23            & 02:11:38.3                                 & 17/05/14 & 620                             & 559                \\
     &           &        & F2 [52]    & 15:19:00.34            & 02:04:36.4                                 & 08/04/14 & 621                             & 559                \\
13   & NGC5927   & 3      & F1 [100]   & 15:28:28.73            & -50:45:30.4                                & 01/02/14 & 603                             & 559                \\
     &           &        & F2 [189]   & 15:27:28.26            & -50:44:48.7                                & 19/05/14 & 603                             & 559                \\
     &           &        &            &                        &                                            & 27/05/15 & 603                             & 559                \\
14   & NGC5986   & 3      & F1 [92]    & 15:46:28.87            & -37:51:38.6                                & 27/01/14 & 676                             & 603                \\
     &           &        & F2 [180]   & 15:45:40.83            & -37:52:21.9                                & 17/05/14 & 603                             & 559                \\
     &		 &	  &    	       &                        &                                            & 10/05/15 & 603                             & 559                \\
15   & NGC6093   & 5      & F1 [255]   & 16:16:35.04            & -22:55:45.9                                & 09/06/12 & 5x760;5x845                     & 5x539              \\
16   & NGC6101   & 5      & F1 [147]   & 16:25:36.08            & -72:18:35.2                                & 04/04/14 & 762                             & 686                \\
     &           &        & F2 [190]   & 16:24:40.57            & -72:16:07.7                                & 25/05/14 & 762                             & 686                \\
     &           &        & F3 [235]   & 16:24:23.48            & -72:11:18.6                                & 29/06/14 & 800                             & 724                \\
     &           &        & F4 [282]   & 16:24:58.26            & -72:06:50.7                                & 14/08/13 & 851                             & 775                \\
     &           &        & F5 [101]   & 16:26:39.71            & -72:17:19.9                                & 28/02/14 & 800                             & 724                \\
17   & NGC6121   & 2      & F1 [272]   & 16:23:12.57            & -26:27:02.1                                & 06/07/14 & 739                             & 666                \\
     &           &        & F2 [98]    & 16:23:55.64            & -26:36:33.4                                & 17/02/15 & 666                             & 593                \\
18   & NGC6144   & 2      & F1 [83]    & 16:27:39.42            & -26:05:00.9                                & 28/02/14 & 679                             & 606                \\
     &           &        & F2 [174]   & 16:26:57.33            & -26:07:05.2                                & 27/05/14 & 679                             & 606                \\
19   & NGC6171   & 4      & F1 [342]   & 16:32:41.94            & -12:56:56.1                                & 31/05/14 & 830;833                         & 2x761              \\
     &           &        & F2 [72]    & 16:32:42.20            & -12:57:07.5                                & 25/03/14 & 800;802                         & 731;730            \\
20   & NGC6218   & 2      & F1 [276]   & 16:46:55.21            & -01:52:03.7                                & 16/08/13 & 721                             & 648                \\
     &           &        & F2 [6]     & 16:47:33.42            & -01:52:07.7                                & 27/05/14 & 645                             & 572                \\
21   & NGC6254   & 2      & F1 [276]   & 16:56:50.04            & -04:01:10.1                                & 16/08/13 & 721                             & 648                \\
     &           &        & F2 [7]     & 16:57:28.66            & -04:01:19.2                                & 27/05/14 & 644                             & 571                \\
22   & NGC6304   & 2      & F1 [184]   & 17:14:10.53            & -29:32:35.1                                & 07/06/14 & 624                             & 559                \\
     &           &        & F2 [274]   & 17:14:09.53            & -29:23:04.8                                & 26/08/13 & 731                             & 658                \\
23   & NGC6341   & 2      & F1 [230]   & 17:16:30.21            & +43:08:03.0                                & 22/10/13 & 638                             & 565                \\
     &           &        & F2 [319]   & 17:17:06.19            & +43:14:56.0                                & 03/08/14 & 750                             & 677                \\
24   & NGC6352   & 2      & F1 [161]   & 17:25:15.00            & -48:31:41.7                                & 27/05/14 & 637                             & 564                \\
     &           &        & F2 [251]   & 17:24:51.02            & -48:22:52.5                                & 13/08/13 & 731                             & 658                \\

   \end{tabular}
   \raggedright{\it Continued on Tab. \ref{tab:logb}}
\end{table*}%
\begin{table*}
  \footnotesize
  \caption{\normalsize Table \ref{tab:log} (continued)}
  \label{tab:logb}
  \begin{tabular}{c c c l c c c l l}
     \hline
     \#   & CLUSTER   & ORBITS & FIELD      & RA (J2000)           & DEC (J2000)                          & EPOCH    & EXP. TIME             & EXP. TIME    \\
          &           &        & [PA (deg)] & $(^{h}\, ^{m}\, ^{s})$ & $(^{\circ}\, ^{\prime}\, ^{\prime\prime})$ &          & F475W (s)             & F814W (s)    \\
      \hline
%
25   & NGC6362   & 2      & F1 [125]   & 17:32:14.07            & -67:09:25.3                                & 30/03/14 & 651                             & 578                \\
     &           &        & F2 [215]   & 17:30:47.67            & -67:04:37.3                                & 01/07/14 & 760                             & 687                \\
26   & NGC6366   & 2      & F1 [293]   & 17:27:31.94            & -04:58:44.4                                & 26/08/13 & 726                             & 653                \\
     &           &        & F2 [351]   & 17:27:58.09            & -04:58:57.1                                & 07/06/15 & 644                             & 571                \\
27   & NGC6388   & 4      & F1 [238]   & 17:35:40.80            & -44:43:02.9                                & 05/07/14 & 865;906                         & 796;834            \\
     &           &        & F2 [141]   & 17:36:17.23            & -44:50:53.5                                & 12/05/14 & 793;795                         & 724;723            \\
28   & NGC6397   & 2      & F1 [90]    & 17:41:17.47            & -53:44:46.1                                & 27/03/14 & 683                             & 610                \\
     &           &        & F2 [180]   & 17:40:12.50            & -53:45:38.3                                & 11/06/14 & 640                             & 567                \\
29   & NGC6441   & 4      & F1 [102]   & 17:50:34.23            & -37:08:21.5                                & 26/03/14 & 833;835                         & 764;763            \\
     &           &        & F2 [192]   & 17:49:46.24            & -37:07:12.6                                & 15/06/14 & 2x794                           & 725;722            \\
30   & NGC6496   & 2      & F1 [157]   & 17:58:53.12            & -44:22:28.2                                & 30/05/14 & 638                             & 656                \\
     &           &        & F2 [247]   & 17:58:27.50            & -44:13:57.0                                & 12/08/13 & 731                             & 658                \\
31   & NGC6535   & 2      & F1 [319]   & 18:03:49.95            & +00:11:04.6                                & 19/07/14 & 724                             & 651                \\
     &           &        & F2 [50]    & 18:04:17.63            & +00:17:50.2                                & 26/05/14 & 644                             & 571                \\
32   & NGC6541   & 2      & F1 [80]    & 18:08:35.16            & -43:46:11.5                                & 14/02/14 & 689                             & 616                \\
     &           &        & F2 [170]   & 18:07:44.04            & -43:48:48.8                                & 11/06/14 & 639                             & 566                \\
33   & NGC6584   & 2      & F1 [155]   & 18:18:26.74            & -52:19:31.3                                & 30/05/14 & 640                             & 567                \\
     &           &        & F2 [245]   & 18:17:54.89            & -52:11:09.9                                & 18/08/13 & 726                             & 653                \\
34   & NGC6624   & 2      & F1 [264]   & 18:23:14.50            & -30:17:51.1                                & 03/09/13 & 731                             & 658                \\
     &           &        & F2 [174]   & 18:23:23.29            & -30:27:20.2                                & 27/06/14 & 638                             & 565                \\
35   & NGC6637   & 4      & F1 [84]    & 18:31:49.55            & -32:24:40.9                                & 18/02/14 & 827;840                         & 758;768            \\
     &           &        & F2 [174]   & 18:31:05.49            & -32:26:31.5                                & 30/06/14 & 792;794                         & 723;722            \\
36   & NGC6652   & 3      & F1 [281]   & 18:35:25.09            & -32:54:08.8                                & 30/08/13 & 707                             & 633                \\
     &           &        & F2 [238]   & 18:35:14.87            & -32:58:25.1                                & 14/08/13 & 733                             & 658                \\
     &           &        & F3 [192]   & 18:35:20.06            & -33:03:36.1                                & 01/07/14 & 621                             & 548                \\
37   & NGC6656   & 4      & F1 [266]   & 18:36:00.25            & -23:50:16.5                                & 23/09/10 & 656;644                         & 2x389              \\ 
     &           &        & F2 [85]    & 18:36:47.76            & -23:58:08.0                                & 17/03/11 & 2x656                           & 2x389              \\
38   & NGC6681   & 2      & F1 [271]   & 18:42:48.78            & -32:13:03.7                                & 05/09/13 & 730                             & 658                \\
     &           &        & F2 [183]   & 18:42:51.19            & -32:22:33.1                                & 29/06/14 & 637                             & 564                \\
39   & NGC6715   & 6      & F1 [153]   & 18:54:56.04            & -30:35:08.9                                & 29/06/14 & 2x736;2x737;2x734               & 3x370              \\
     &           &        & F2 [243]   & 18:54:33.96            & -30:27:06.5                                & 05/09/13 & 4x819;2x817                     & 3x390              \\
40   & NGC6717   & 3      & F1 [80]    & 18:55:31.98            & -22:45:22.7                                & 06/05/14 & 619                             & 535                \\
     &           &        & F2 [125]   & 18:55:12.41            & -22:48:27.0                                & 03/07/14 & 619                             & 544                \\
     &           &        & F3 [171]   & 18:54:51.25            & -22:47:57.0                                & 03/07/14 & 617                             & 544                \\
41   & NGC6723   & 3      & F1 [103]   & 18:59:54.19            & -36:43:18.0                                & 03/04/14 & 666                             & 592                \\
     &           &        & F2 [148]   & 18:59:27.49            & -36:44:21.9                                & 15/06/14 & 626                             & 551                \\
     &           &        & F3 [193]   & 18:59:06.07            & -36:41:59.9                                & 05/07/14 & 624                             & 551                \\
42   & NGC6779   & 2      & F1 [241]   & 19:16:04.77            & +30:12:18.9                                & 13/10/13 & 731                             & 658                \\
     &           &        & F2 [333]   & 19:16:42.17            & +30:17:38.2                                & 19/07/14 & 637                             & 564                \\
43   & NGC6791   & 2      & F1 [322]   & 19:20:53.69            & +37:53:04.8                                & 17/08/13 & 631                             & 559                \\
     &           &        & F2 [50]    & 19:21:27.31            & +37:46:16.6                                & 26/04/14 & 638                             & 565                \\
44   & NGC6809   & 2      & F1 [262]   & 19:39:32.79            & -30:54:19.4                                & 21/08/14 & 753                             & 680                \\
     &           &        & F2 [82]    & 19:40:26.65            & -31:01:26.4                                & 29/03/14 & 677                             & 604                \\
45   & NGC6838   & 2      & F1 [244]   & 19:53:18.63            & +18:48:20.3                                & 23/10/13 & 723                             & 650                \\
     &           &        & F2 [75]    & 19:54:12.48            & +18:43:53.9                                & 03/05/14 & 681                             & 608                \\
46   & NGC6934   & 2      & F1 [245]   & 20:33:44.84            & +07:25:55.5                                & 08/10/13 & 723                             & 650                \\
     &           &        & F2 [334]   & 20:34:17.59            & +07:30:52.3                                & 18/08/14 & 644                             & 571                \\
47   & NGC6981   & 2      & F1 [289]   & 20:53:13.45            & -12:26:24.9                                & 13/08/13 & 641                             & 568                \\
     &           &        & F2 [19]    & 20:53:51.29            & -12:28:39.5                                & 03/08/14 & 624                             & 551                \\
48   & NGC7089   & 3      & F1 [237]   & 21:32:59.85            & +00:48:42.0                                & 18/10/13 & 717                             & 643                \\
     &           &        & F2 [281]   & 21:33:10.93            & +00:44:15.0                                & 29/08/13 & 668                             & 593                \\
     &           &        & F3 [327]   & 21:33:29.92            & +00:42:39.1                                & 14/08/13 & 611                             & 534                \\
49   & NGC7099   & 2      & F1 [92]    & 21:40:44.34            & -23:15:15.1                                & 08/06/14 & 656                             & 583                \\
     &           &        & F2 [182]   & 21:40:02.30            & -23:15:49.0                                & 19/08/14 & 656                             & 583                \\

      \hline
   \end{tabular}
\end{table*}%

Tables \ref{tab:log} and \ref{tab:logb} report the log of {\rm ACS/WFC} observations used to construct the catalogues.
For each target, we indicate the total number of orbits assigned to each observed field separated by the different position angles (PAs) of the V3 axis of the {\it HST} focal plane. Typically, one F475W and one F814W image were taken each orbit, with a dither between the two dictated by primary {\rm WFC3} observing strategy.
For each field, right ascension and declination of the centre of {\rm ACS/WFC} are provided along with exposure time in each filter.

\begin{figure*}
  \includegraphics[width=8.5 cm]{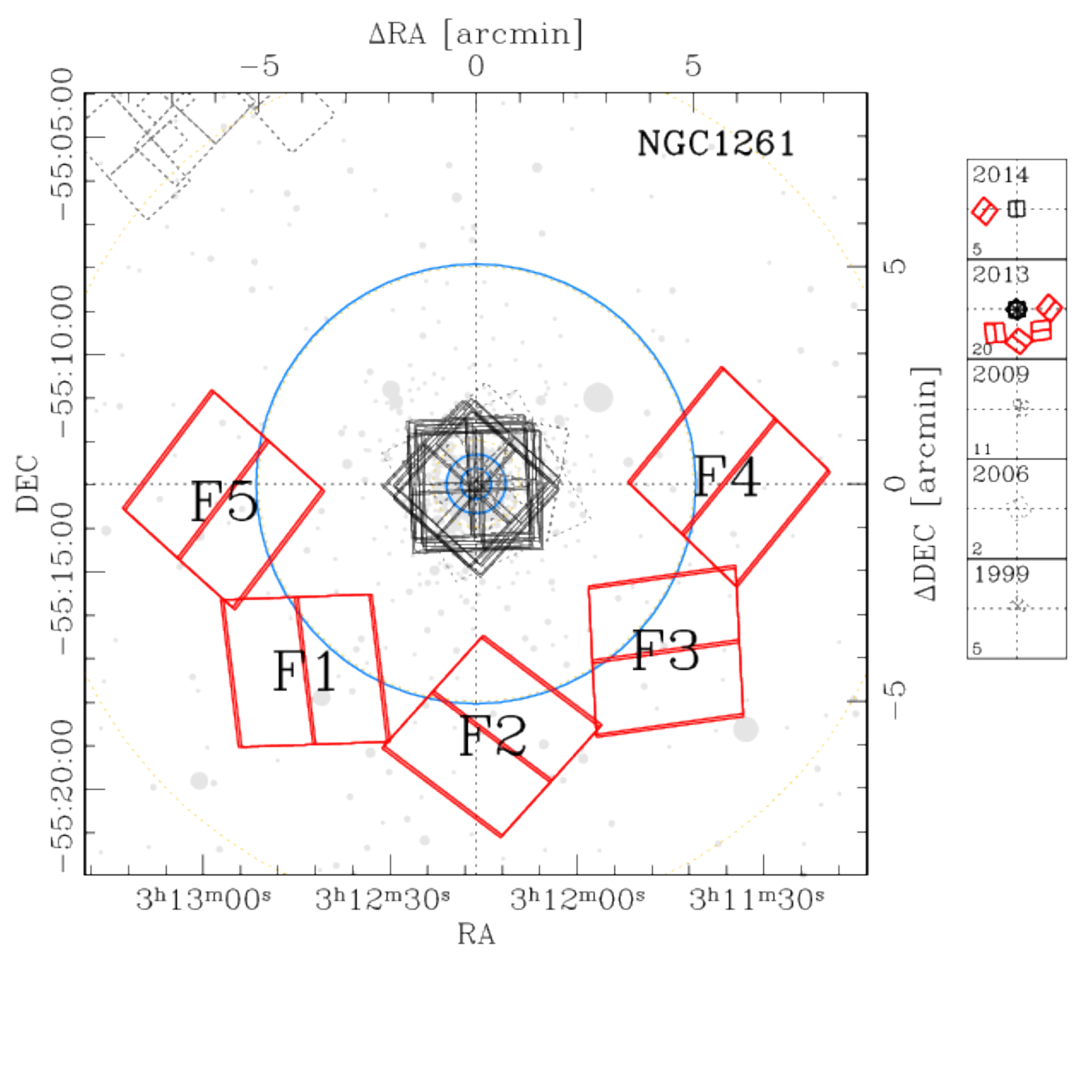}
  \includegraphics[width=8.5 cm]{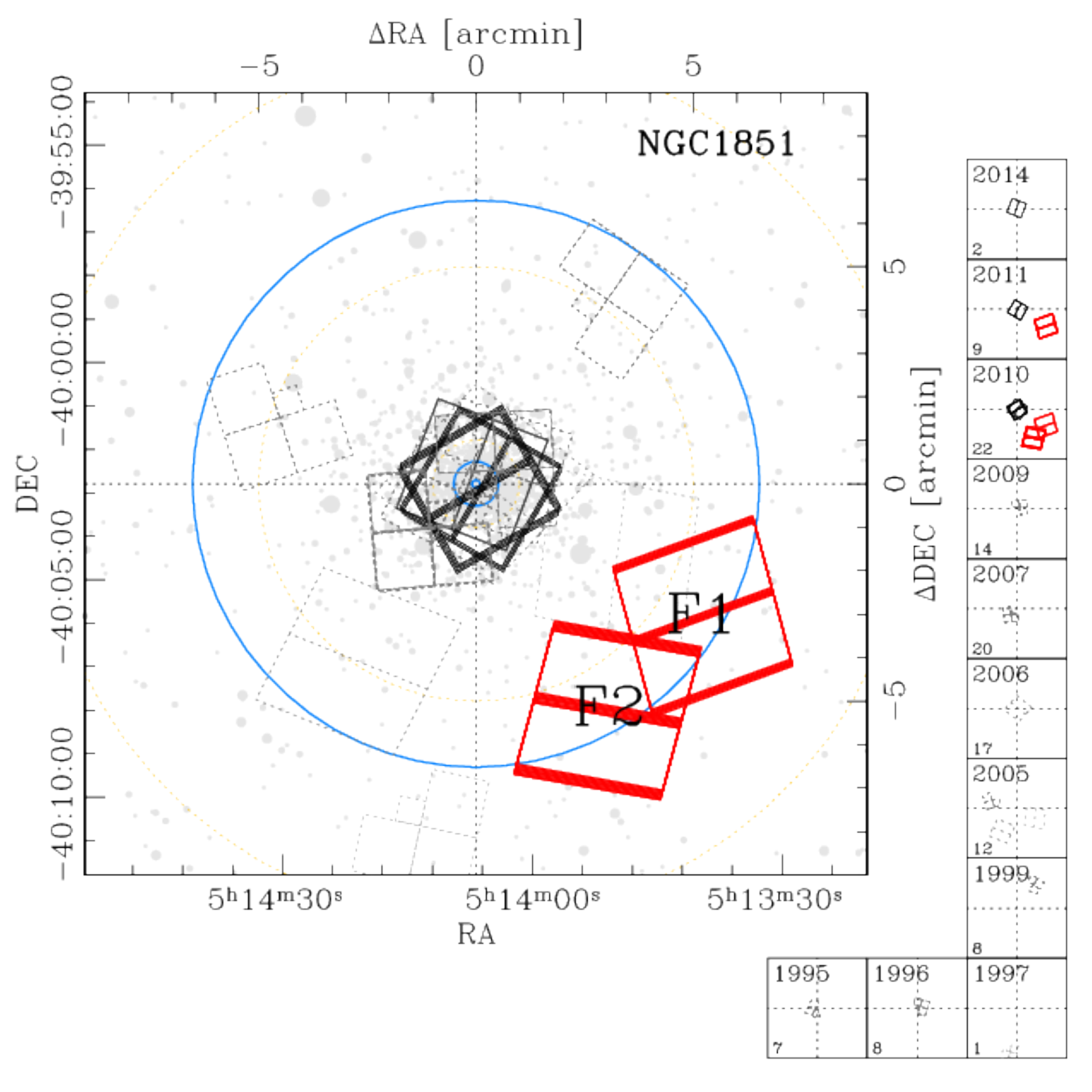}\\
  \includegraphics[width=8.5 cm]{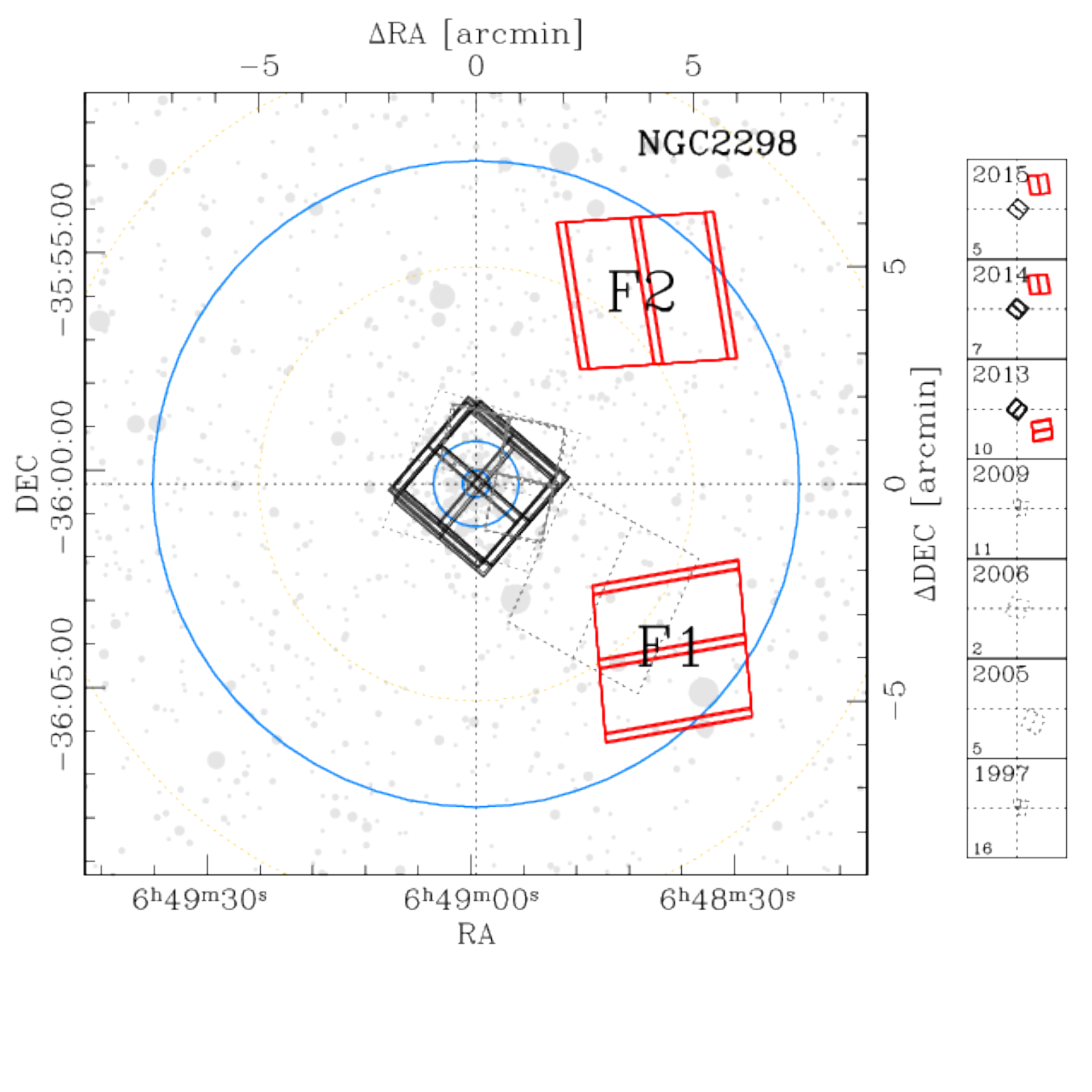}
  \includegraphics[width=8.5 cm]{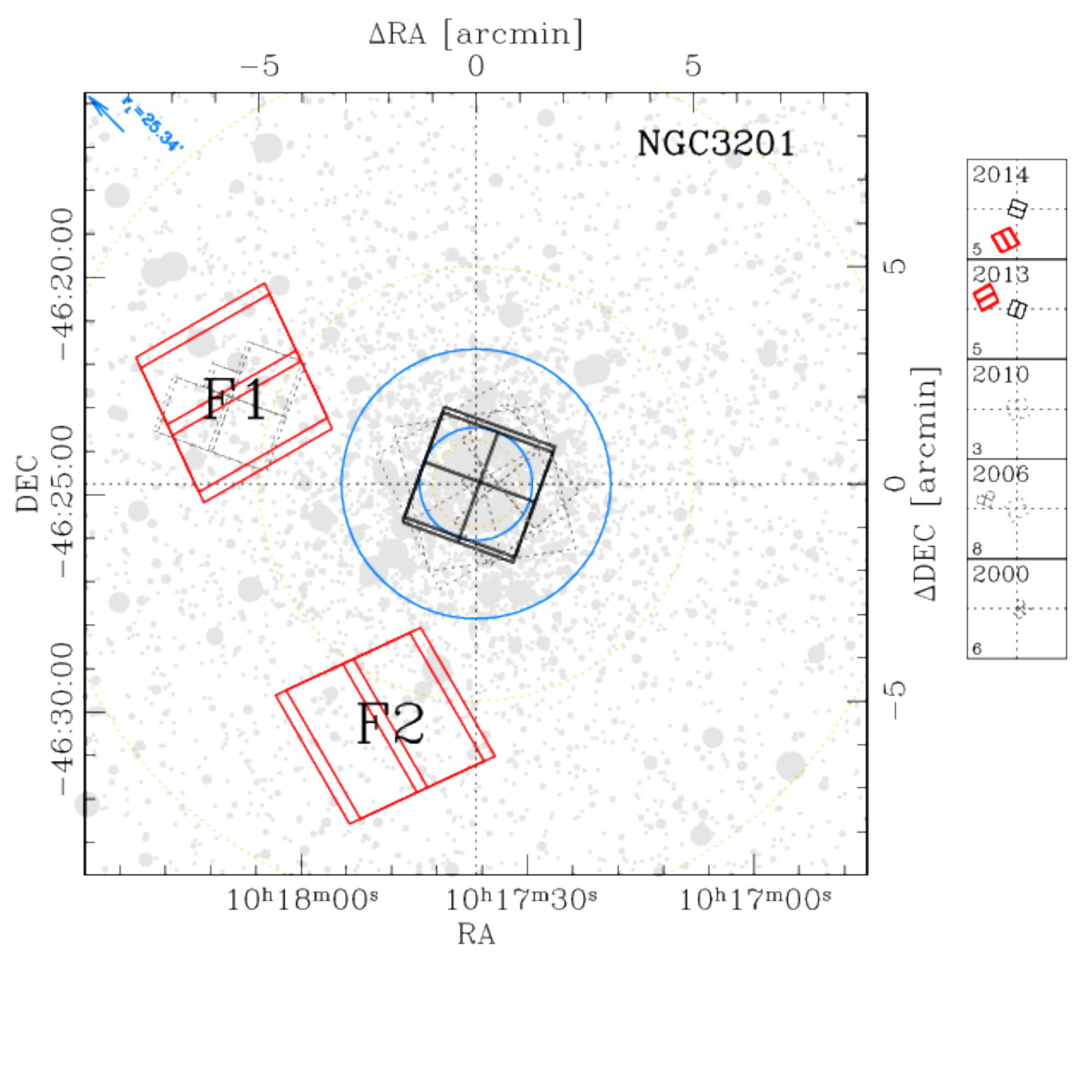}\\
  \caption{Finding charts for NGC\,1261, NGC\,1851, NGC\,2298 and NGC\,3201. Red outlines, represent {\rm ACS/WFC} parallel observations of GO-13297, black outlines refer to {\rm WFC3/UVIS} observations. Dark gray, dashed outlines represent archive {\it HST} observations in the same regions. Footprints are labeled as in Table \ref{tab:log}. Observations are subdivided by epoch in the smaller panels. 
    Blue circles mark the position of core radius, half-light radius and tidal radius for each cluster from \citet{1996AJ....112.1487H}. Where the tidal radius could not be included in the image, its value has been indicated in the upper-let corner of the image. Yellow, dashed circles mark the distances of $1$, $5$ and $10$ arcminutes from cluster centres. Gray dots corresponds to 2MASS sources, with brighter sources being larger.}
  \label{fig:crt}
\end{figure*}

The physical position of {\rm ACS/WFC} detectors in the focal plane of {\it HST} is such that its projected field of view (FoV) in the sky is located at a distance of about $6.5$ arcminutes from the centre of the {\rm WFC3} FoV.
Depending on the number of orbits allocated to each GC, from $2$ to a maximum of $5$ non-overlapping fields were observed. This is because, in order to secure a good handling of charge-transfer-efficiency (CTE) systematic errors, primary {\rm WFC3} observations were taken by applying a different telescope rotation at each orbit (\citetalias{2015AJ....149...91P}, Section 4).
For the majority of clusters, which were allocated $2$ orbits, a rotation of about $90^{\circ}$ was performed between the first and second orbit; for clusters observed for more than $2$ orbits, a minimum difference of $\sim45^{\circ}$ between the V3 PA of each orbit was required.
Five distinct pointings were obtained for 3 clusters, namely NGC\,1261, NGC\,5053 and NGC\,6101. Three distinct fields were obtained for 4 clusters: NGC\,6652, NGC\,6717, NGC\,6723 and NGC\,7089. For the other clusters, only 2 pointings were planned.
M80 is an exception and was not observed as part of the Program GO-13297. For it, we make use of archival {\it HST} data from GO-12311.   
When possible, {\rm ACS} parallel observations targeted pre-existing {\it HST} observations.

Figures \ref{fig:crt} and \ref{fig:xcrt1} -- \ref{fig:xcrt8} display all {\it HST} observations that sample the sky area in the vicinity of those covered in this survey. Three cameras onboard {\it HST} were considered in order to enhance the probability of an overlap between observations: Wide Field and Planetary Camera 2 (WFPC2), ACS and {\rm WFC3} (both UVIS and IR channels). 
Taking 2 images per orbit, one in F475W and one in F814W filters, the typical exposure times for both filters are of the order of $700 {\rm s}$.

All exposures have been corrected for CTE effects using the method described in \citet{2010PASP..122.1035A}.
Photometric measurements of stellar objects in each field have been performed using a suite of {\sc fortran} programs based on {\tt img2xym } \citep{2006acs..rept....1A} and {\tt kitchen\_sync} presented in \citet{2008AJ....135.2055A}.
The spatial variation of the PSF has been taken into account adopting a grid of $9\times10$ model PSFs distributed along each image. However focus changes/breathing of the telescope, imperfect guiding, residual noise related to CTE can produce image-to-image variations of the PSF. To mitigate these sources of systematic errors, we derived a set of spatially varying perturbations of the PSF models for each calibrated, non-drizzled ({\tt flc}) image. Adopting the procedure presented in \citet{2013ApJ...769L..32B}, each image is divided into a grid with a number of cells changing from $2 \times 2$ to $5\times5$. In each cell, a subset of well measured stars is used to locally adjust the PSF models to the stellar profiles.
Using an updated version of \citet{2006acs..rept....1A} software, in combination with the newly created PSF models, we extracted raw catalogues of stellar positions and magnitudes in each image.
We choose the grid refinement (between $2\times2$ to $5\times5$) that produces the best results, inspecting the distribution of the quality of fit parameters as a function of magnitude, and taking into account the number of reference stars used to tailor the PSF perturbations in each cell.

Each exposure related to the same field has been subsequently referred to a common reference frame.
Since we are mainly interested in the faint, red part of the cluster CMDs, we used the F814W images to construct the reference frames.
Catalogues are finally produced using a version of the software presented in \citet{2008AJ....135.2055A}, specifically tuned for this project. In particular, the {\tt kitchen\_sync} routine has been modified in order to work properly with only one image per filter. 
In addition, the method presented in \citet{2004acs..rept...17G} was applied in order to provide reliable photometry also for saturated stars.

The raw, instrumental magnitudes have been zero-pointed into {\rm ACS/WFC} Vega-mag photometric system following the prescriptions of \citet{2005MNRAS.357.1038B}. The zero-points and aperture corrections from $0.5$ arcseconds to infinity of \citet{2016AJ....152...60B} have been used.

Especially in the present case, the photometric calibration plays a critical role, and we put strong efforts to precisely evaluate zero-point differences between various observations. Crowding is not a serious problem in the outer cluster regions and high-precision photometric measurements are relatively easy to obtain.
But field-to-field zero-point variations must be accounted for to have the external fields of the same cluster on the same photometric scale.
The main source of this photometric offsets between catalogues of distinct fields, is related to PSF modelling. The PSF model by construction is normalized to a surface flux of unity within a radius of $10$ {\rm ACS/WFC} pixels ($0.5$ arcsecond).  Only the inner $5\times5$-pixel region of sources was used to fit to the PSF model in order to minimize the contaminating impact of nearby neighbors, but any mismatch between the adopted PSF model and the pixels beyond the small square fitting aperture would result in a slight zeropoint shift. The fact that we perturb the library PSFs (above) minimizes this even further. However, the best way to regularize the photometry is empirically: following \citet{2005MNRAS.357.1038B} and \citet{2008AJ....135.2055A}, we measured aperture corrections inside this aperture using the calibrated and drizzled ({\tt drc}) images as reference.

\begin{table}
   \centering
   \normalsize
   \caption{Tabulated $\Delta m_{\rm PSF-AP(0''.5)}$ values used in the aperture correction}
   \label{tab:apcor}
   \begin{tabular}{c c c l l}
   \hline
   \# & CLUSTER & FIELD & $\Delta m_{\rm PSF-AP(0''.5)}^{\rm F475W}$ & $\Delta m_{\rm PSF-AP(0''.5)}^{\rm F814W}$ \\
      &         &       & mag                                 & mag                                  \\
   \hline
   \input{./apcor_sub_hr21.log}
   \end{tabular}
   \raggedright{\it Continued on Tab. \ref{tab:apcorb}}
\end{table}
\begin{table}
   \centering
   \normalsize
   \caption{Table \ref{tab:apcor} (continued)}
   \label{tab:apcorb}
   \begin{tabular}{c c c l l}
   \hline
   \# & CLUSTER & FIELD & $\Delta m_{\rm PSF-AP(0''.5)}^{\rm F475W}$ & $\Delta m_{\rm PSF-AP(0''.5)}^{\rm F814W}$ \\
      &         &       & mag                                 & mag                                  \\
   \hline
   \input{./apcor_sub_hr22.log}
   \hline
   \end{tabular}
\end{table}
\begin{figure*}
  \includegraphics[width=\textwidth]{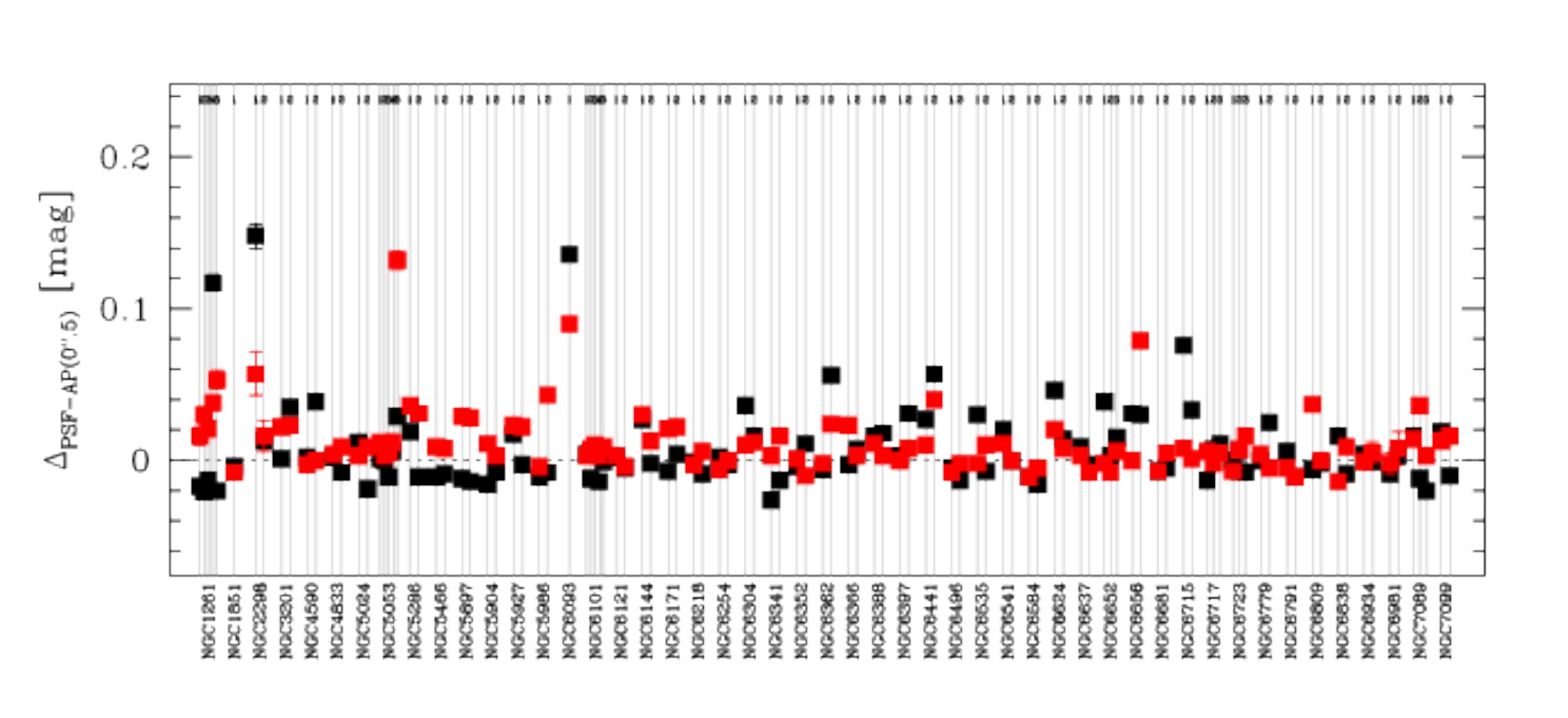}
  \caption{Aperture corrections measured in each field for each cluster. Red squares refer to F475W observations, while black squares refer to F814W ones.}
  \label{fig:cal}
\end{figure*}

These corrections were sufficient to properly take into account the majority of the photometric biases in our catalogues. 
The obtained values of the aperture correction for an aperture of $0.5$ arcseconds are listed in Tables \ref{tab:apcor} and \ref{tab:apcorb}. We maintain the same nomenclature as in \citet{2005MNRAS.357.1038B}. The aperture correction values, along with their associated uncertainties are also reported in Figure \ref{fig:cal}. Black dots are referred to F814W observations, red dots to F475W ones. It can be noted that for the majority of cases corrections are small: typically smaller than $0.04$ magnitudes, but in the most severe cases, they can reach values as high as $0.15$ mag. 

\begin{table}
  \centering
   \caption{Precision reached with the re-derived astrometric solution using {\it Gaia} catalogues; for completeness we report the number of reference stars used. We also provide measures of the accuracy of the original STScI astrometric solution in the form of differences between RA and DEC.}
   \label{tab:astro}
   \normalsize
   \begin{tabular}{p{.03cm} p{1.2cm} p{.15cm} p{.4cm} p{.95cm} p{1.7cm} p{1.7cm}} 
   \hline
   \# & CLUSTER & F & stars & precision    & $\Delta{\rm RA}$ & $\Delta{\rm DEC}$\\
      &         &   &       & mas    & WFC px           & WFC px           \\  
   \hline
   01 & NGC1261 & 1   & $  32$ & $15.5$ & $  1.66 \pm 0.34$ & $ -4.50 \pm 0.16$ \\
   &         & 2   & $  41$ & $10.0$ & $ -1.89 \pm 0.08$ & $  4.31 \pm 0.21$ \\
   &         & 3   & $  37$ & $ 8.0$ & $ -0.51 \pm 0.06$ & $  1.61 \pm 0.18$ \\
   &         & 4   & $  36$ & $11.0$ & $ -0.50 \pm 0.14$ & $  1.03 \pm 0.31$ \\
   &         & 5   & $  46$ & $10.0$ & $  0.07 \pm 0.05$ & $ -4.81 \pm 0.19$ \\
02 & NGC1851 & 1+2 & $ 232$ & $10.0$ & $  3.54 \pm 0.20$ & $ -4.25 \pm 0.19$ \\
03 & NGC2298 & 1   & $  79$ & $16.5$ & $ -0.17 \pm 0.04$ & $  5.62 \pm 0.24$ \\
   &         & 2   & $  53$ & $14.0$ & $ -0.11 \pm 0.06$ & $ -1.37 \pm 0.25$ \\
04 & NGC3201 & 1   & $ 635$ & $13.0$ & $  0.14 \pm 0.32$ & $ -0.57 \pm 0.23$ \\
   &         & 2   & $ 653$ & $10.0$ & $  4.33 \pm 0.33$ & $  8.39 \pm 0.20$ \\
05 & NGC4590 & 1   & $ 110$ & $11.0$ & $ -0.78 \pm 0.10$ & $  4.90 \pm 0.15$ \\
   &         & 2   & $ 111$ & $11.5$ & $ -0.12 \pm 0.04$ & $ -3.86 \pm 0.17$ \\
06 & NGC4833 & 1   & $ 492$ & $ 7.0$ & $ -3.53 \pm 0.35$ & $  6.86 \pm 0.22$ \\
   &         & 2   & $ 570$ & $ 6.5$ & $  1.75 \pm 0.15$ & $ -0.02 \pm 0.20$ \\
07 & NGC5024 & 1   & $ 173$ & $11.5$ & $ -0.14 \pm 0.23$ & $  3.09 \pm 0.20$ \\
   &         & 2   & $ 142$ & $10.0$ & $-45.45 \pm 0.77$ & $  1.35 \pm 0.18$ \\
08 & NGC5053 & 1   & $  44$ & $ 9.5$ & $  0.72 \pm 0.16$ & $ -0.62 \pm 0.16$ \\
   &         & 2   & $  41$ & $ 8.0$ & $ -0.07 \pm 0.15$ & $ -1.86 \pm 0.18$ \\
   &         & 3   & $  38$ & $ 8.5$ & $  0.22 \pm 0.13$ & $  0.29 \pm 0.23$ \\
   &         & 4   & $  42$ & $10.0$ & $  0.43 \pm 0.09$ & $  0.23 \pm 0.19$ \\
   &         & 5   & $  52$ & $12.0$ & $  0.57 \pm 0.16$ & $ -0.65 \pm 0.16$ \\
09 & NGC5286 & 1   & $ 215$ & $17.0$ & $  1.63 \pm 0.20$ & $  1.85 \pm 0.25$ \\
   &         & 2   & $ 217$ & $16.5$ & $  1.61 \pm 0.20$ & $  1.10 \pm 0.21$ \\
10 & NGC5466 & 1   & $  56$ & $ 7.5$ & $ -0.13 \pm 0.27$ & $  1.37 \pm 0.12$ \\
   &         & 2   & $  72$ & $ 8.0$ & $ -0.79 \pm 0.29$ & $ -2.22 \pm 0.20$ \\
11 & NGC5897 & 1   & $ 164$ & $ 9.0$ & $  0.12 \pm 0.22$ & $ -0.37 \pm 0.20$ \\
   &         & 2   & $ 154$ & $ 7.5$ & $  2.58 \pm 0.17$ & $ -2.51 \pm 0.18$ \\
12 & NGC5904 & 1   & $ 579$ & $ 7.0$ & $ -1.29 \pm 0.19$ & $ -2.69 \pm 0.15$ \\
   &         & 2   & $ 630$ & $ 7.0$ & $ -0.16 \pm 0.15$ & $ -1.08 \pm 0.22$ \\
13 & NGC5927 & 1   & $ 848$ & $ 8.5$ & $  3.65 \pm 0.30$ & $ -0.38 \pm 0.20$ \\
   &         & 2   & $ 985$ & $ 8.0$ & $ -9.20 \pm 0.35$ & $ 11.91 \pm 0.21$ \\
14 & NGC5986 & 1   & $ 278$ & $ 8.0$ & $ -0.54 \pm 0.30$ & $  1.36 \pm 0.22$ \\
   &         & 2   & $ 268$ & $ 6.0$ & $ -6.46 \pm 0.26$ & $  4.12 \pm 0.18$ \\
15 & NGC6093 & 1   & $ 236$ & $22.0$ & $  1.81 \pm 0.19$ & $ -4.47 \pm 0.20$ \\
16 & NGC6101 & 1   & $ 235$ & $ 7.0$ & $ -1.85 \pm 0.76$ & $  1.09 \pm 0.21$ \\
   &         & 2   & $ 232$ & $ 6.5$ & $ -0.80 \pm 0.61$ & $  2.90 \pm 0.20$ \\
   &         & 3   & $ 229$ & $ 8.0$ & $-15.82 \pm 0.73$ & $  0.55 \pm 0.25$ \\
   &         & 4   & $ 263$ & $11.0$ & $ -1.83 \pm 0.89$ & $ -0.52 \pm 0.27$ \\
   &         & 5   & $ 204$ & $ 8.5$ & $ 12.13 \pm 0.85$ & $ -3.52 \pm 0.24$ \\
17 & NGC6121 & 1   & $ 633$ & $ 7.5$ & $ -4.81 \pm 0.37$ & $  7.07 \pm 0.23$ \\
   &         & 2   & $ 649$ & $ 7.5$ & $ -0.69 \pm 0.14$ & $  1.90 \pm 0.22$ \\
18 & NGC6144 & 1   & $ 147$ & $ 7.0$ & $ -4.48 \pm 0.18$ & $ -0.10 \pm 0.19$ \\
   &         & 2   & $ 147$ & $ 9.5$ & $ -3.09 \pm 0.15$ & $ -0.76 \pm 0.20$ \\
19 & NGC6171 & 1   & $ 172$ & $ 9.0$ & $  4.44 \pm 0.20$ & $ -2.36 \pm 0.16$ \\
   &         & 2   & $ 196$ & $ 7.5$ & $ -3.26 \pm 0.19$ & $ -1.51 \pm 0.21$ \\
20 & NGC6218 & 1   & $ 300$ & $10.0$ & $  0.40 \pm 0.07$ & $ -4.05 \pm 0.17$ \\
   &         & 2   & $ 362$ & $ 6.5$ & $ -0.72 \pm 0.09$ & $  0.13 \pm 0.19$ \\
21 & NGC6254 & 1   & $ 525$ & $ 7.5$ & $ -0.94 \pm 0.11$ & $  2.02 \pm 0.20$ \\
   &         & 2   & $ 507$ & $ 5.0$ & $ -0.08 \pm 0.12$ & $ -0.52 \pm 0.18$ \\
22 & NGC6304 & 1   & $1021$ & $11.0$ & $  0.00 \pm 0.09$ & $  7.29 \pm 0.22$ \\
   &         & 2   & $1137$ & $11.0$ & $ -2.63 \pm 0.15$ & $  3.66 \pm 0.28$ \\
23 & NGC6341 & 1   & $ 272$ & $ 7.0$ & $  3.31 \pm 0.13$ & $ -1.46 \pm 0.17$ \\
   &         & 2   & $ 276$ & $ 6.5$ & $  1.30 \pm 0.16$ & $  2.01 \pm 0.14$ \\

   \end{tabular}
   \normalsize
   \raggedright{\it continued on Table \ref{tab:astrob}}
\end{table}
\begin{table}
   \centering
   \caption{Table \ref{tab:astro} (continued)}
   \label{tab:astrob}
   \normalsize
   \begin{tabular}{p{.03cm} p{1.2cm} p{.05cm} p{.4cm} p{.95cm} p{1.7cm} p{1.7cm}} 
   \hline
   \# & CLUSTER & F & stars & precision    & $\Delta{\rm RA}$ & $\Delta{\rm DEC}$\\
      &         &   &       & mas    & WFC px           & WFC px           \\
   \hline
   24 & NGC6352 & 1   & $ 676$ & $ 8.0$ & $  0.44 \pm 0.09$ & $  6.24 \pm 0.18$ \\
   &         & 2   & $ 658$ & $ 9.5$ & $  0.20 \pm 0.11$ & $  0.84 \pm 0.21$ \\
25 & NGC6362 & 1   & $ 399$ & $ 6.0$ & $ -1.98 \pm 0.18$ & $ -1.19 \pm 0.20$ \\
   &         & 2   & $ 412$ & $ 6.0$ & $ -0.15 \pm 0.23$ & $  2.51 \pm 0.17$ \\
26 & NGC6366 & 1   & $ 265$ & $ 7.0$ & $  1.07 \pm 0.11$ & $ -4.03 \pm 0.22$ \\
   &         & 2   & $ 263$ & $ 6.0$ & $  2.89 \pm 0.19$ & $  4.66 \pm 0.18$ \\
27 & NGC6388 & 1   & $ 812$ & $ 6.5$ & $  1.68 \pm 0.24$ & $  4.01 \pm 0.26$ \\
   &         & 2   & $ 873$ & $ 7.0$ & $ -0.86 \pm 0.22$ & $  5.67 \pm 0.22$ \\	
28 & NGC6397 & 1   & $ 605$ & $ 8.0$ & $ 16.46 \pm 0.47$ & $ 12.42 \pm 0.26$ \\
   &         & 2   & $ 602$ & $ 7.0$ & $ -4.15 \pm 0.33$ & $  4.79 \pm 0.30$ \\
29 & NGC6441 & 1   & $1075$ & $10.0$ & $ 34.98 \pm 0.54$ & $ -3.77 \pm 0.26$ \\
   &         & 2   & $1114$ & $ 9.5$ & $  4.46 \pm 0.31$ & $  3.80 \pm 0.27$ \\
30 & NGC6496 & 1   & $ 482$ & $ 6.0$ & $  1.85 \pm 0.30$ & $  7.03 \pm 0.21$ \\
   &         & 2   & $ 525$ & $ 9.0$ & $  8.00 \pm 0.28$ & $  1.65 \pm 0.22$ \\
31 & NGC6535 & 1   & $ 304$ & $ 6.5$ & $ -4.97 \pm 0.19$ & $  0.75 \pm 0.18$ \\
   &         & 2   & $ 275$ & $ 7.5$ & $ -1.78 \pm 0.18$ & $ -9.16 \pm 0.20$ \\
32 & NGC6541 & 1   & $ 385$ & $ 7.5$ & $-11.12 \pm 0.28$ & $  7.28 \pm 0.23$ \\
   &         & 2   & $ 481$ & $ 9.0$ & $ -7.23 \pm 0.32$ & $  7.72 \pm 0.22$ \\
33 & NGC6584 & 1   & $ 119$ & $ 7.5$ & $  1.82 \pm 0.19$ & $  6.15 \pm 0.24$ \\
   &         & 2   & $ 141$ & $10.0$ & $  2.67 \pm 0.14$ & $  2.71 \pm 0.25$ \\
34 & NGC6624 & 1   & $ 360$ & $ 8.5$ & $  0.83 \pm 0.12$ & $  1.93 \pm 0.23$ \\
   &         & 2   & $ 774$ & $ 9.5$ & $  0.61 \pm 0.15$ & $  6.08 \pm 0.21$ \\
35 & NGC6637 & 1   & $ 620$ & $10.0$ & $ -3.66 \pm 0.20$ & $  1.83 \pm 0.25$ \\
   &         & 2   & $ 615$ & $11.5$ & $ -2.54 \pm 0.23$ & $ -0.80 \pm 0.29$ \\
36 & NGC6652 & 1   & $ 608$ & $12.0$ & $  0.35 \pm 0.09$ & $  5.17 \pm 0.26$ \\
   &         & 2   & $ 624$ & $ 9.5$ & $ -0.06 \pm 0.06$ & $  2.40 \pm 0.25$ \\
   &         & 3   & $ 682$ & $ 7.0$ & $  0.49 \pm 0.15$ & $  5.78 \pm 0.21$ \\
37 & NGC6656 & 1   & $ 734$ & $31.0$ & $ -1.20 \pm 0.21$ & $  3.03 \pm 0.25$ \\
   &         & 2   & $1052$ & $27.5$ & $  1.21 \pm 0.30$ & $  6.93 \pm 0.31$ \\
38 & NGC6681 & 1   & $ 450$ & $14.0$ & $ -4.20 \pm 0.25$ & $  1.76 \pm 0.29$ \\
   &         & 2   & $ 385$ & $15.5$ & $ -2.37 \pm 0.20$ & $  1.75 \pm 0.32$ \\
39 & NGC6715 & 1   & $ 484$ & $16.0$ & $ -0.69 \pm 0.25$ & $ -4.67 \pm 0.28$ \\
   &         & 2   & $ 505$ & $15.5$ & $  3.28 \pm 0.18$ & $ -1.79 \pm 0.27$ \\
40 & NGC6717 & 1   & $ 439$ & $14.0$ & $  3.48 \pm 0.21$ & $  0.07 \pm 0.30$ \\
   &         & 2   & $ 406$ & $13.0$ & $  0.49 \pm 0.20$ & $  1.11 \pm 0.28$ \\
   &         & 3   & $ 445$ & $12.0$ & $  0.11 \pm 0.18$ & $  4.99 \pm 0.26$ \\
41 & NGC6723 & 1   & $ 256$ & $14.5$ & $  0.14 \pm 0.20$ & $  0.95 \pm 0.30$ \\
   &         & 2   & $ 144$ & $11.5$ & $ -0.25 \pm 0.19$ & $  4.59 \pm 0.22$ \\
   &         & 3   & $ 111$ & $ 9.0$ & $ -0.33 \pm 0.16$ & $  3.99 \pm 0.18$ \\
42 & NGC6779 & 1   & $ 345$ & $ 7.5$ & $  0.86 \pm 0.07$ & $  1.07 \pm 0.21$ \\
   &         & 2   & $ 333$ & $ 6.5$ & $  3.24 \pm 0.16$ & $ -8.92 \pm 0.19$ \\
43 & NGC6791 & 1   & $ 288$ & $ 9.0$ & $ -1.92 \pm 0.26$ & $ -7.50 \pm 0.21$ \\
   &         & 2   & $ 274$ & $ 5.5$ & $ -4.98 \pm 0.25$ & $ -6.18 \pm 0.21$ \\
44 & NGC6809 & 1   & $ 584$ & $ 7.0$ & $ -1.58 \pm 0.19$ & $  0.99 \pm 0.22$ \\
   &         & 2   & $ 590$ & $ 7.5$ & $ -9.19 \pm 0.21$ & $ -0.68 \pm 0.23$ \\
45 & NGC6838 & 1   & $ 816$ & $ 7.0$ & $ -8.81 \pm 0.24$ & $-12.24 \pm 0.22$ \\
   &         & 2   & $ 800$ & $ 6.5$ & $ -3.96 \pm 0.22$ & $ -8.28 \pm 0.21$ \\
46 & NGC6934 & 1   & $  87$ & $11.5$ & $  3.41 \pm 0.20$ & $  2.62 \pm 0.24$ \\
   &         & 2   & $  84$ & $ 5.5$ & $  1.41 \pm 0.10$ & $-12.00 \pm 0.17$ \\
47 & NGC6981 & 1   & $  31$ & $14.5$ & $  1.29 \pm 0.26$ & $ -9.37 \pm 0.19$ \\
   &         & 2   & $  27$ & $ 9.5$ & $  1.73 \pm 0.23$ & $ -8.07 \pm 0.23$ \\
48 & NGC7089 & 1   & $  92$ & $ 5.5$ & $ -0.55 \pm 0.12$ & $ -6.84 \pm 0.15$ \\
   &         & 2   & $  77$ & $10.5$ & $  2.52 \pm 0.15$ & $ -1.40 \pm 0.18$ \\
   &         & 3   & $  82$ & $11.0$ & $  1.85 \pm 0.12$ & $ -0.54 \pm 0.20$ \\
49 & NGC7099 & 1   & $  79$ & $ 4.5$ & $  0.93 \pm 0.08$ & $ -0.89 \pm 0.17$ \\
   &         & 2   & $  89$ & $ 6.5$ & $  0.66 \pm 0.08$ & $  1.75 \pm 0.21$ \\

   \hline
   \end{tabular}
\end{table}

Astrometrised, stacked images of each observed field have also been produced for both filters with a $1\times1$ pixel sampling. These have been created, for each field, combining all overlapping {\tt flc} images using the same coordinate transformations that define the common reference frames.

\begin{figure*}
  \includegraphics[width=\textwidth]{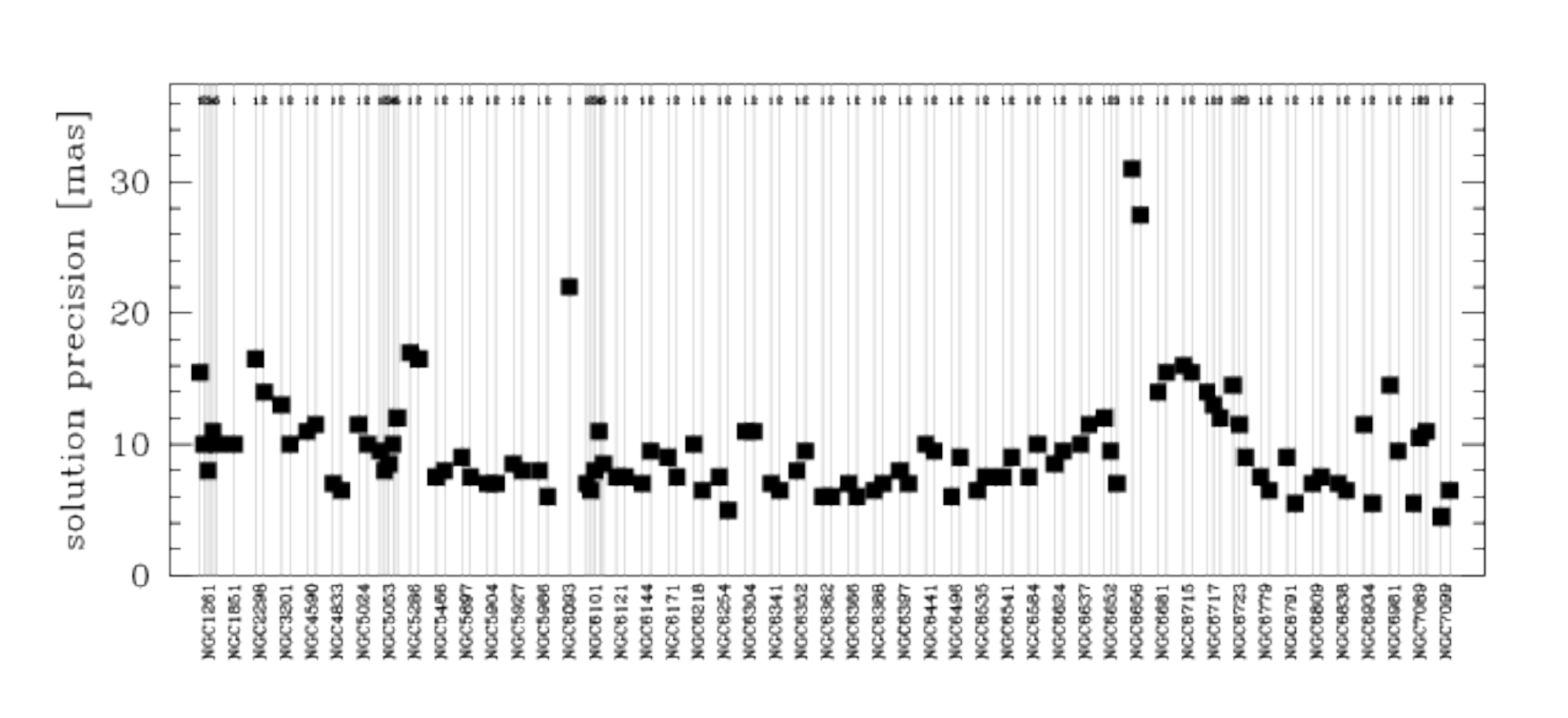}
  \caption{Precision of the redefined astrometric solution for each observed field, see text for details.}
  \label{fig:solpx}
\end{figure*}
Astrometric solutions have been independently derived using the {\it Gaia} DR1 catalogue as a reference \citep{2016A&A...595A...2G}. As a consequence, positions are given for Equinox J2000 at epoch 2015. Table \ref{tab:astro} reports the precision reached by the new astrometric solution in the fifth column, which is the root mean square error of the offset between {\it Gaia} positions and those derived, for the same stars, in our astrometrised stacked images. The measured average value is $0.2$ pixel, or $\sim10$ milliarcseconds. These values are also visualised in Figure \ref{fig:solpx}.

\begin{figure}
  \includegraphics[width=\columnwidth]{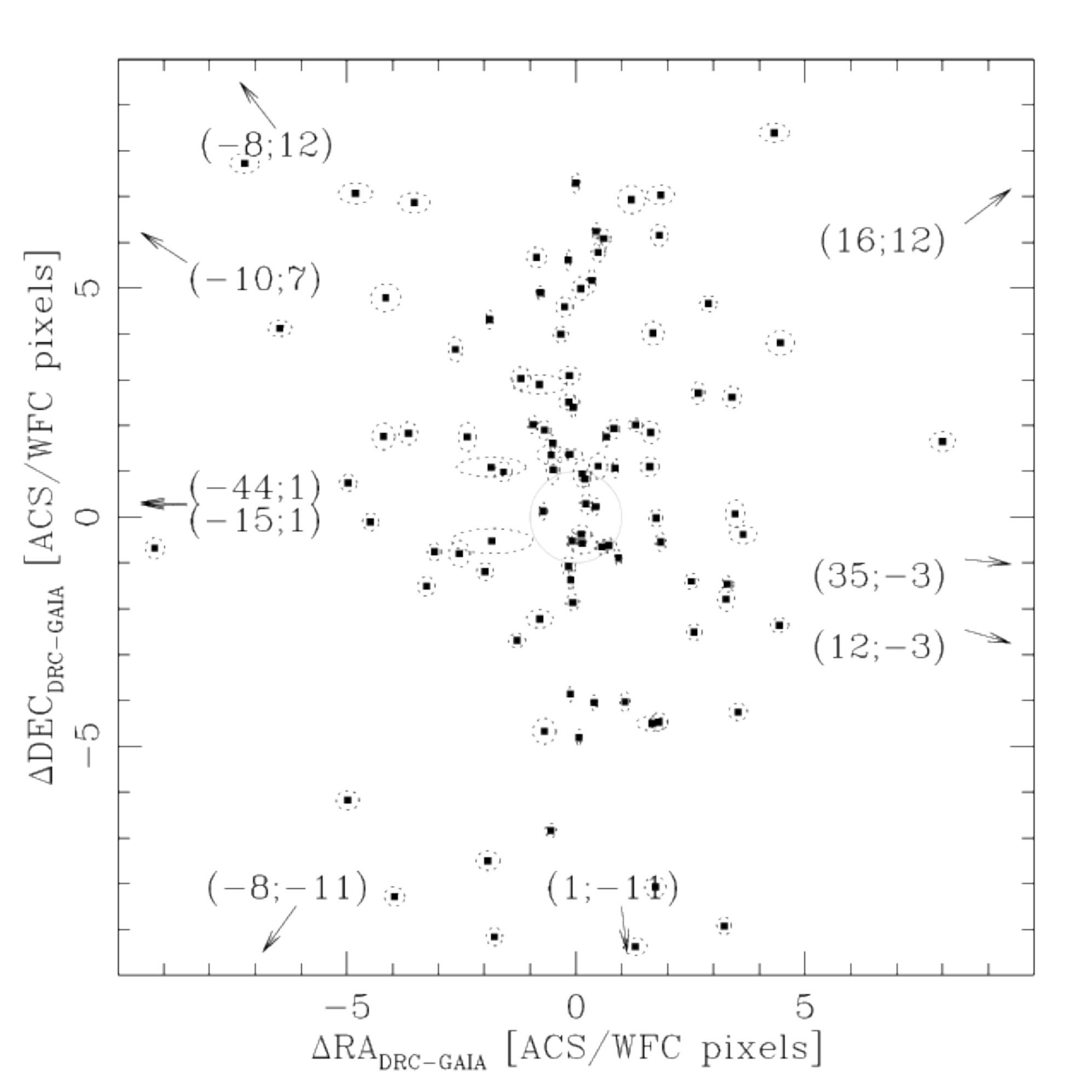}
  \caption{Offsets between star positions obtained using the original STScI astrometric solution (header of {\tt drc} images) and {\it Gaia} ones. Points represent single fields; the semi-axes of each ellipse have length equal to the measured standard deviation. The gray circle represents an offset of $1$ {\rm ACS/WFC} pixel, which corresponds to $\sim 0.05$ arcseconds. Some points falls outside the used limits; we indicate their position along the edges of the figure, the arrows point in their direction.}
  \label{fig:dradec}
\end{figure}
For completeness, we also measure the astrometric precision of the original astrometric solution of {\tt drc} images. The position offset between common sources in the {\it Gaia} DR1 catalogue and the {\tt drc} images is used to define this quantity. Offset values are referred to RA and DEC distances in image pixels and are reported in columns six and seven of Table \ref{tab:astro}. The associated errors corresponds to the measured standard deviations of each sample. A visual representation is also given in Figure \ref{fig:dradec}. It can be noted that, in general, offsets are lower than $5$ {\rm ACS/WFC} pixels ($0.25$ arcseconds), and, in many cases, below $1$ pixel, with some notable exceptions. 

\section{The Colour-Magnitude Diagrams and trichromatic stacked images}\label{sec:cmd}

\begin{figure*}
  \includegraphics[height=22 cm]{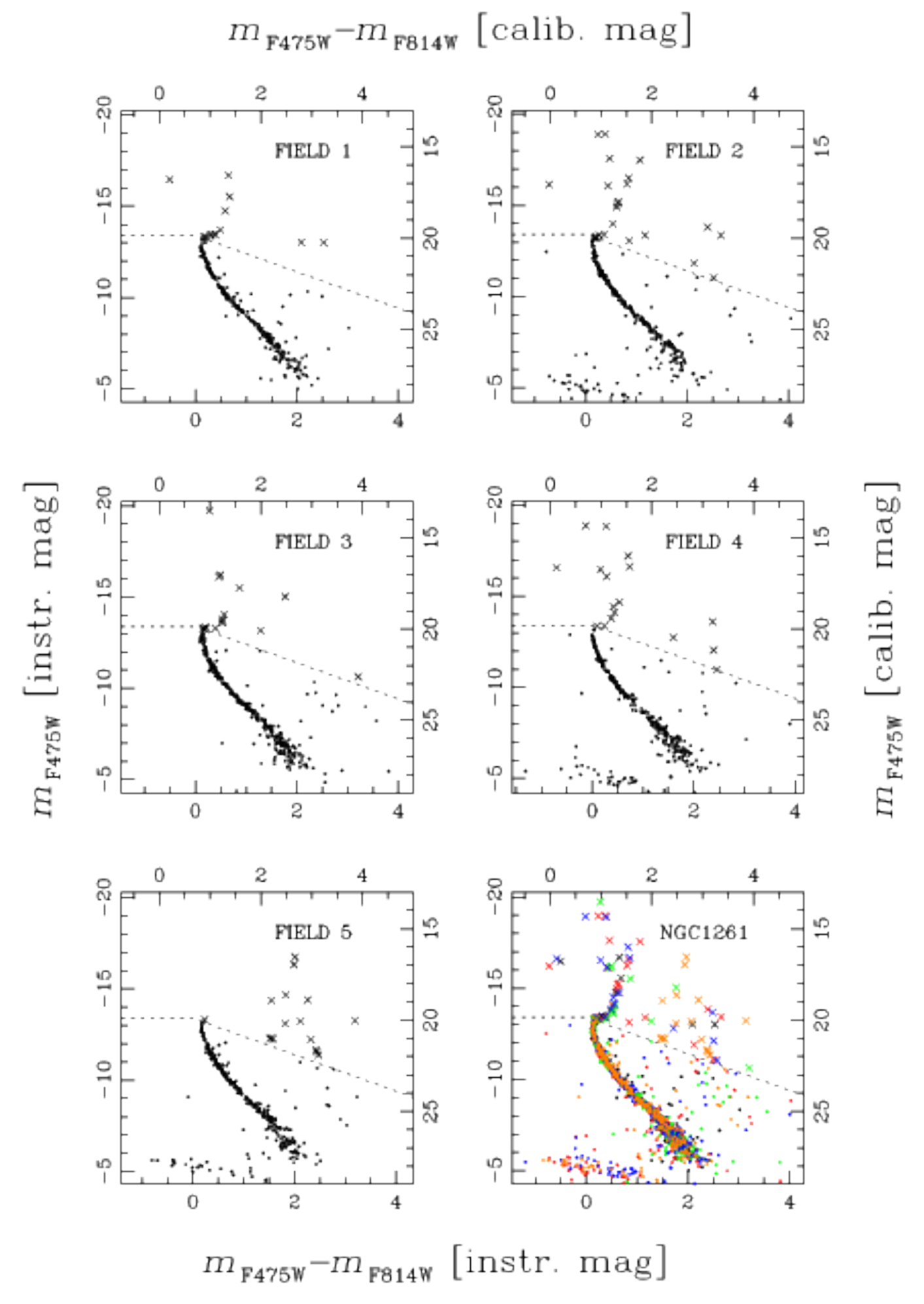}
  \caption{CMDs of the parallel fields of NGC\,1261. Both instrumental and calibrated magnitude scales are shown. Dashed lines represent the saturation levels. Saturated stars are marked with crosses.
    The bottom-right panel collects all the stars present in all the five fields. Stars are here colour-coded as follows. Black dots represent stars measured in F1, red dots stars of F2, green dots stars of F3, blue dots stars of F4, and orange dots represent stars measured in F5. The instrumental magnitude scale and the saturation limit refer to F1 only.}
  \label{fig:cmd}
\end{figure*}
In Figure \ref{fig:cmd} we report the CMDs obtained for all 5 fields of NGC\,1261. Magnitudes are given both in the instrumental and Vega-mag photometric systems. Dashed lines represent the saturation limit, black dots represent unsaturated stars, saturated stars are represented with crosses. The five CMDs are all merged in the bottom-right panel, where a different colour have been assigned to each different field. In this case, the instrumental magnitude scale, along with saturation level (dashed line), refer only to Field 1.
      
The final CMDs for all other targets are presented in Figures \ref{fig:xcmd1} - \ref{fig:xcmd8}.
The presented CMDs have been obtained by selecting only high-quality stars, according to quality parameters presented in \citet{2008AJ....135.2055A}. In Section \ref{sec:ex} we describe the adopted selection procedure.
It is important to mention here that, since in the majority of the cases only one image per filter has been taken, artifact rejection is not an easy task.
In order to include faint sources, we have chosen not to limit in flux our raw catalogues. We nonetheless restricted the detections only to those sources observed in at least one F814W and one F475W image simultaneously, with positions in the common reference frame consistent within $0.8$ pixels.
  
No rejection of foreground/background contamination has been performed, nor any correction for differential reddening. The homogeneity of the data guarantees very similar results in every case. Nonetheless, it is out of the scope of the present work to characterize in detail the obtained results, which require taking into consideration several issues. For example, the number density of cluster members present in the observed fields depends on the properties of each GC: in some cases, especially for bulge clusters, or those that appear projected in this dense Galactic region, it is difficult to identify the cluster sequence. 
We recall that another interesting application of the present data is the study of the stellar populations, external to the clusters, that contaminate the observed fields.
In particular, for at least 6 GCs, namely NGC\,6624, NGC\,6637, NGC\,6652, NGC\,6681, NGC\,6715 and NGC\,6809, traces of the Sagittarius Stream are visible in the obtained CMDs \citep{2011ApJ...743...20S}.
  
\begin{figure*}
  \includegraphics[width=\textwidth]{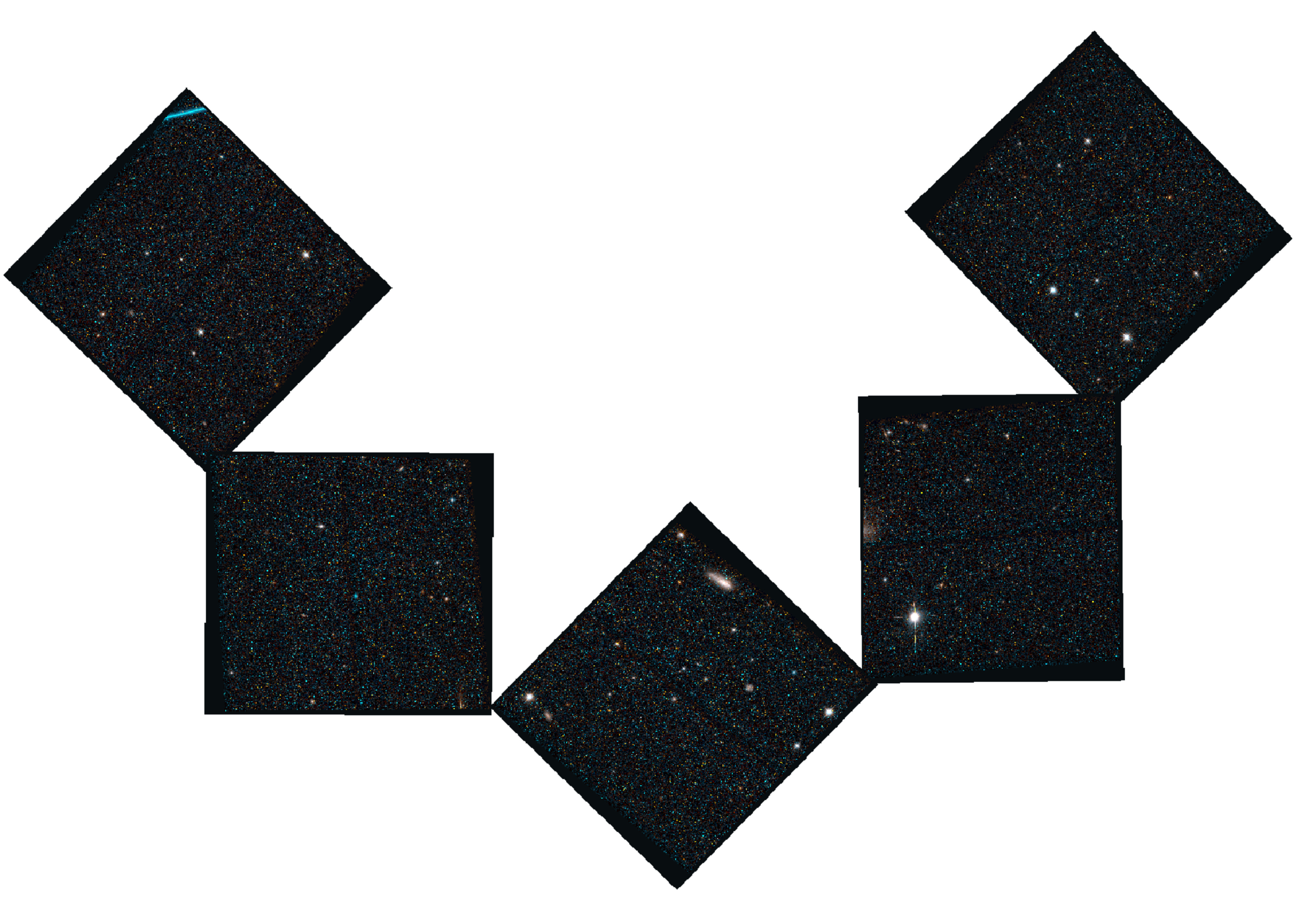}
  \caption{Mosaic of all trichromatic images created for NGC\,1261; see text for details.}
  \label{fig:stack}
\end{figure*}
As an additional tool to explore and characterize observations, we have created colour images of each observed field, combining the astrometrised stacked images: an example is shown in Figure \ref{fig:stack}.
The F814W stacked images have been associated with red channel while the F475W images are associated to the blue channel. The images associated to the green channel have been obtained as a result of a 3:1 weighted mean of F475W and F814W counts respectively.

\section{Selection of well-measured stars}\label{sec:ex}

This section describes the procedure adopted to reject spurious or poorly measured sources in the catalogues and to obtain a sample of bona-fide stellar sources. In this example we refer to the catalogue associated to Field 1 of NGC\,6121 (M4). The V3 PA is 272 degrees and, for this field, 2 images were collected, one in F814W, and the other in F475W (666 and 739 seconds respectively).

\begin{figure*}
  \includegraphics[width=\textwidth]{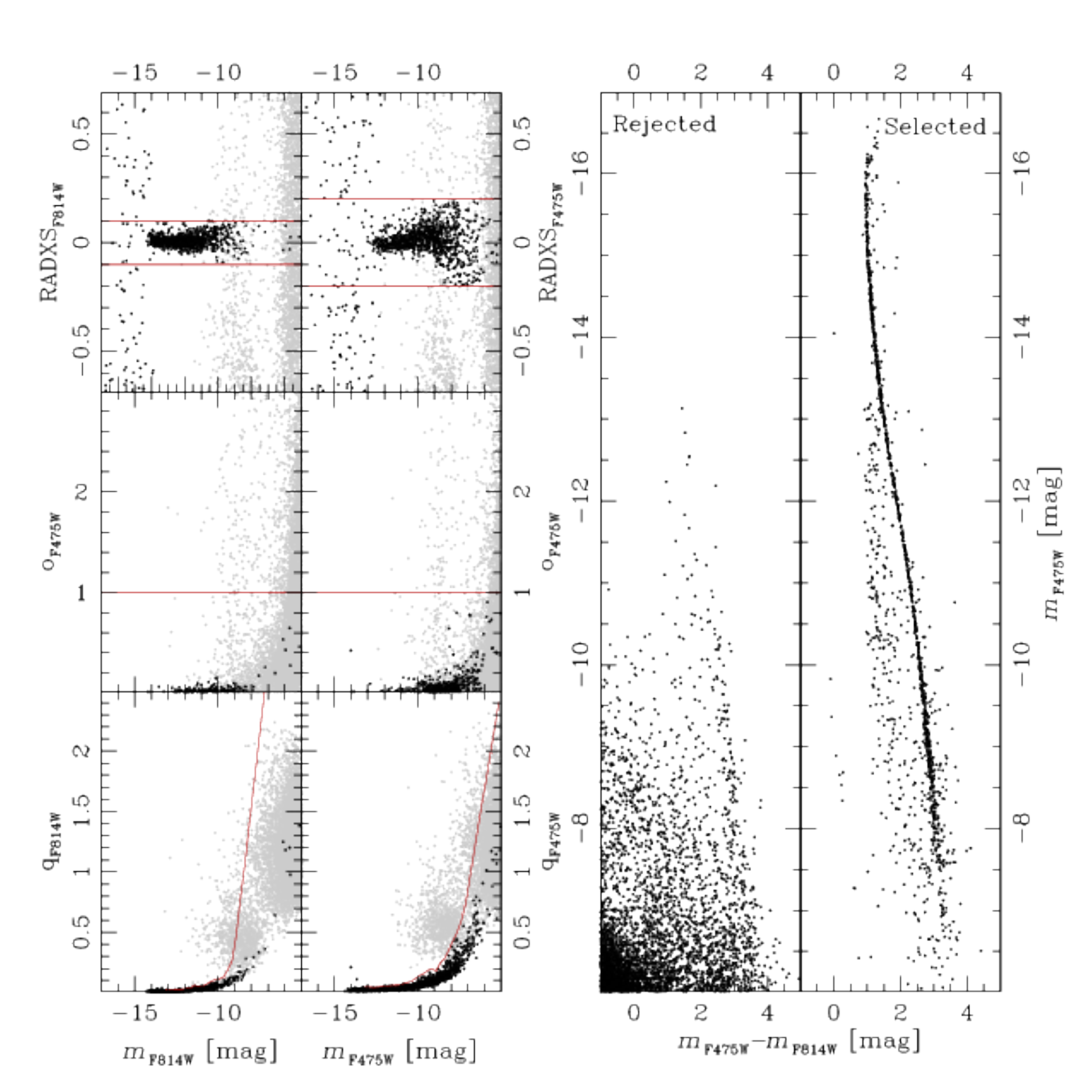}
  \caption{Selection procedure for Field 1 of NGC\,6121. Left panels: black points mark the stars that were selected as well-measured. Red lines mark the limits chosen for the {\it q}, {\it o} and {\it RADXS} parameters. Second panel from right: CMD of rejected sources. Right panel: CMD with only sources that survived the selection. All plots refer to instrumental magnitudes.}
  \label{fig:sel}
\end{figure*}

For the selection, we have adopted a procedure similar to that described in \citet{2012A&A...540A..16M}, defining limits in both {\it q} and {\it o} (quality) parameters. In addition, we also made use of the {\it RADXS} parameter (\citealt{2008ApJ...678.1279B}, \citealt{2009ApJ...697..965B}, \citealt{2010ApJ...708L..32B}).

  As shown in the lower left panels of Figure \ref{fig:sel}, the {\it q} parameter displays a characteristic trend with magnitude. This parameter is defined as the absolute value of the subtraction between the PSF model and the spatial distribution of light of a particular detection in the image, inside the fitting radius. For a perfectly modeled source, the {\it q} parameter assumes value $0$.

The {\it o} parameter quantifies the amount of light that falls on the aperture used for PSF fitting, due to neighboring sources \citep{2008AJ....135.2055A}. Unlike the {\it q} parameter, it does not shows a clear trend with magnitude (middle-left panels of Figure \ref{fig:sel}), for this reason we have used a fixed limit \citep{2012A&A...540A..16M}.
  
Finally, the {\it RADXS} parameter is related to the spatial extent of the sources, and it is used to distinguish between point sources and extended sources. It is defined as the flux in excess of that predicted from the PSF fitting just outside the core of each source \citep{2008ApJ...678.1279B}. Positive values are expected for extended objects, while negative values indicate detections that are sharper than stellar.
The introduction of this parameter in the selection process for the present case is particularly necessary: since only one image per filter is available for most cases, spurious detections due to cosmic rays are present in the catalogues. Moreover, these observations cover the external regions of GCs, where the stellar density is not as high as in the central regions. The observed fields are thus relatively populated by extra-galactic, non-stellar objects.

We started by taking a more stringent limit in position-consistency for each source. We selected only sources with an rms error in position less than $0.3$ pixel. We adopted fixed limits in the {\it o} and {\it RADXS} parameters for both filters simultaneously. Finally, we measured, for the sources that survived this first selection, the median trend in the plane defined by magnitude and {\it q} parameter values. In this way, we removed from the sample non-stellar and poorly-measured sources, while not a priori losing faint sources.
Red lines in the left panels of Figure \ref{fig:sel} represent the limits used for the case of Field 1 of NGC\,6121 (M4). Black points represent sources that passed the selection process. The CMD corresponding to rejected sources (grey dots in the left panels) is shown in the second panel from the right. 

The resulting CMD is shown in the right panel of Figure \ref{fig:sel}. Note how the cleaning process allows the clear detection of the white-dwarf cooling sequence.
Saturated stars have been excluded from the selection procedure, but their fluxes have been recovered using the procedure described in \citet{2004acs..rept...17G}. 

\section{Released Electronic Material}\label{sec:cat}

\begin{table*}
   \caption{\normalsize First ten rows extracted from the catalogue referring to Field 1 of NGC\,6121 (M4). A detailed description of each column is given in Table \ref{tab:catinfo}.}
   \label{tab:cat}
   \fontsize{6.3pt}{8pt}\selectfont
   \begin{verbatim}
#id      ra(2015)     dec(2015)     x         dx    y         dy    F814W  di    F475W  db     qi    qb    oi    ob    RADXSi  RADXSb  ni nb wi wb good
00000001 245.78962440  -26.47823845   549.416 0.353   688.174 0.224 28.832 9.900 27.078 9.900 1.620 0.760 1.526 0.115 -0.9900 -0.9157  1  1  1  1 0
00000002 245.78961100  -26.47812901   557.336 0.452   687.260 0.382 26.948 9.900 29.764 9.900 1.384 1.274 0.772 9.900  6.2647 -0.9900  1  1  1  1 0
00000003 245.78968050  -26.47823281   549.844 0.632   691.807 0.152 27.164 9.900 28.433 9.900 1.285 1.234 0.179 0.998 -0.9900 -0.9900  1  1  1  1 0
00000004 245.78969050  -26.47817087   554.332 0.391   692.424 0.136 27.335 9.900 30.545 9.900 1.229 1.715 1.014 2.384  0.9430 -0.9900  1  1  1  1 0
00000005 245.78969230  -26.47815496   555.485 0.426   692.537 0.076 27.109 9.900 29.691 9.900 1.189 1.274 0.404 9.900  2.2225 -0.9900  1  1  1  1 0
00000006 245.78973530  -26.47813294   557.094 0.032   695.317 0.263 28.421 9.900 28.038 9.900 1.204 1.009 3.708 0.336  5.6081 -0.9900  1  1  1  1 0
00000007 245.78971500  -26.47807749   561.102 0.068   693.975 0.088 27.108 9.900 28.184 9.900 1.187 0.948 0.787 0.228  1.0828 -0.9900  1  1  1  1 0
00000008 245.78970010  -26.47806742   561.826 0.001   693.008 0.354 26.711 9.900 30.436 9.900 1.154 1.722 0.354 2.059 -0.9900 -0.9900  1  1  1  1 0
00000009 245.78977530  -26.47818364   553.437 0.442   697.928 0.434 27.300 9.900 28.887 9.900 1.330 1.371 0.459 1.208 -0.7255  9.9900  1  1  1  1 0
00000010 245.79005820  -26.47833427   542.628 0.483   716.320 0.322 26.991 9.900 28.863 9.900 1.025 1.193 0.699 5.580  1.8430  1.1484  1  1  1  1 0
   \end{verbatim}
   \normalsize
\end{table*}
\begin{table}
  \normalsize
  \begin{centering}
    \caption{Information provided by each catalogue.}
    \label{tab:catinfo}
    \begin{tabular}{ p{.3cm} p{1.1cm} p{6cm}} 
\hline      
Col.   & Name  & Explanation                                                \\
\hline
01     & id    & ID number for each star                                    \\
02     & ra    & Right Ascension for each star (in deg, epoch 2015)              \\
03     & dec   & Declination for each star (in deg, epoch 2015)                  \\
04     & x     & x position of each star on the reference frame (in pixels) \\
05     & dx    & rms errors associated to x position (in pixels)            \\
06     & y     & y position of each star on the reference frame (in pixels) \\
07     & dy    & rms errors associated to y position (in pixels)            \\
08     & F814W & F814W magnitude calibrated into Vegamag system             \\
09     & di    & rms errors associated to F814W magnitude                   \\
10     & F475W & F475W magnitude calibrated into Vegamag system             \\
11     & db    & rms errors associated to F475W magnitude                   \\
12     & qi    & {\it q} parameter for F814W magnitudes                     \\
13     & qb    & {\it q} parameter for F475W magnitudes                     \\
14     & oi    & {\it o} parameter for F814W magnitudes                     \\
15     & ob    & {\it o} parameter for F475W magnitudes                     \\
16     & RADXSi& {\it RADXS} parameter for F814W                            \\
17     & RADXSb& {\it RADXS} parameter for F475W                            \\
18     & ni    & number of F814W images the source has been detected in     \\
19     & nb    & number of F475W images the source has been detected in     \\
20     & wi    & source of F814W photometry                                 \\
       &       & 1: unsaturated in deep; 2: unsaturated in short;           \\
       &       & 3: saturated in short; 4: saturated in deep;               \\
21     & wb    & source of F475W photometry (same as wi)                    \\
22     & good  & the source has passed the selction process                 \\
\hline
    \end{tabular}
  \end{centering}
\end{table}

We release, for each {\rm ACS/WFC} parallel field, the astrometric and photometric catalogues and trichromatic astrometrised stacked images.
All the released material will be available for download at the website of the Exoplanets \& Stellar Populations Group of the Universit\`{a} degli Studi di Padova\footnote{\url{http://groups.dfa.unipd.it/ESPG/treasury.php}}.

Table \ref{tab:cat} shows the first ten rows of the catalogue produced for Field 1 of NGC\,6121. The content of each column is explained in detail in Table \ref{tab:catinfo}. The parameters {\it wi} and {\it wb} are the same as in \citet{2008AJ....135.2055A}. They are records that represent the level of saturation of each source in F814W and F475W images respectively.
The last column includes the results of the selection of well-measured stars presented in Section \ref{sec:ex}. 

Note that for many stars, we report a magnitude rms error of $9.900$. This is because the routine empirically determines errors based on multiple observations in each filter. When there is only one observation per filter, the error is given a high default value.

\section{Summary and Conclusions}\label{sec:sum}

In the context of the {\it Hubble Space Telescope} UV Legacy Survey Treasury program of Galactic Globular Clusters (GO-13297; PI: Piotto, \citetalias{2015AJ....149...91P}), we are releasing the photometric catalogues relative to the {\rm ACS/WFC} parallel observations.
They represent the first {\it HST} photometric survey of external regions of Galactic GCs and consist of $109$ distinct stellar fields of $49$ targets: $48$ GCs and one open cluster, NGC\,6791.

In the majority of cases, only two images per field were taken, one in F814W and one in F475W, centred at about $6.5$ arcminutes from cluster centre. Exposure times were selected in order to obtain reliable photometry of the main sequence of target GCs.

These observations complement the {\rm WFC3} observations of the central regions of the surveyed GCs, and represent valuable tools for different investigations as outlined in \citetalias{2015AJ....149...91P}. 
These data represent a first epoch for future studies aimed at proper motions measurements in these regions. 
Even without proper motions, these catalogues are suitable to various interesting studies such as measurements of mass functions and binaries fractions in external regions of GCs.
Furthermore, crowding is not an issue in these external cluster fields, as a result, this database could be also used as an input list for spectroscopic follow-up, for example for precise chemical tagging of cluster members.

\section*{Acknowledgements}
We thank the anonymous referee for his careful revision that improved the quality of the present manuscript. M.S., A.A. and G.P. acknowledge support from the Spanish Ministry of Economy and Competitiveness (MINECO) under grant AYA2013-42781. M.S. and A.A. acknowledge support from the Instituto de Astrof{\'i}sica de Canarias (IAC) under grant 309403. G.P. acknowledges partial support by the Universit{\`a} degli Studi di Padova Progetto di Ateneo CPDA141214 ``Towards understanding complex star formation in Galactic globular clusters'' and by INAF under the program PRIN-INAF2014. A.P.M. acknowledges support by the Australian Research Council through Discovery Early Carreer Researcher Award DE150101816 and by the ERC-StG 2016 716082 project `GALFOR' funded by the European Research Council.
This work has made use of data from the European Space Agency (ESA) mission {\it Gaia} (\url{http://www.cosmos.esa.int/gaia}), processed by the {\it Gaia} Data Processing and Analysis Consortium (DPAC, \url{http://www.cosmos.esa.int/web/gaia/dpac/consortium}). Funding for the DPAC has been provided by national institutions, in particular the institutions participating in the {\it Gaia} Multilateral Agreement.

\bibliographystyle{}

\appendix
\section{Extra material}\label{sec:a1}

\begin{figure*}
  \includegraphics[width=7.7 cm]{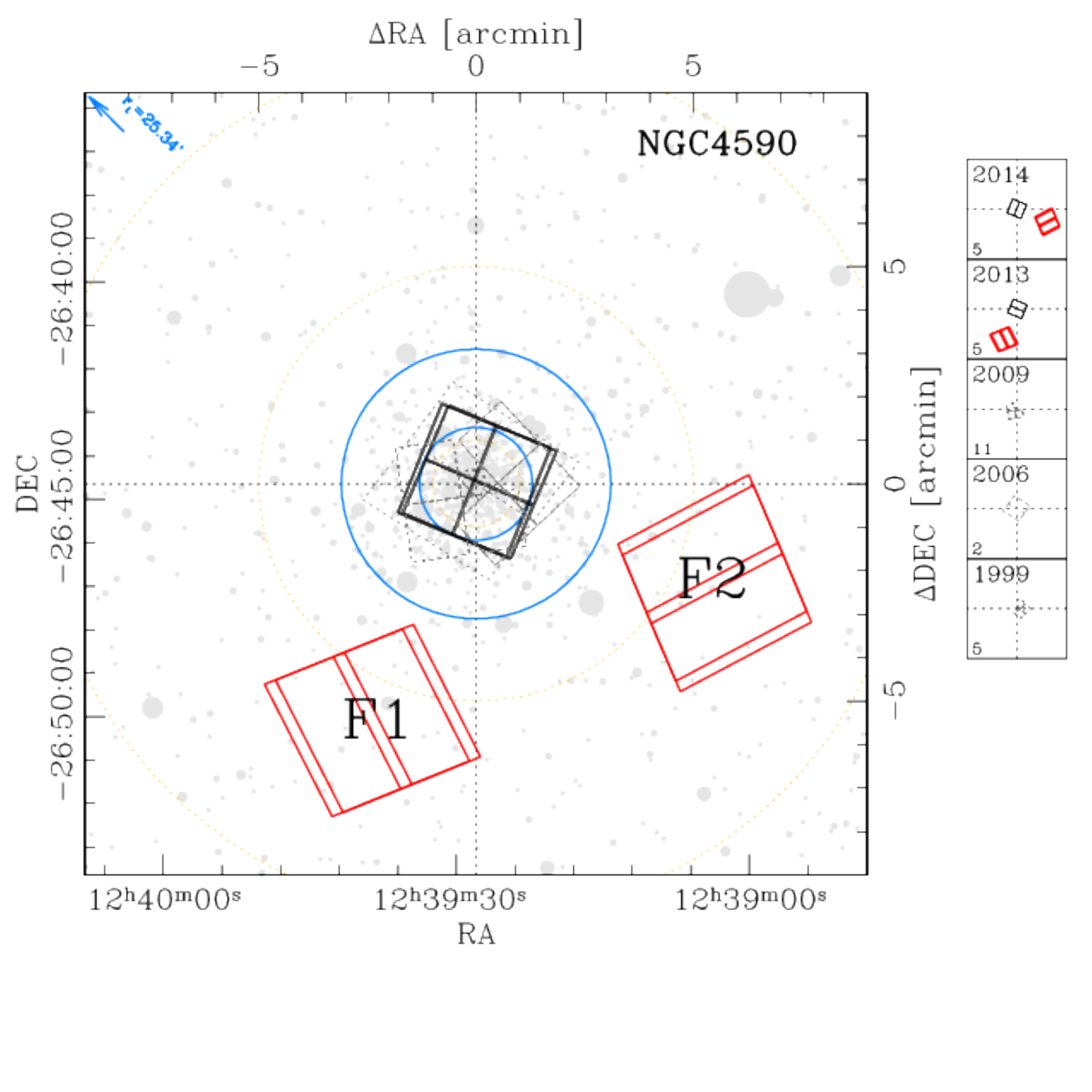}
  \includegraphics[width=7.7 cm]{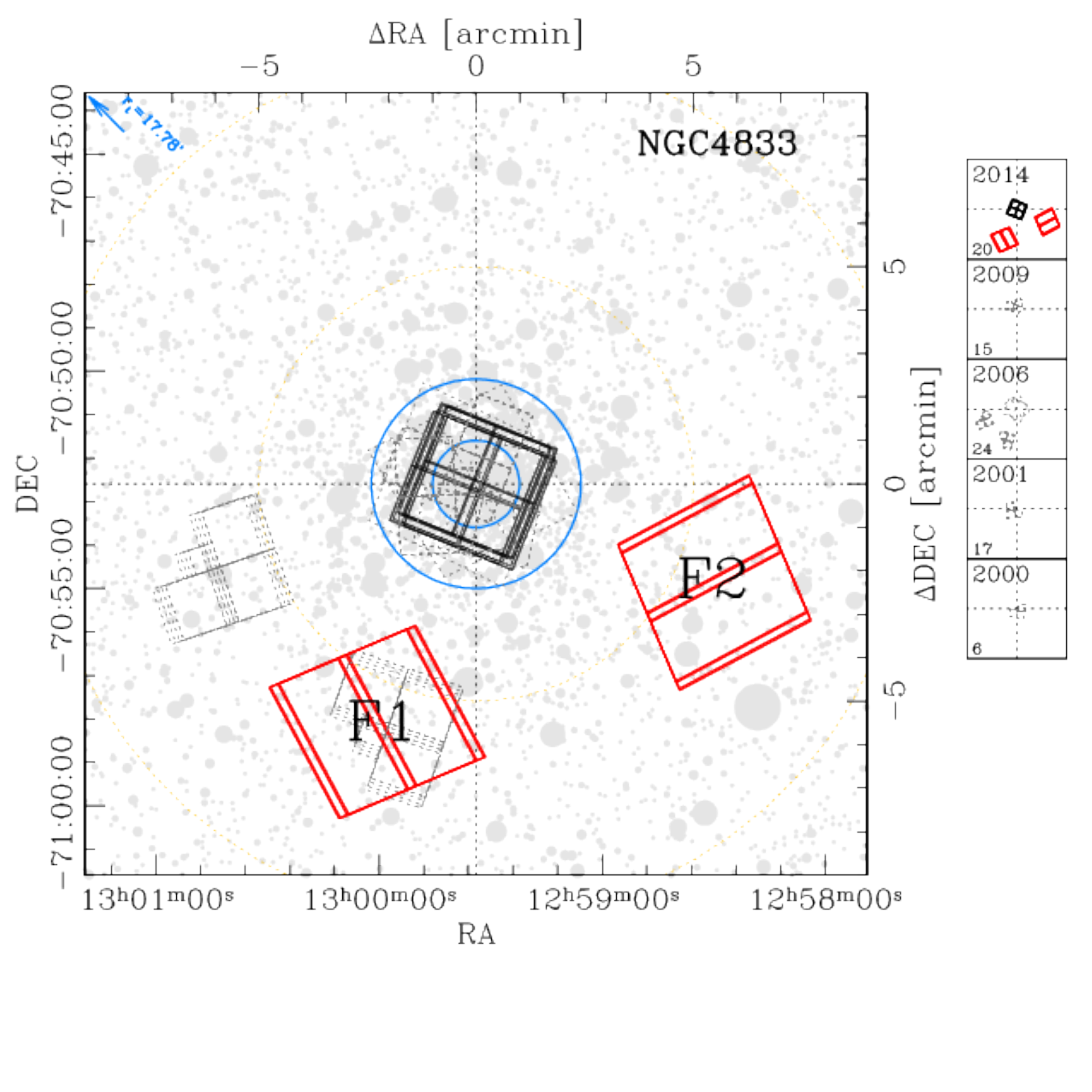}\\
  \includegraphics[width=7.7 cm]{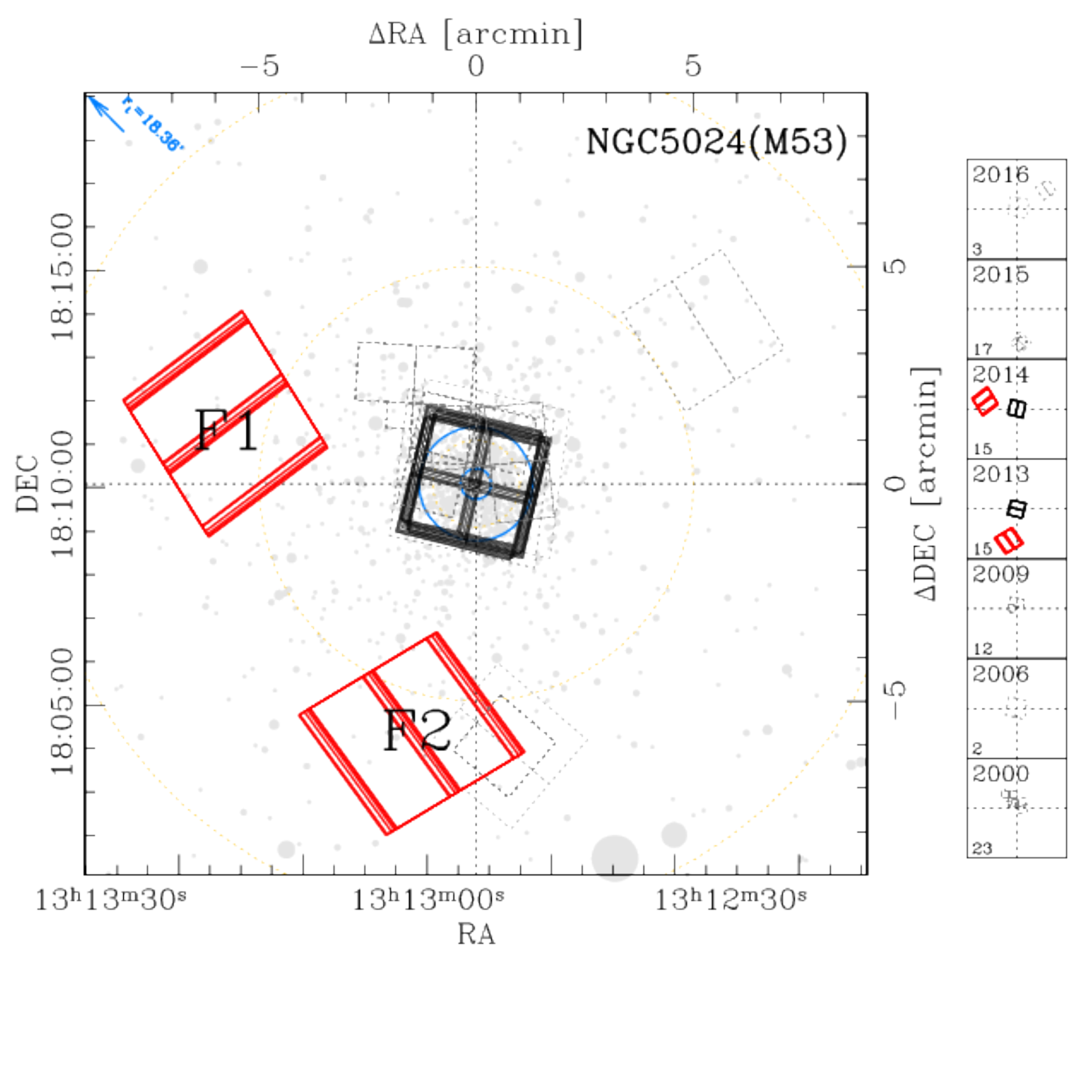}
  \includegraphics[width=7.7 cm]{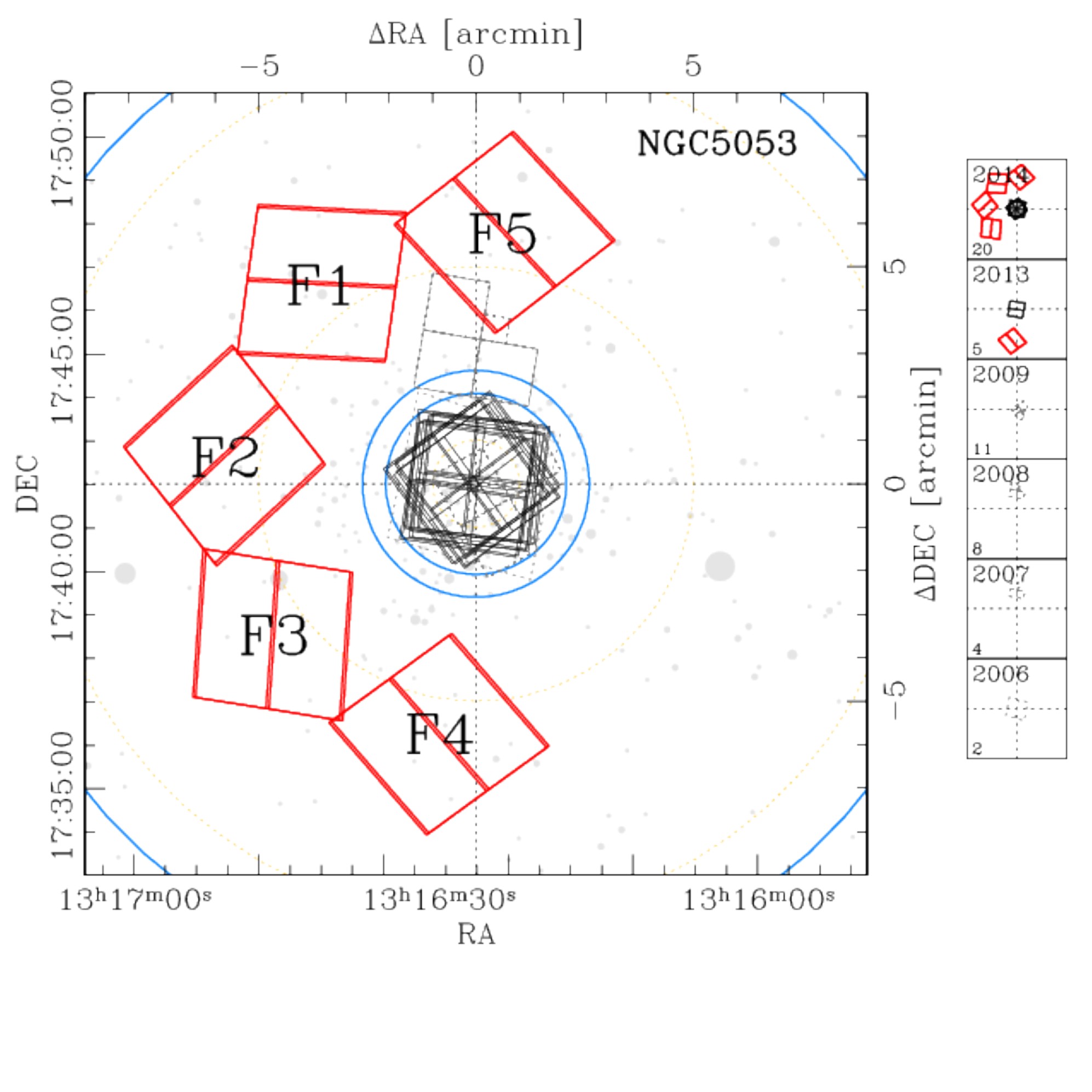}\\
  \includegraphics[width=7.7 cm]{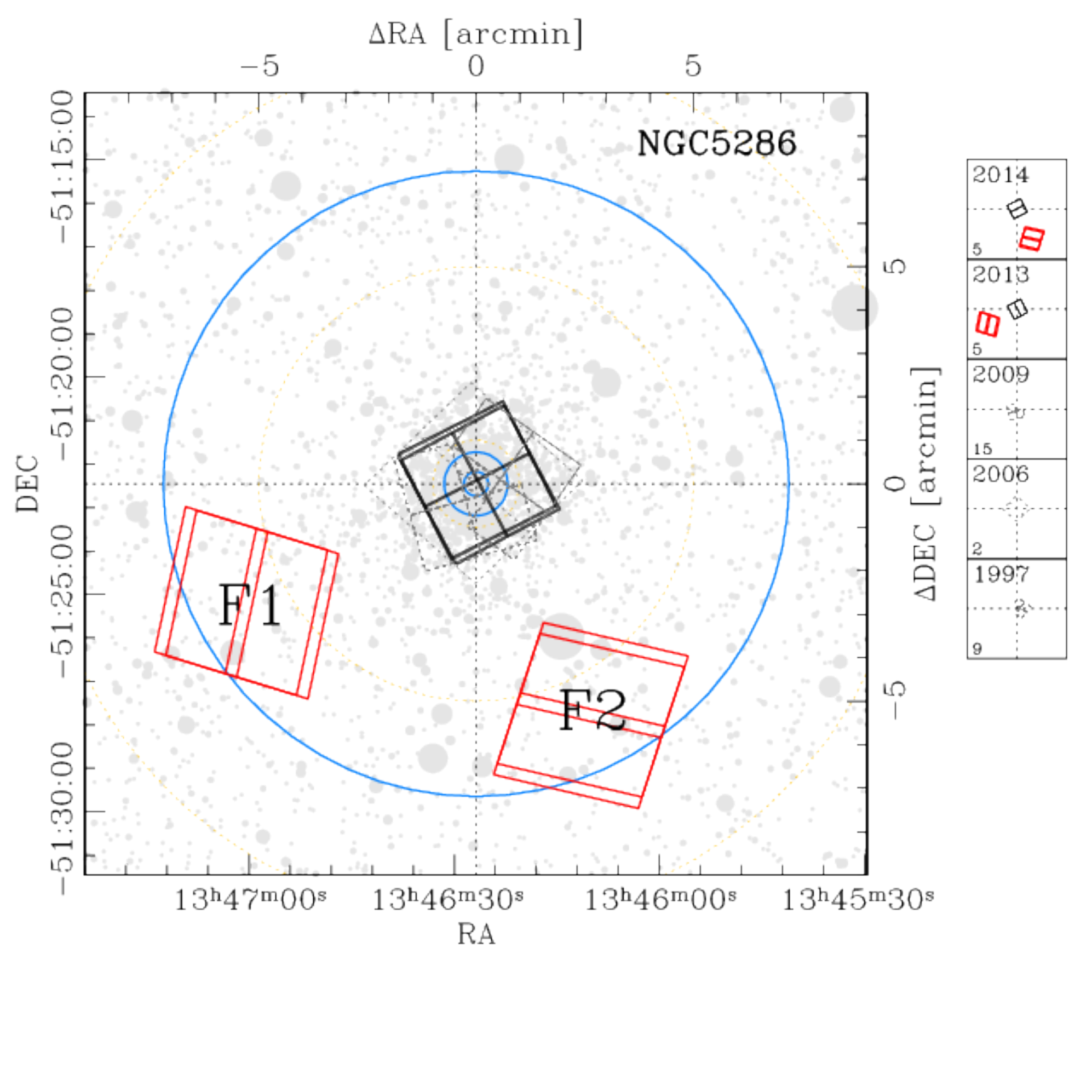}
  \includegraphics[width=7.7 cm]{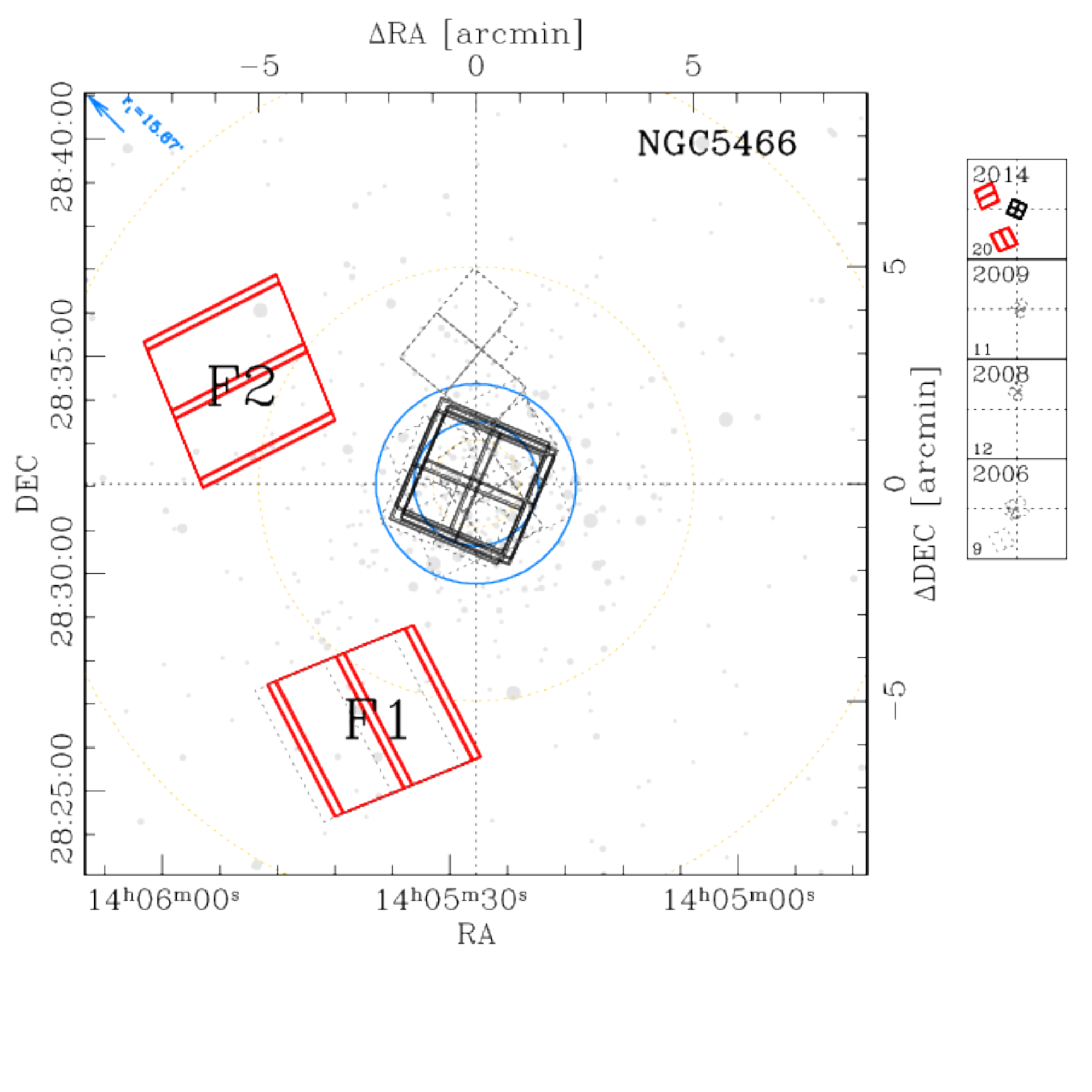}\\
  \caption{As in Figure \ref{fig:crt} but for NGC\,4590, NGC\,4833, NGC\,5024, NGC\,5053, NGC\,5286 and NGC\,5466}
  \label{fig:xcrt1}
\end{figure*}

\begin{figure*}
  \includegraphics[width=7.7 cm]{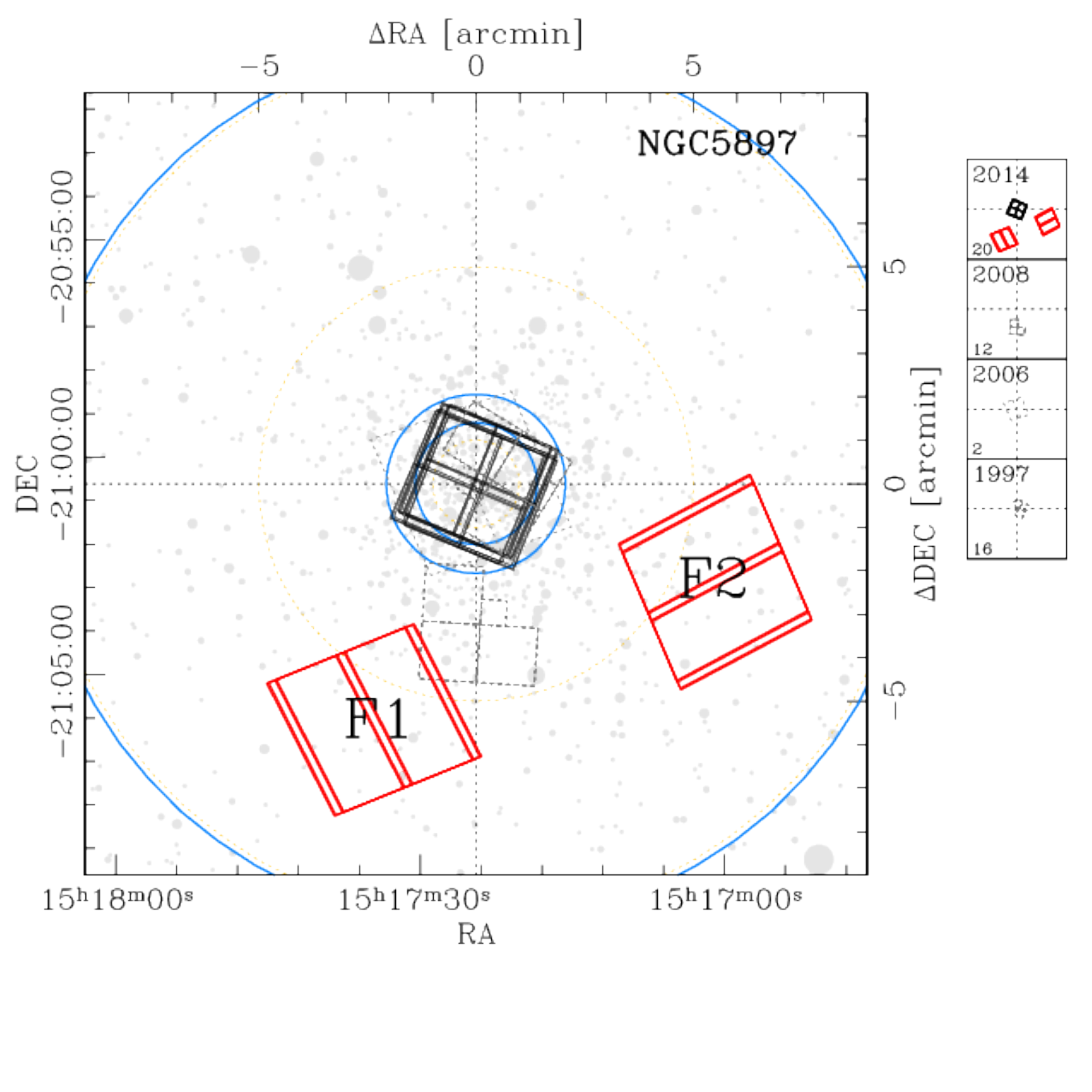}
  \includegraphics[width=7.7 cm]{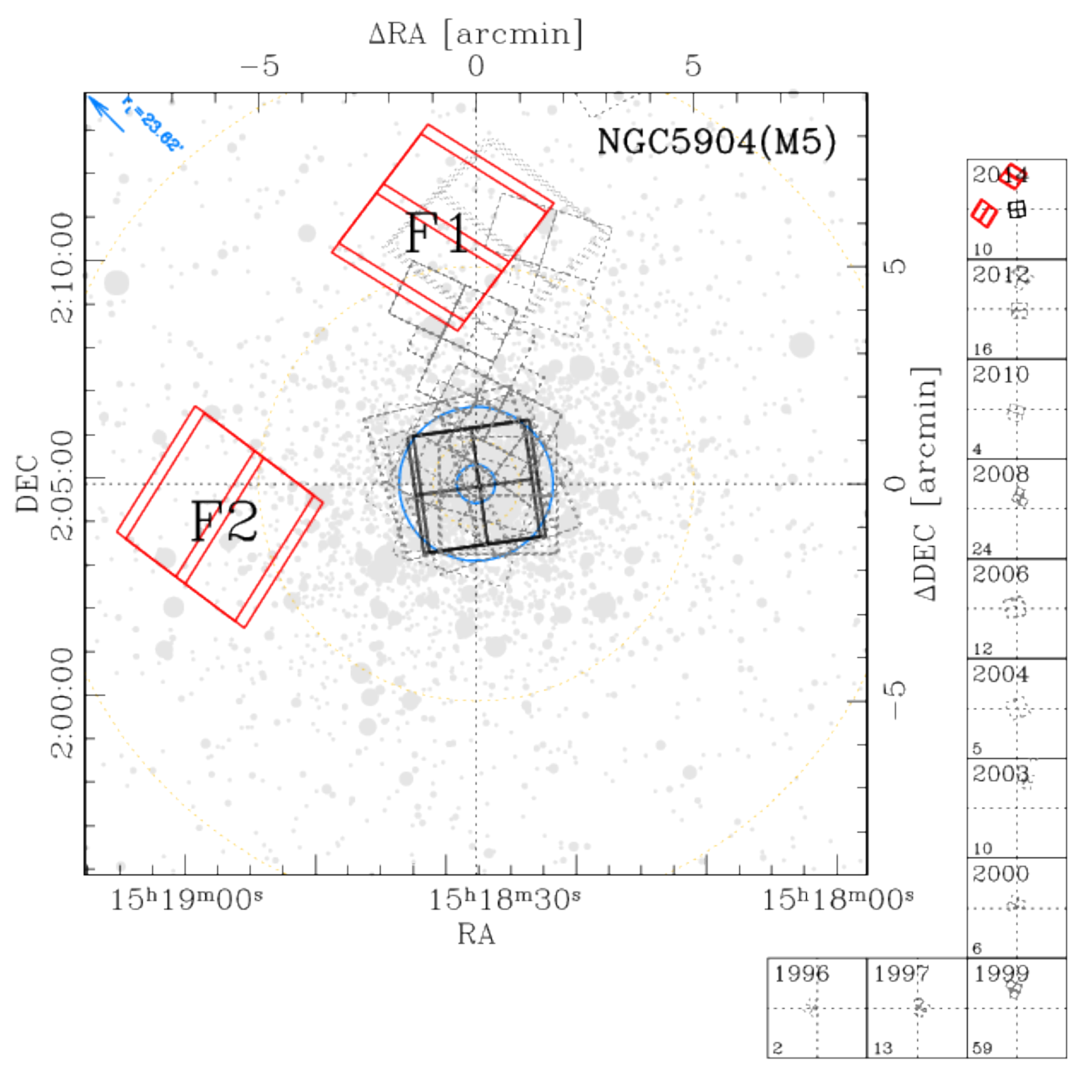}\\
  \includegraphics[width=7.7 cm]{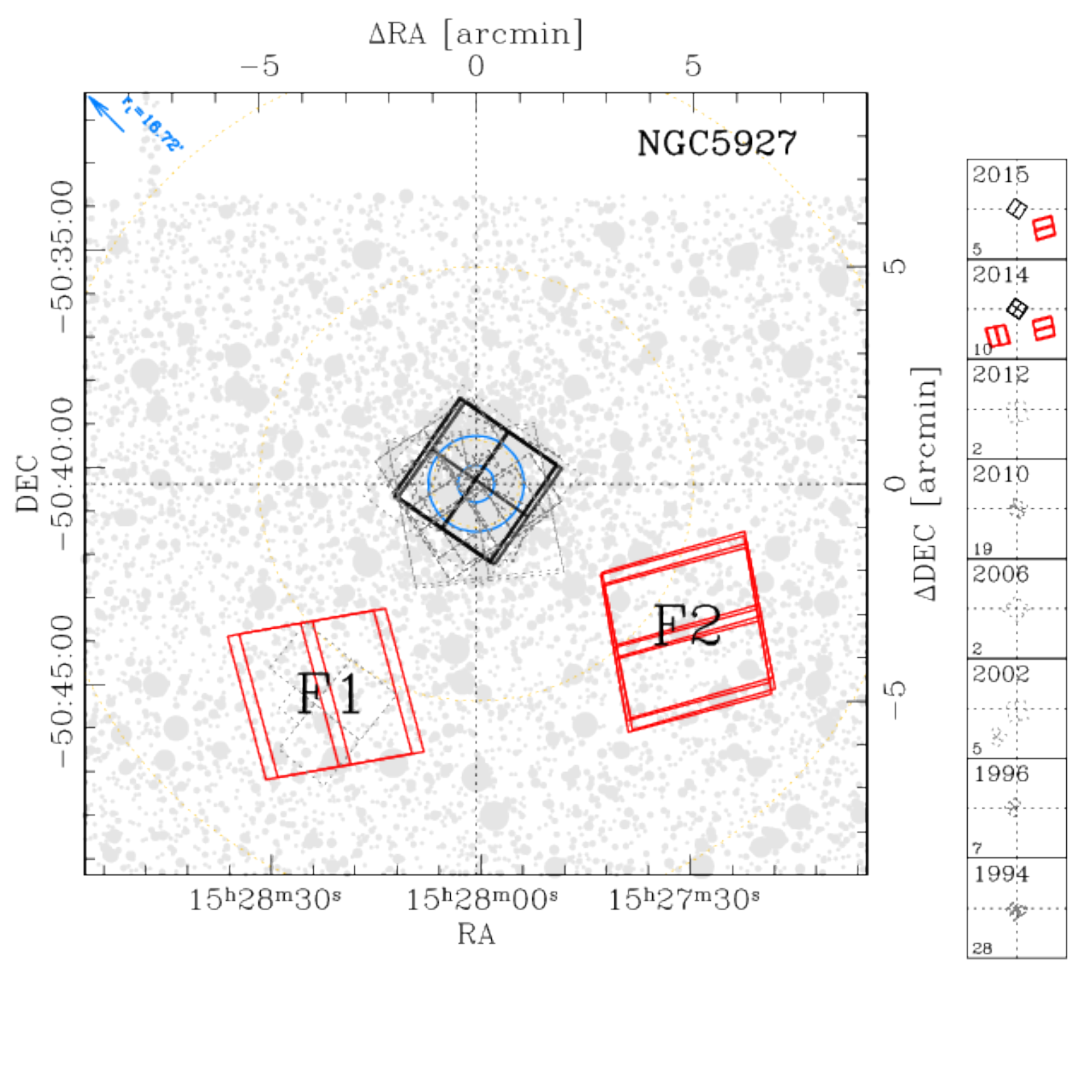}
  \includegraphics[width=7.7 cm]{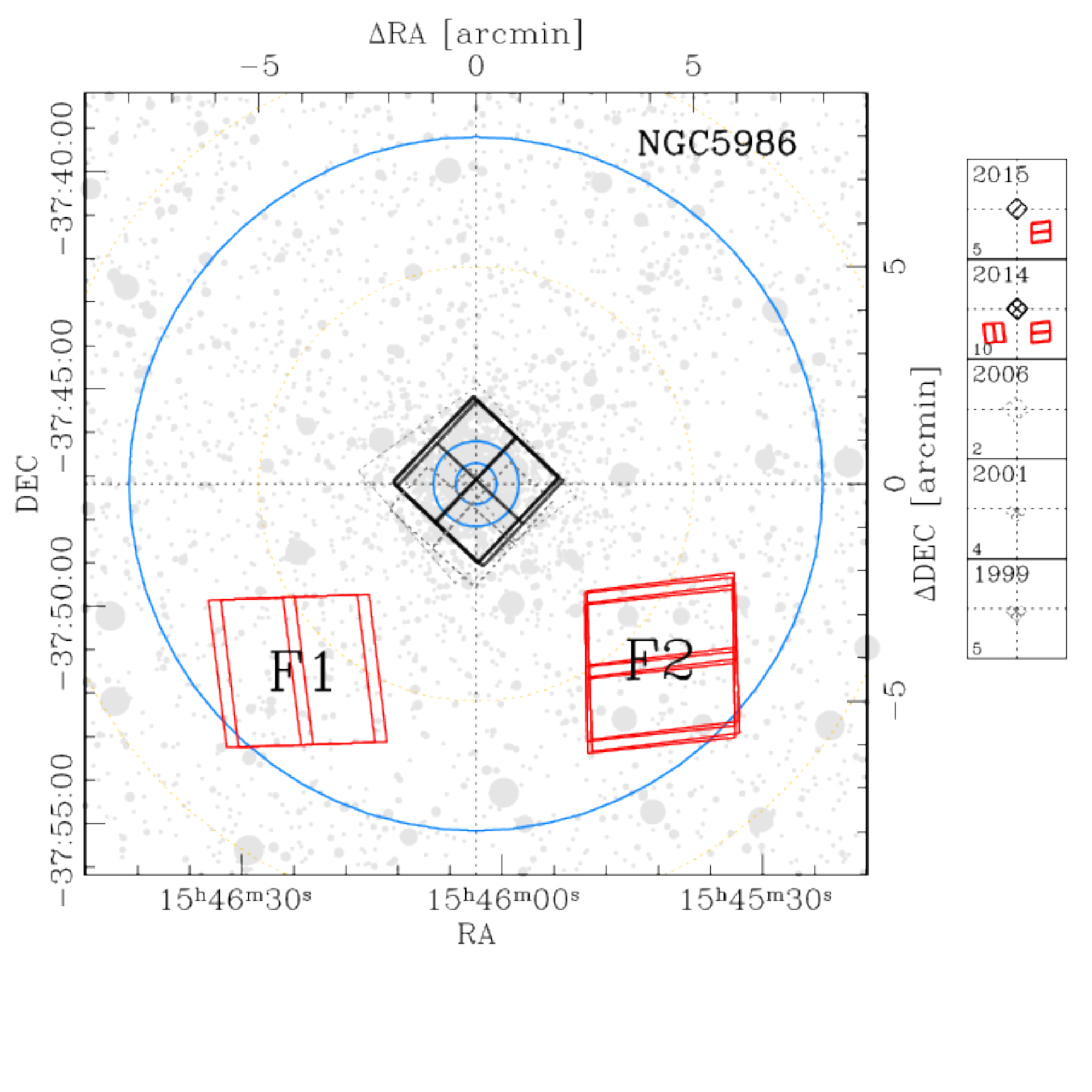}\\
  \includegraphics[width=7.7 cm]{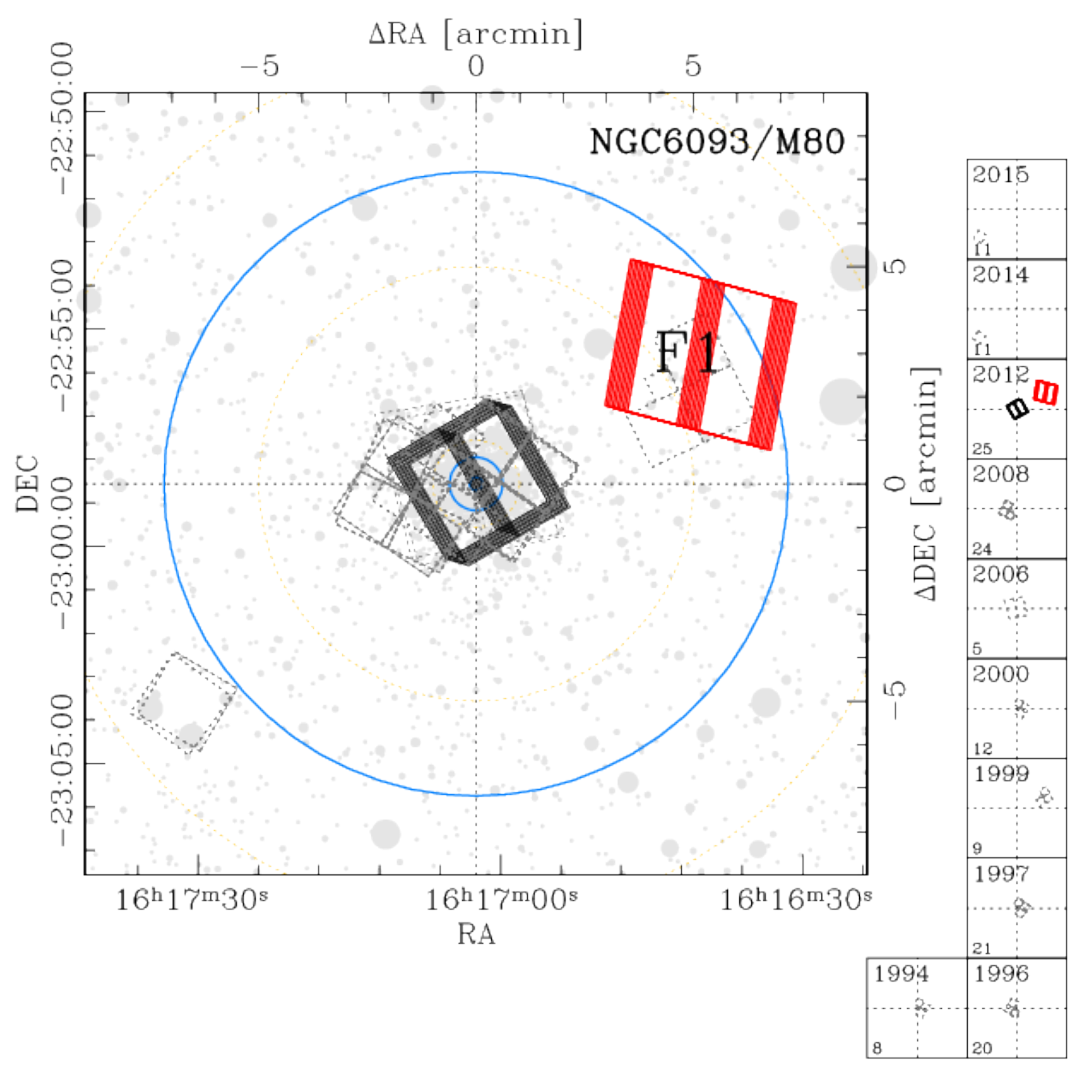}
  \includegraphics[width=7.7 cm]{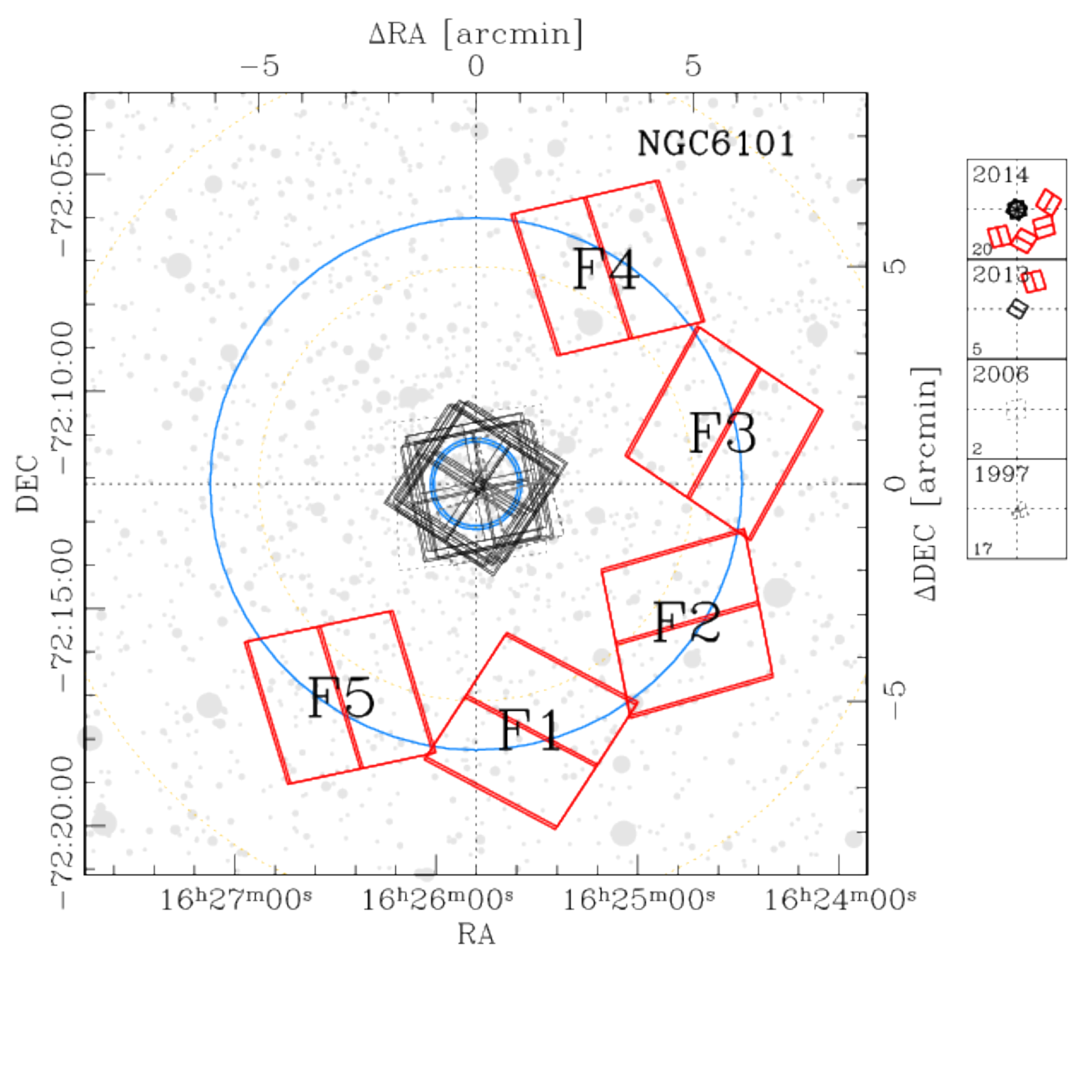}\\
  \caption{As in Figure \ref{fig:crt} but for NGC\,5897, NGC\,5904, NGC\,5927, NGC\,5986, NGC\,6093 and NGC\,6101}
  \label{fig:xcrt2}
\end{figure*}

\begin{figure*}
  \includegraphics[width=7.7 cm]{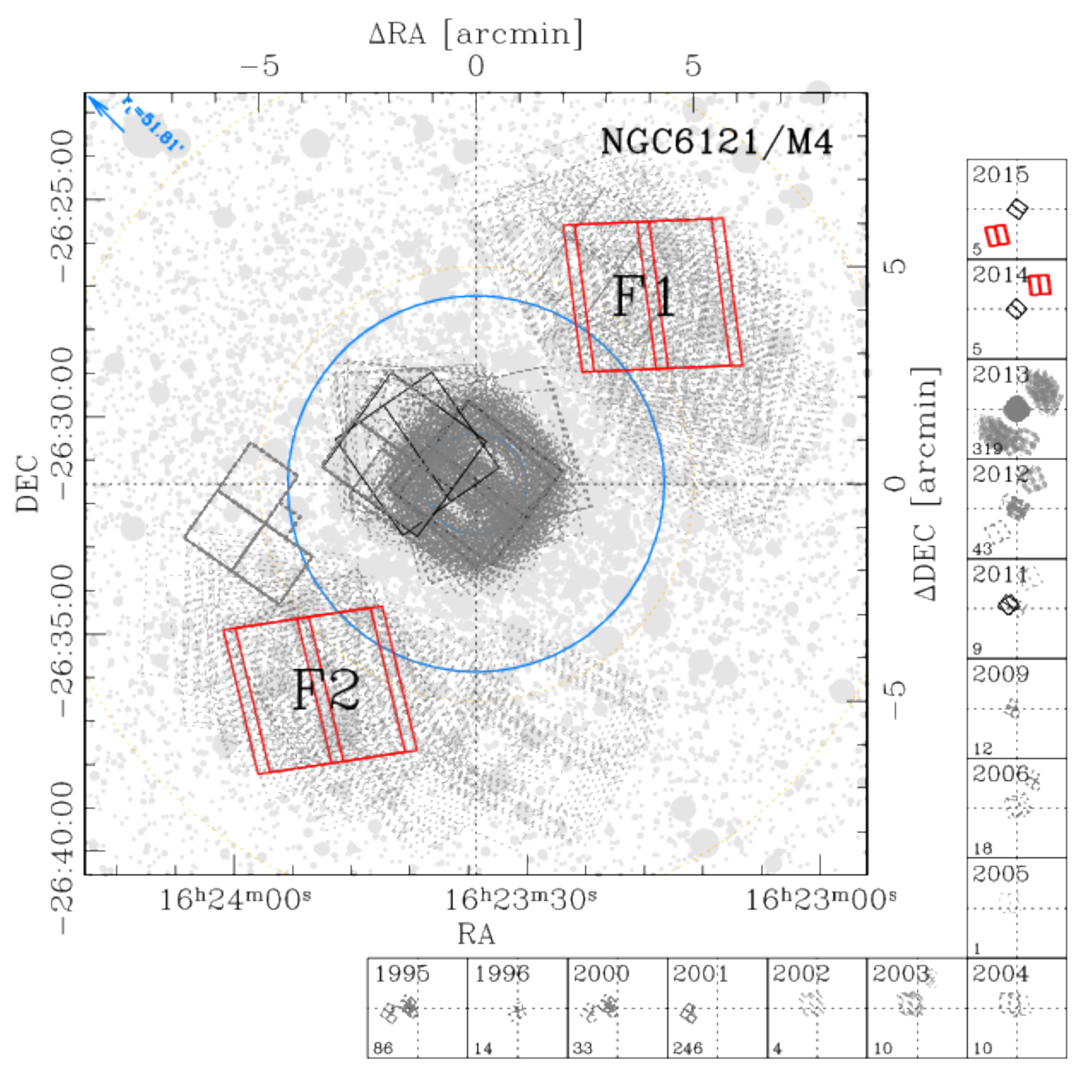}
  \includegraphics[width=7.7 cm]{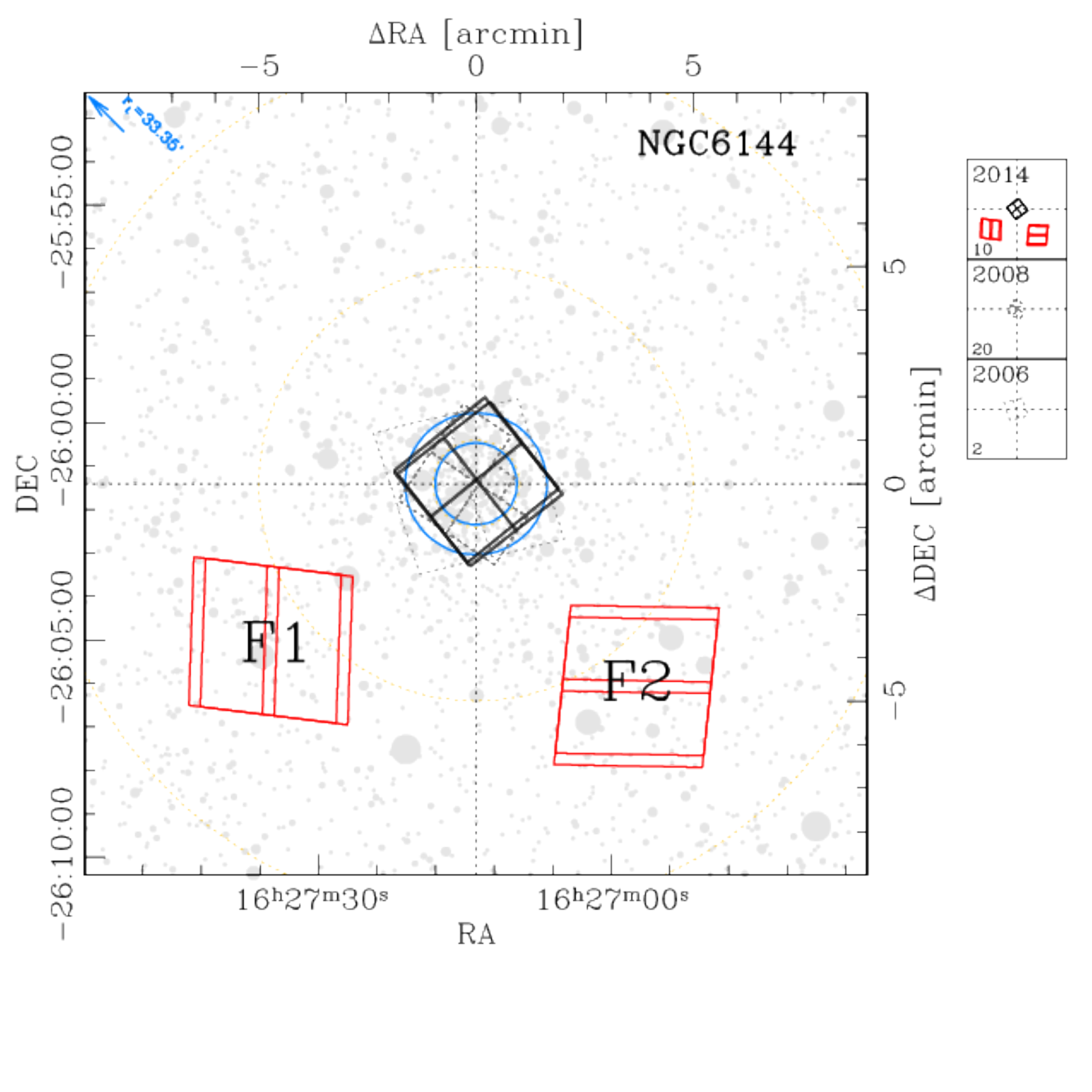}\\
  \includegraphics[width=7.7 cm]{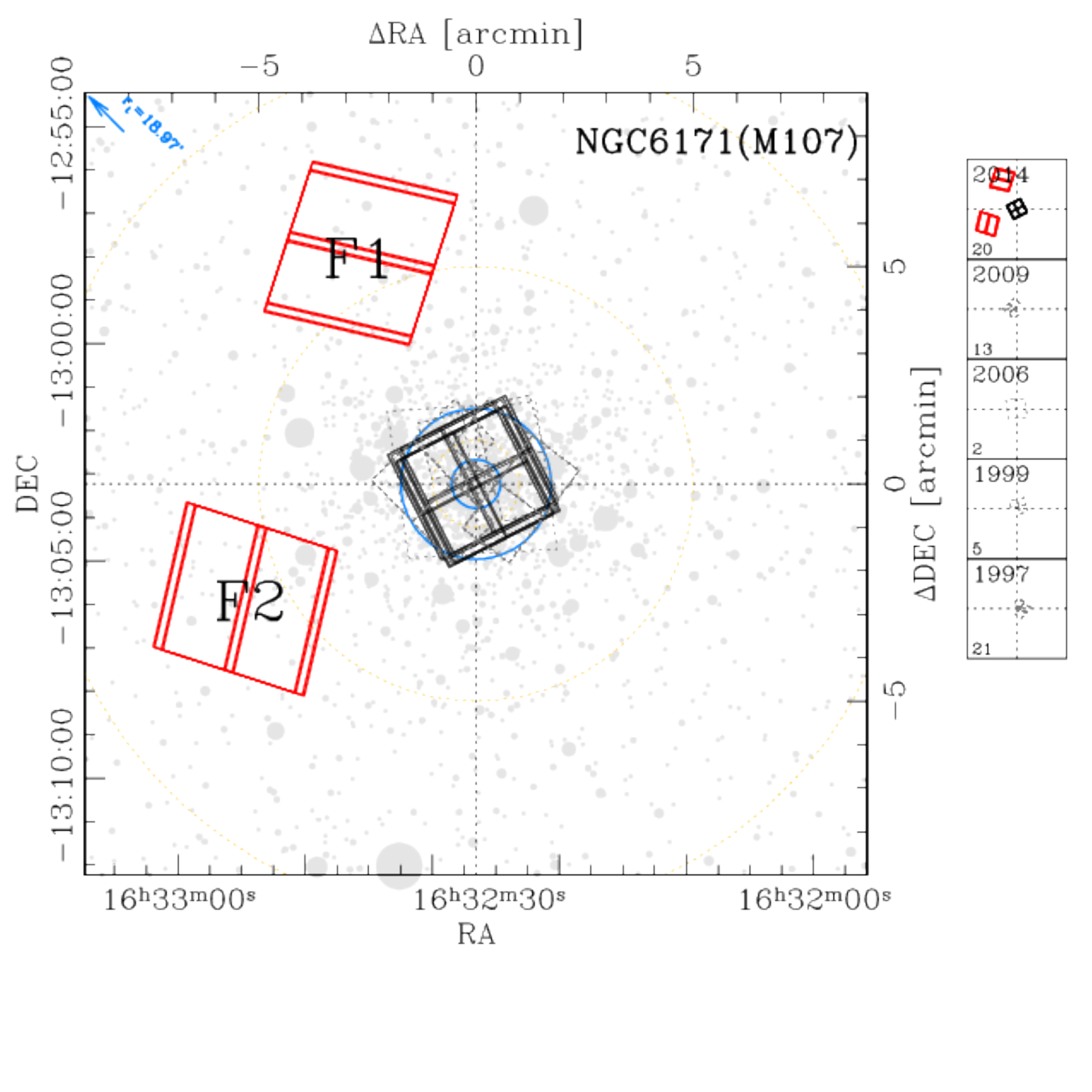}
  \includegraphics[width=7.7 cm]{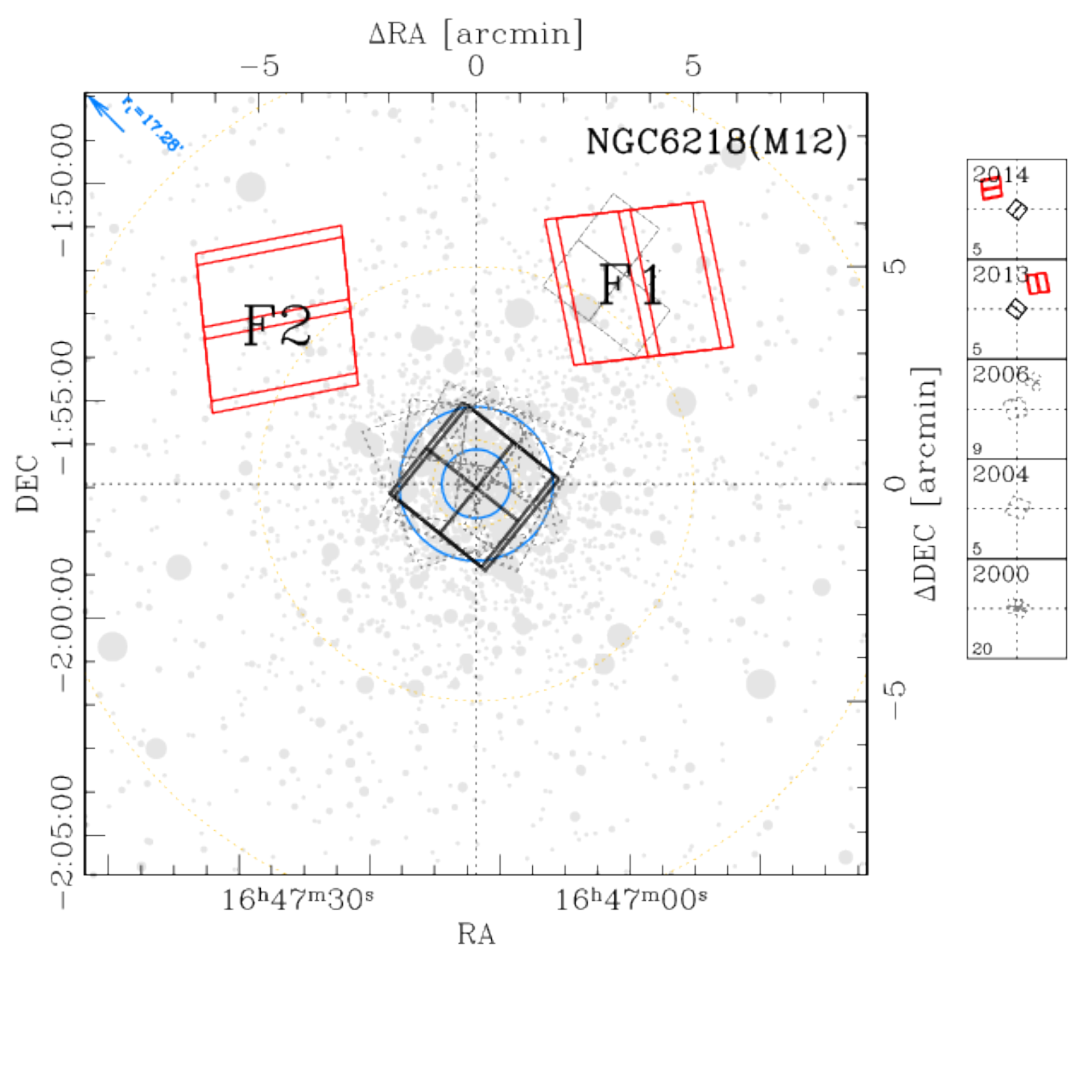}\\
  \includegraphics[width=7.7 cm]{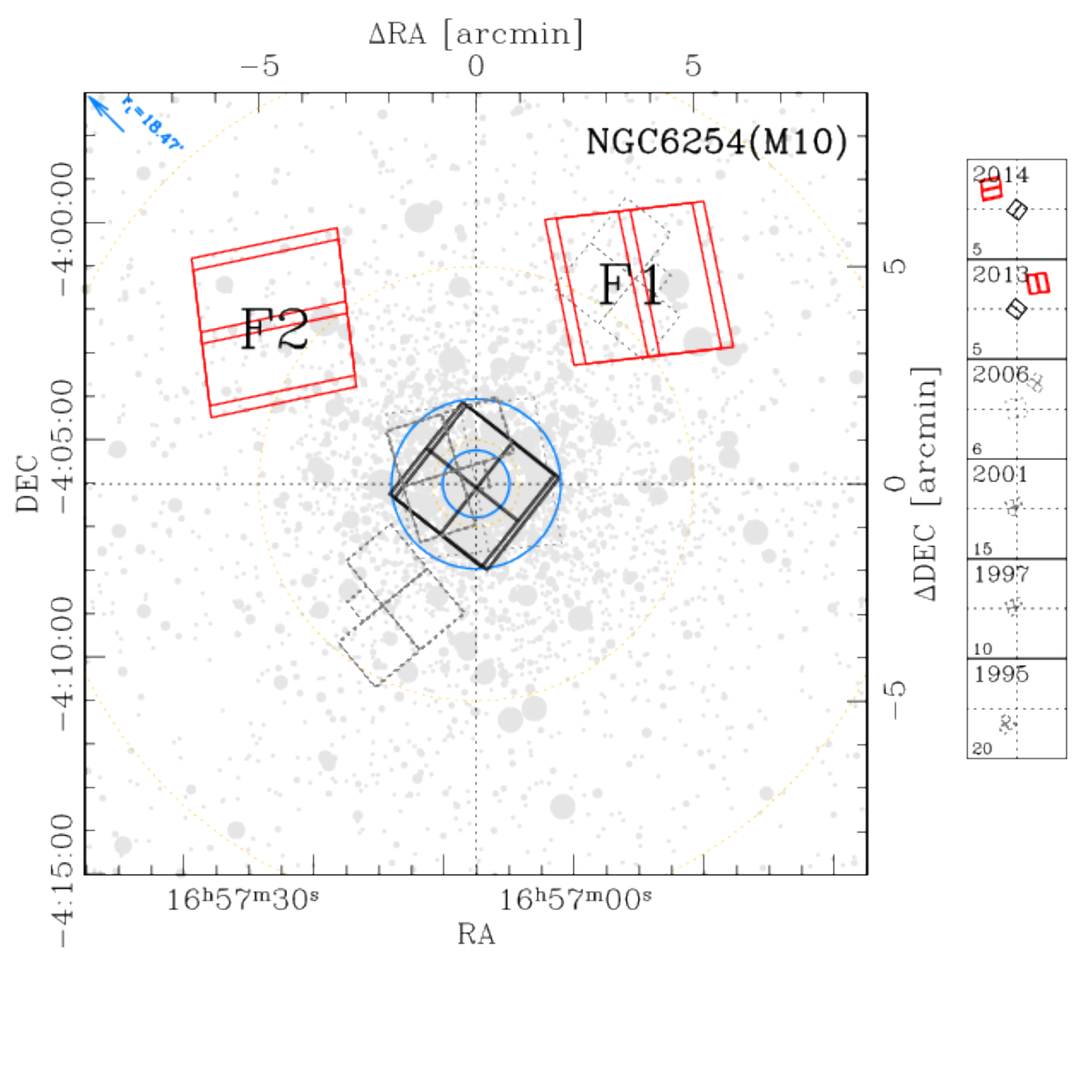}
  \includegraphics[width=7.7 cm]{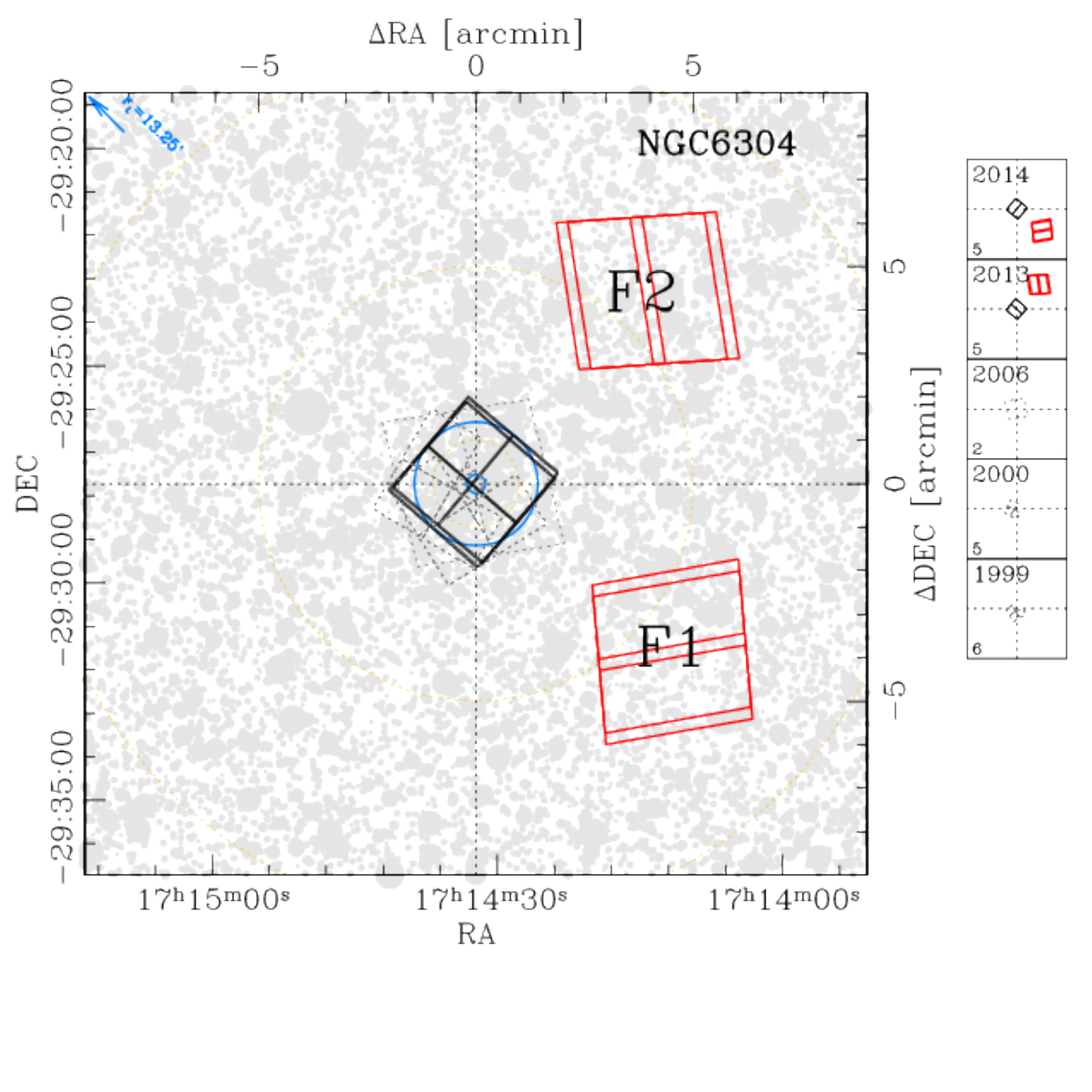}\\
  \caption{As in Figure \ref{fig:crt} but for NGC\,6121, NGC\,6144, NGC\,6171, NGC\,6218, NGC\,6254 and NGC\,6304}
  \label{fig:xcrt3}
\end{figure*}

\begin{figure*}
  \includegraphics[width=7.7 cm]{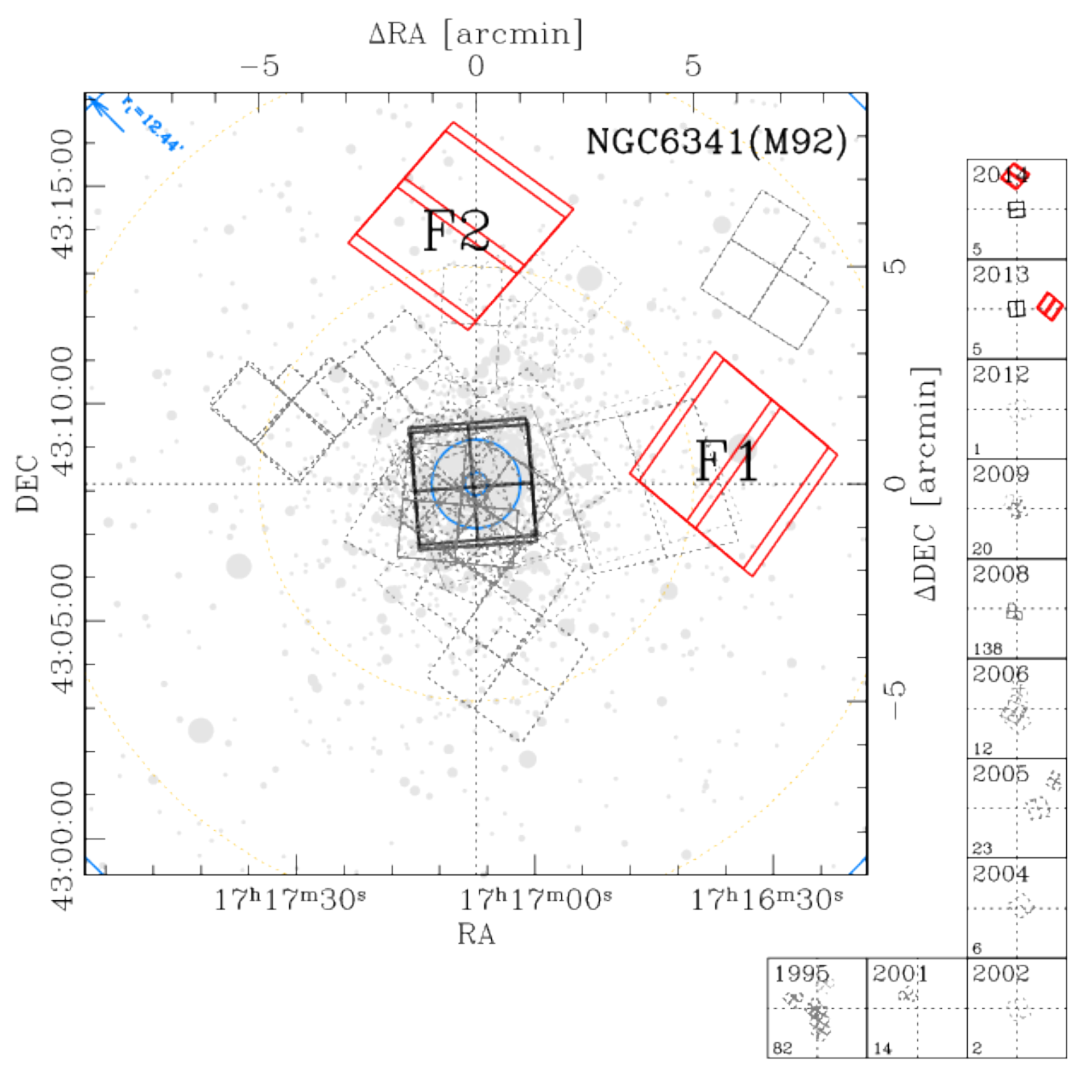}
  \includegraphics[width=7.7 cm]{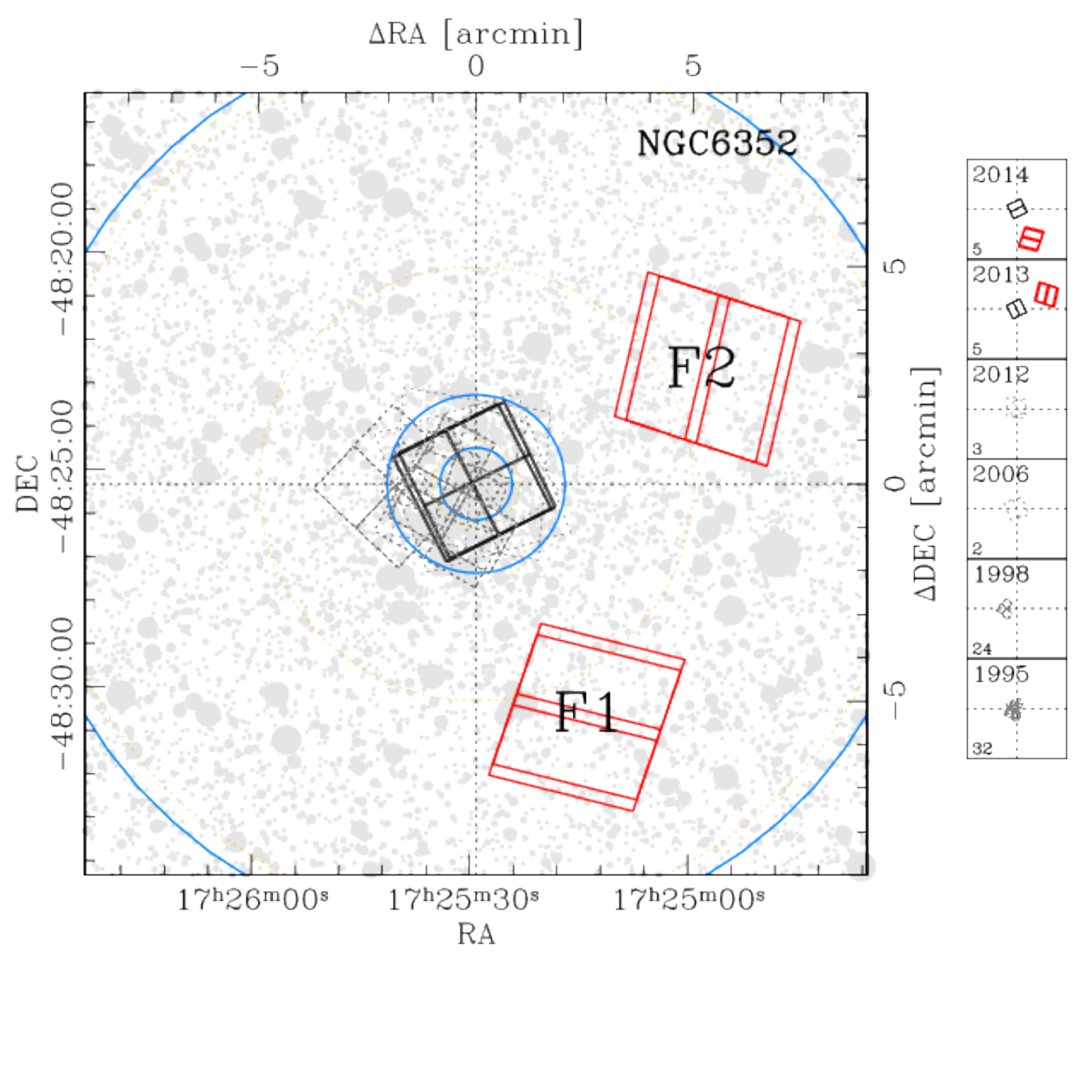}\\
  \includegraphics[width=7.7 cm]{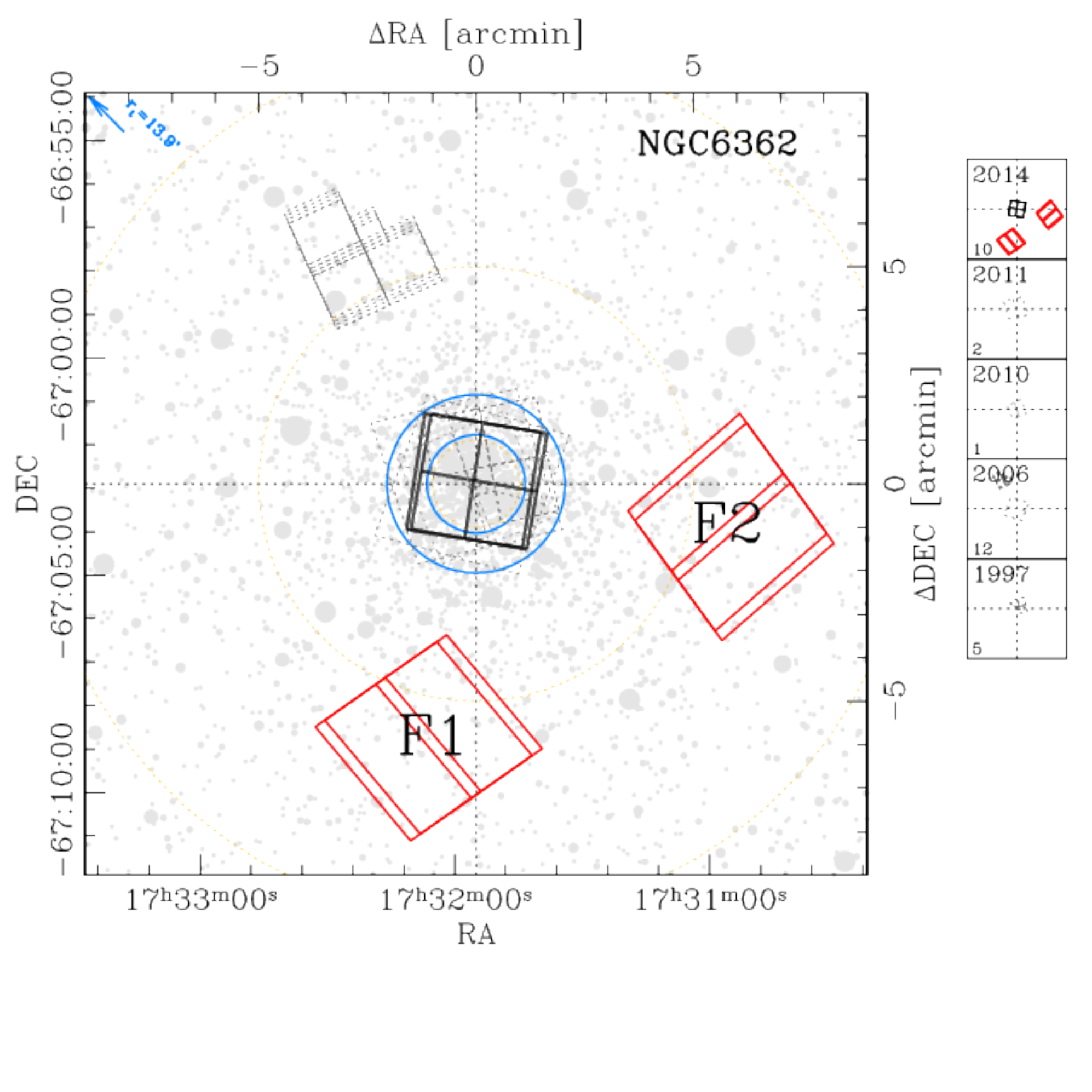}
  \includegraphics[width=7.7 cm]{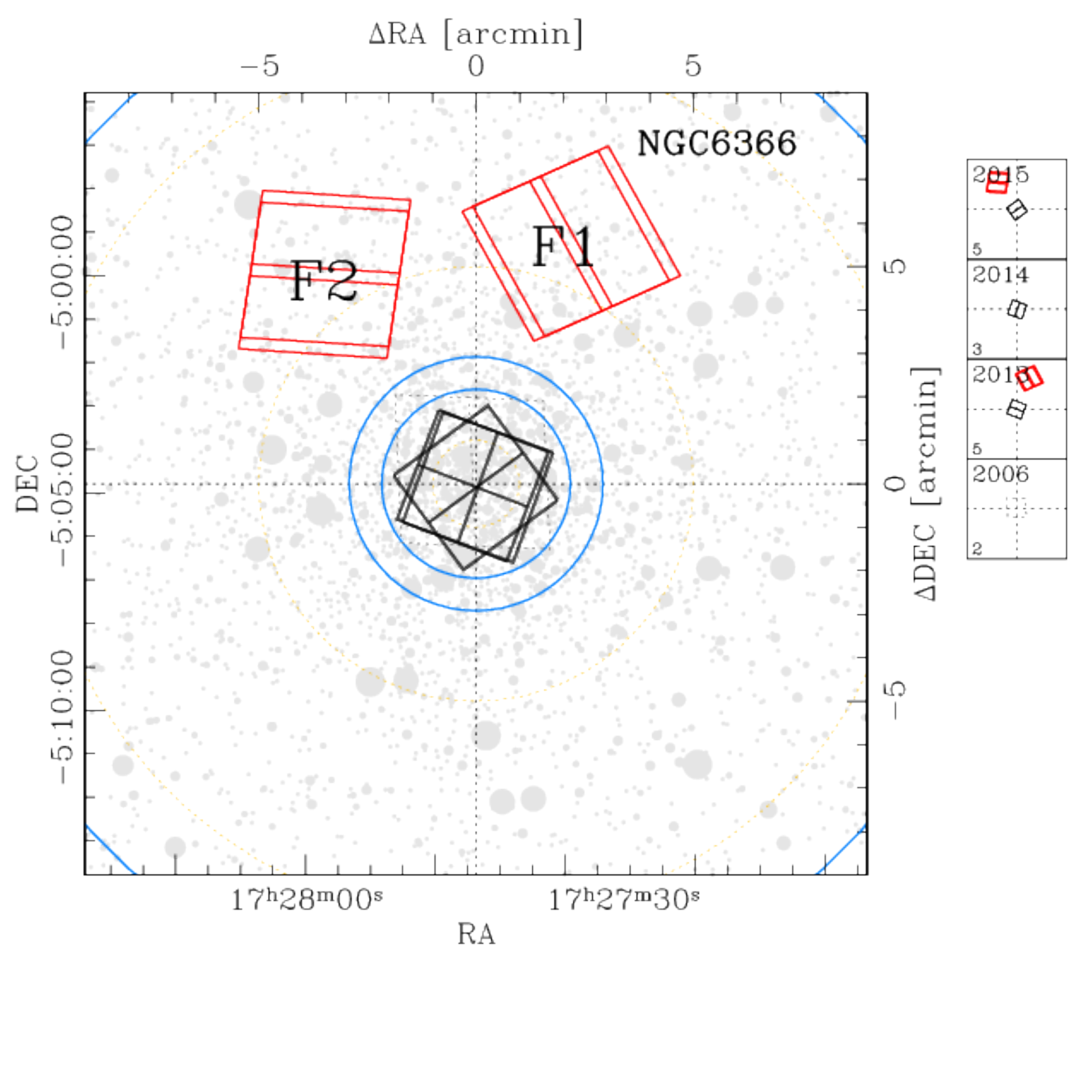}\\
  \includegraphics[width=7.7 cm]{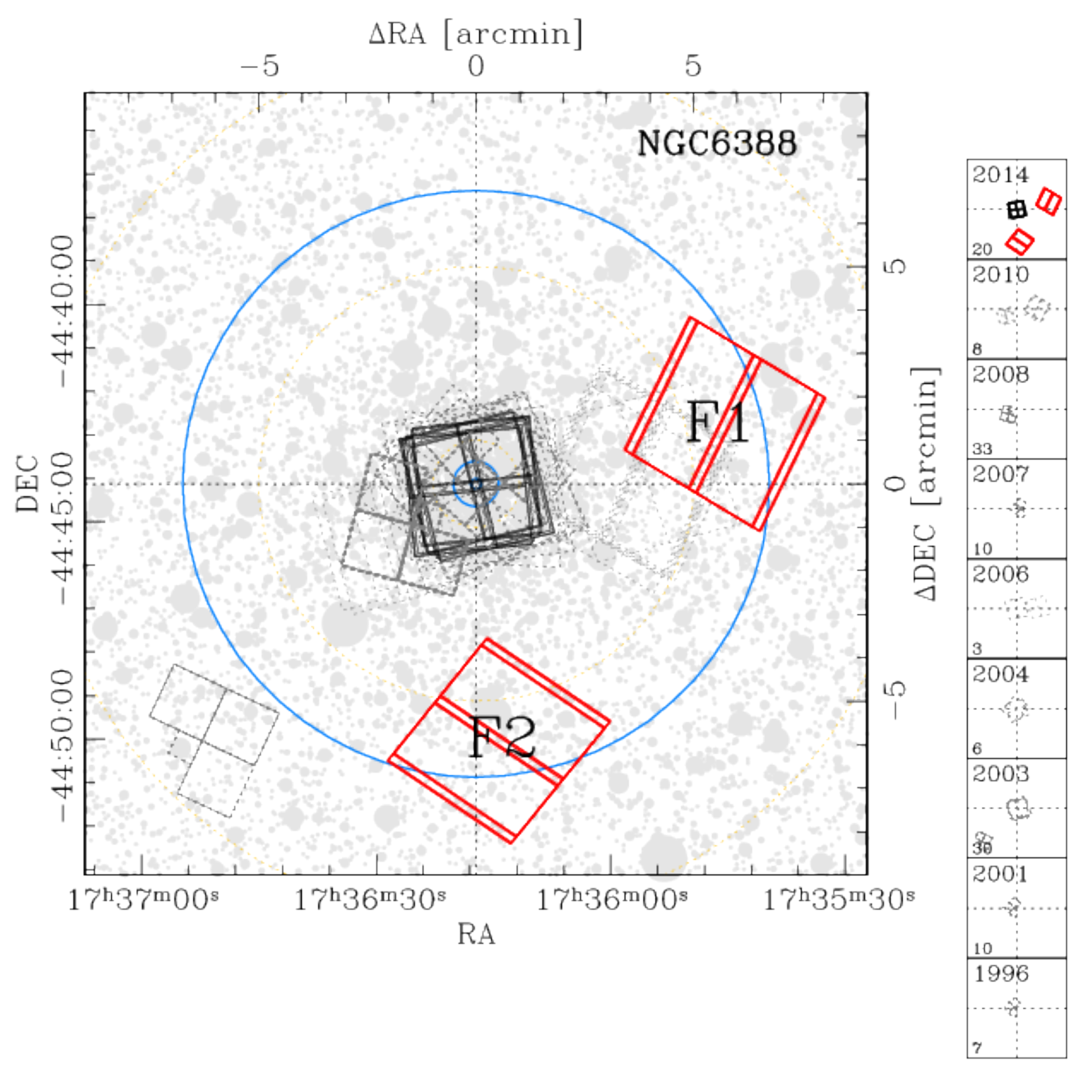}
  \includegraphics[width=7.7 cm]{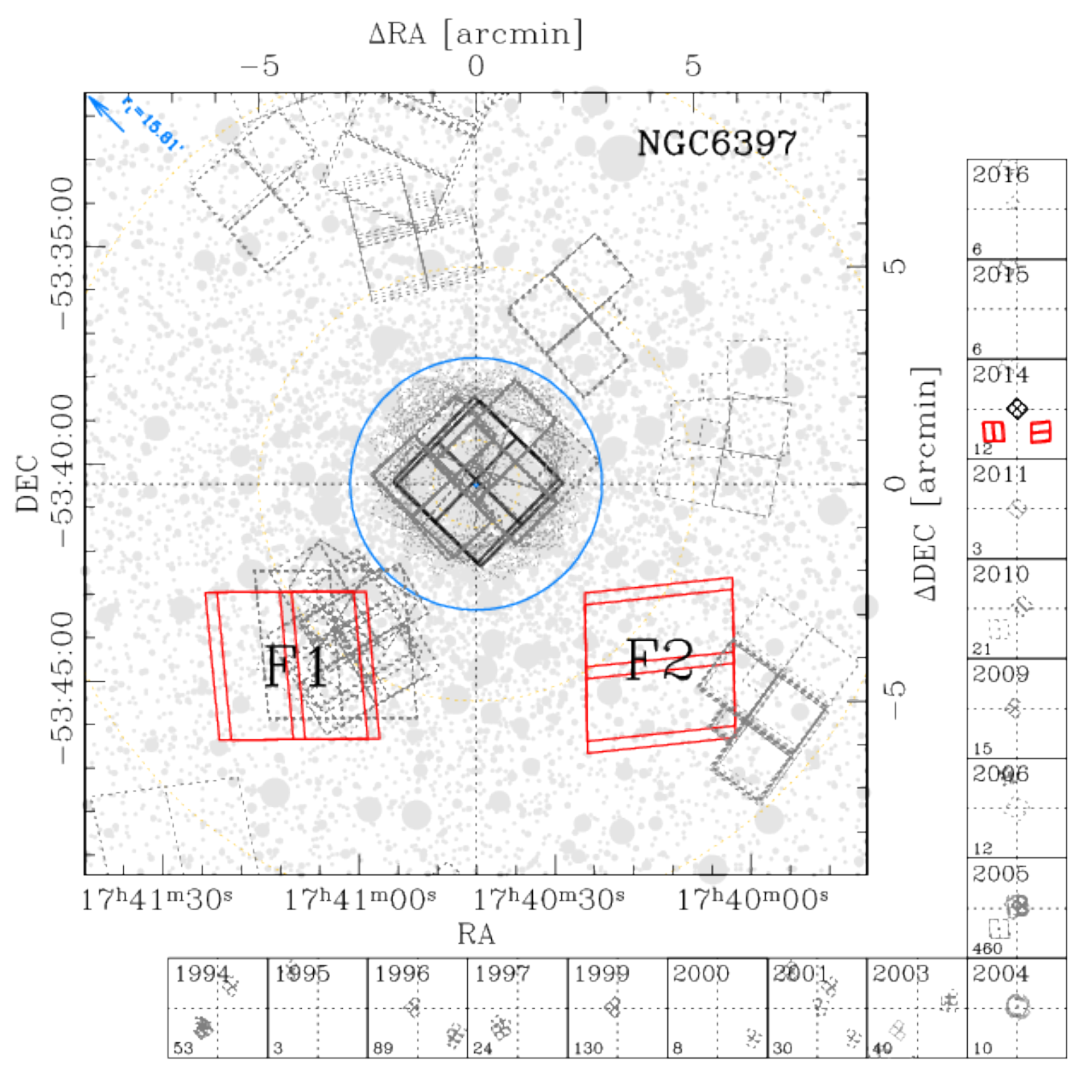}\\
  \caption{As in Figure \ref{fig:crt} but for NGC\,6341, NGC\,6352, NGC\,6362, NGC\,6366, NGC\,6388 and NGC\,6397}
  \label{fig:xcrt4}
\end{figure*}

\begin{figure*}
  \includegraphics[width=7.7 cm]{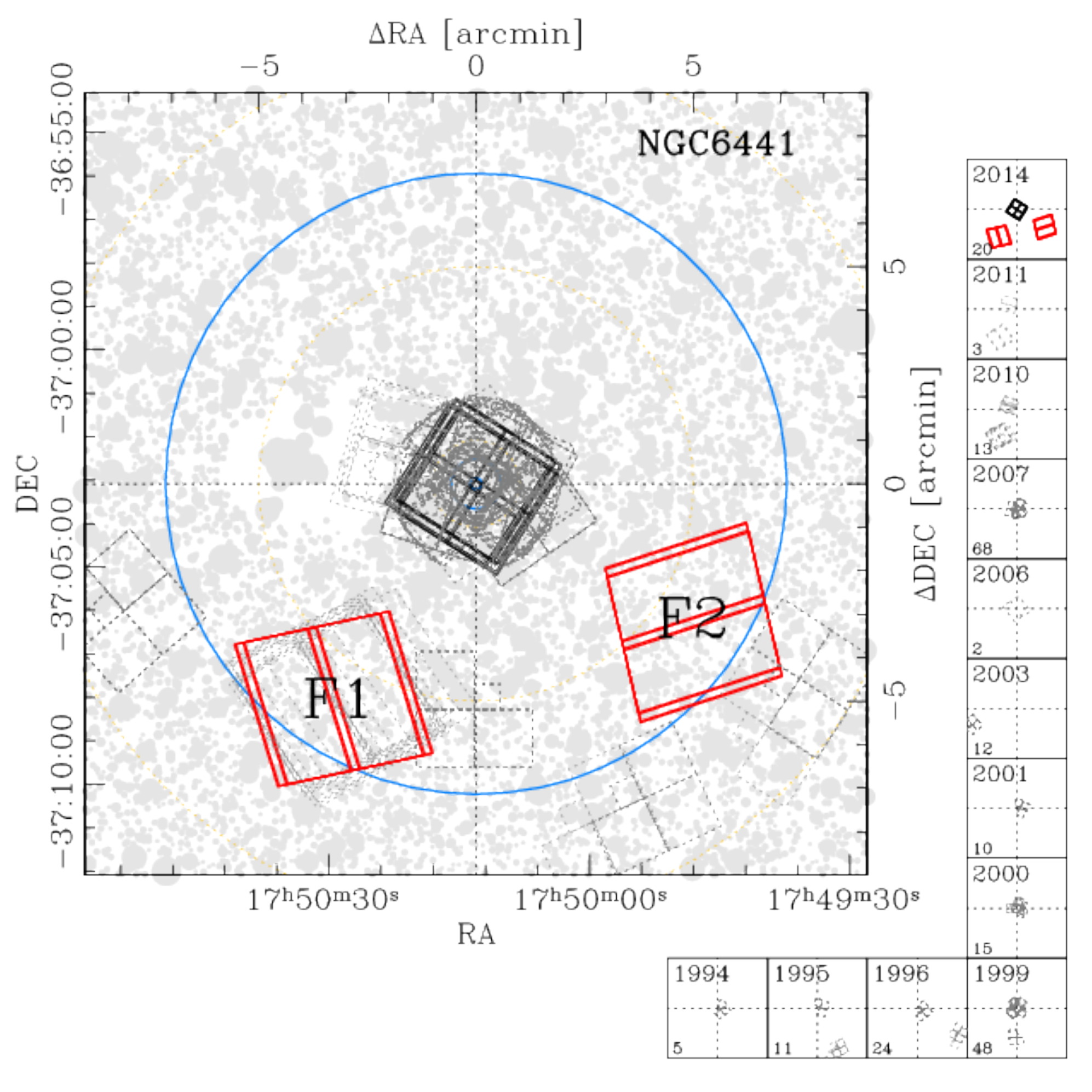}
  \includegraphics[width=7.7 cm]{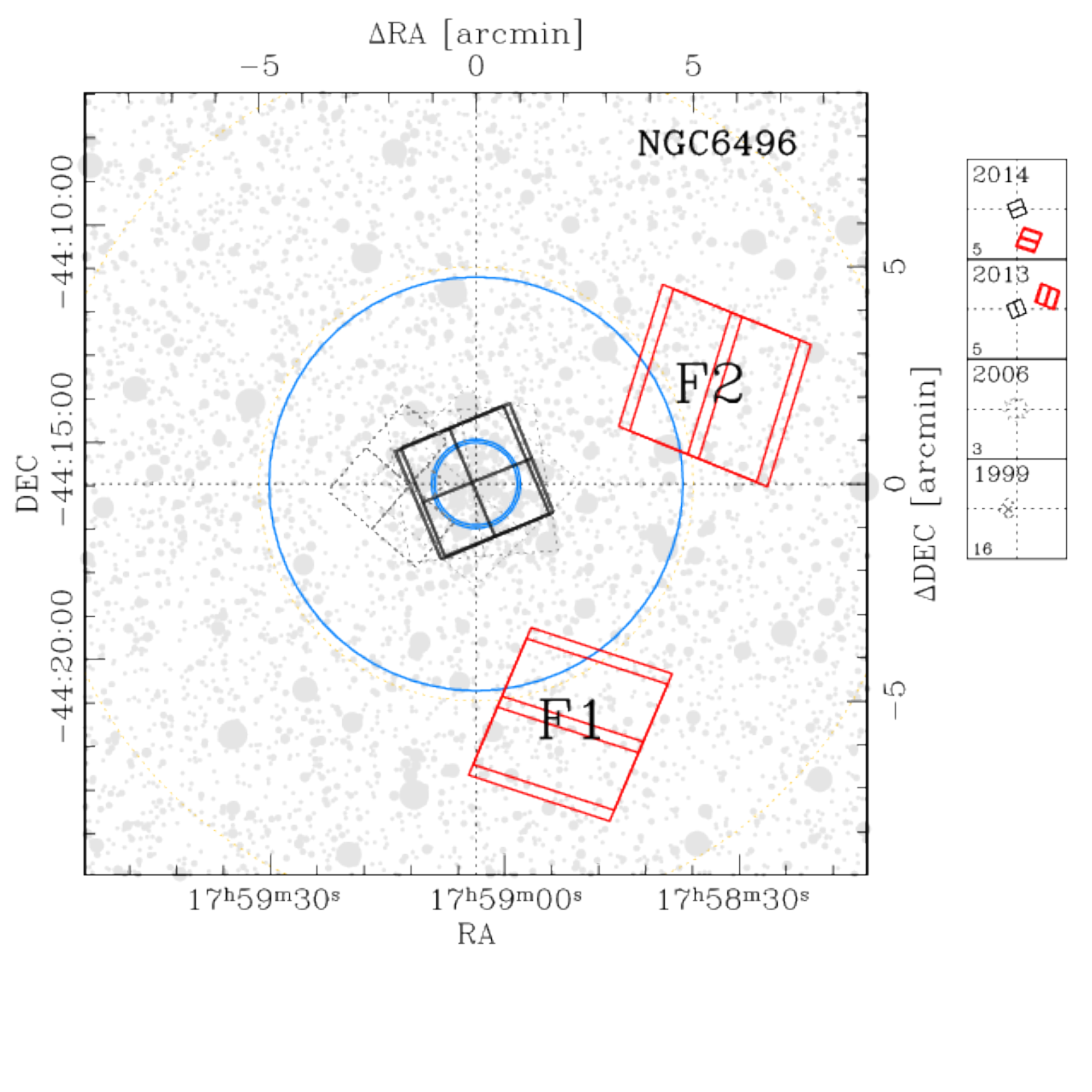}\\
  \includegraphics[width=7.7 cm]{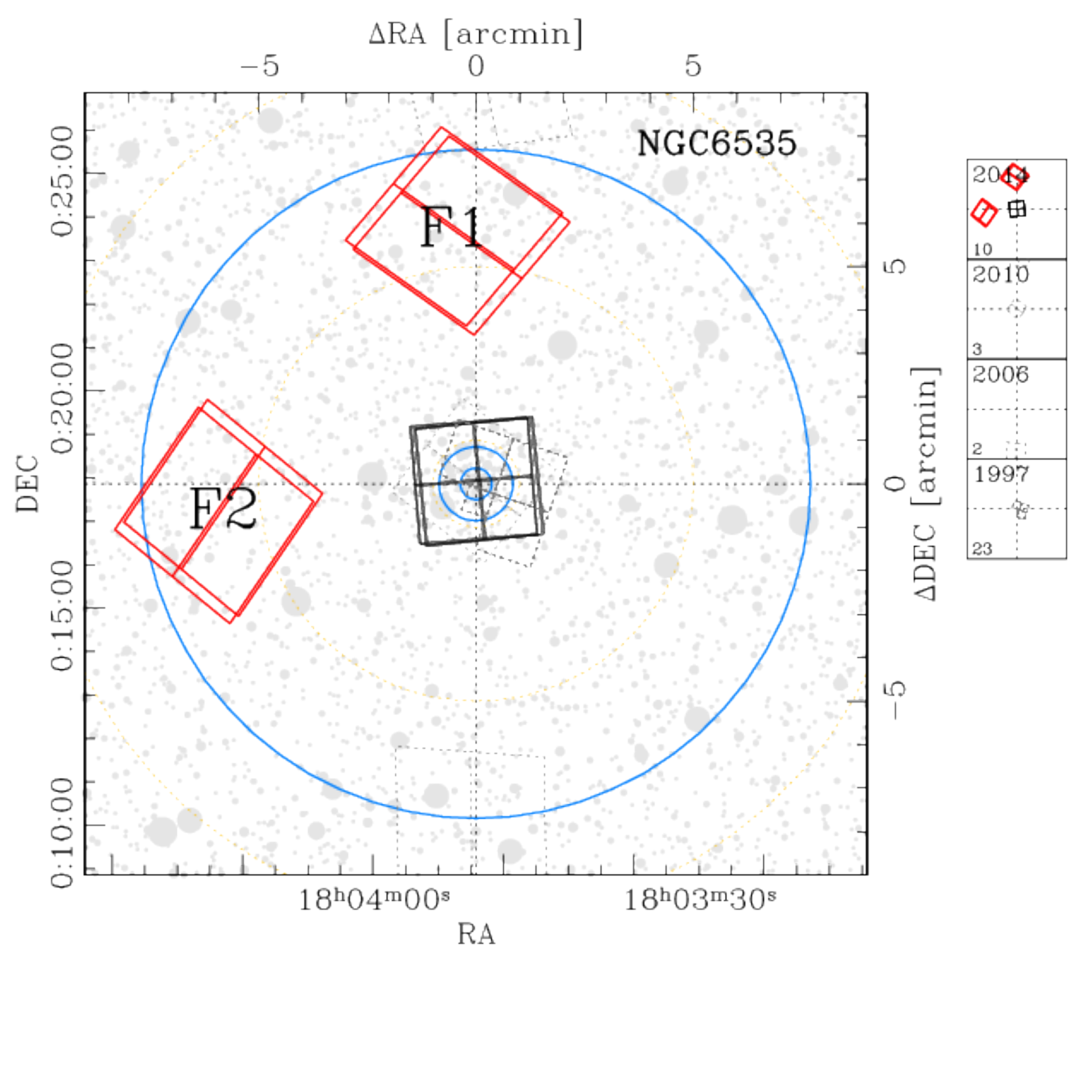}
  \includegraphics[width=7.7 cm]{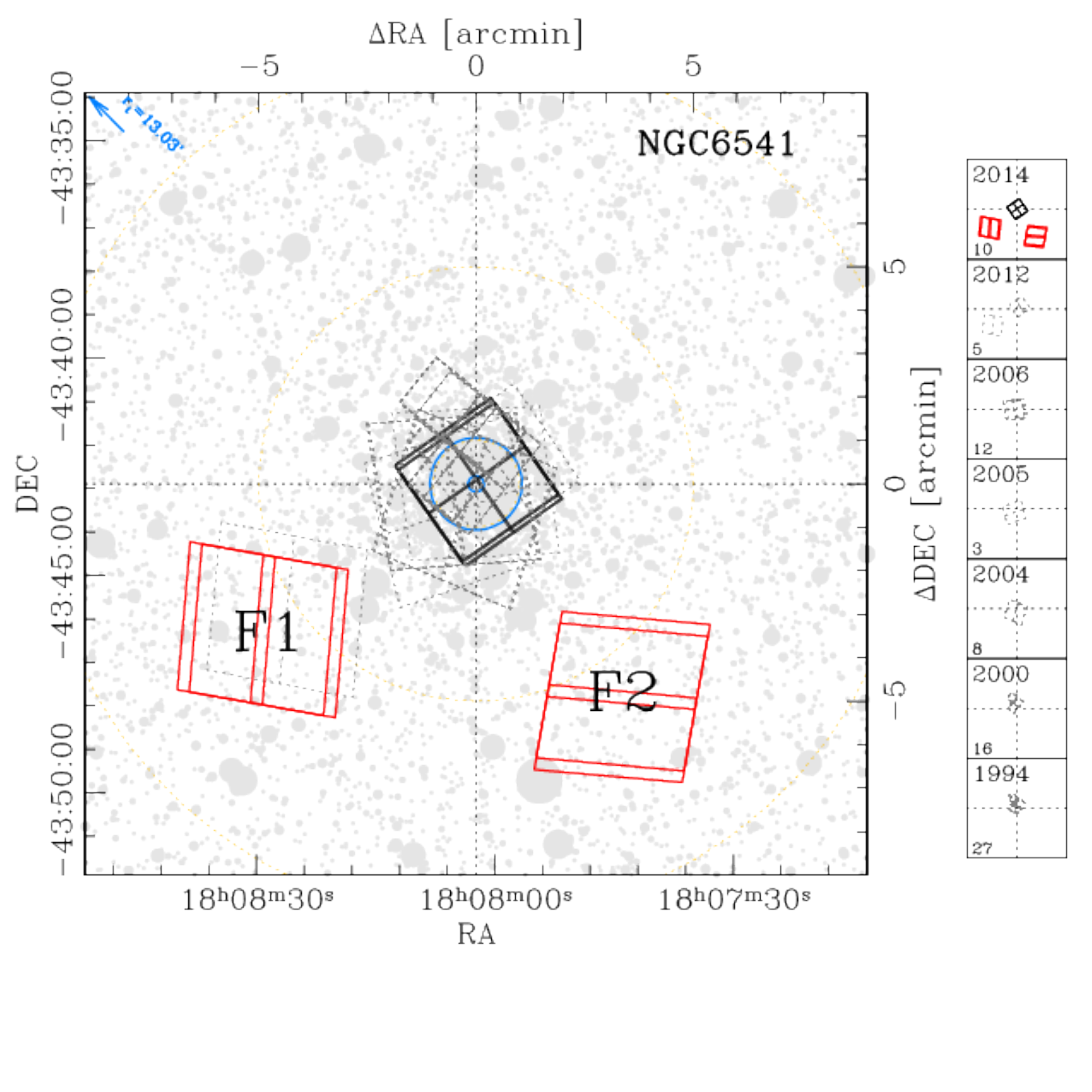}\\
  \includegraphics[width=7.7 cm]{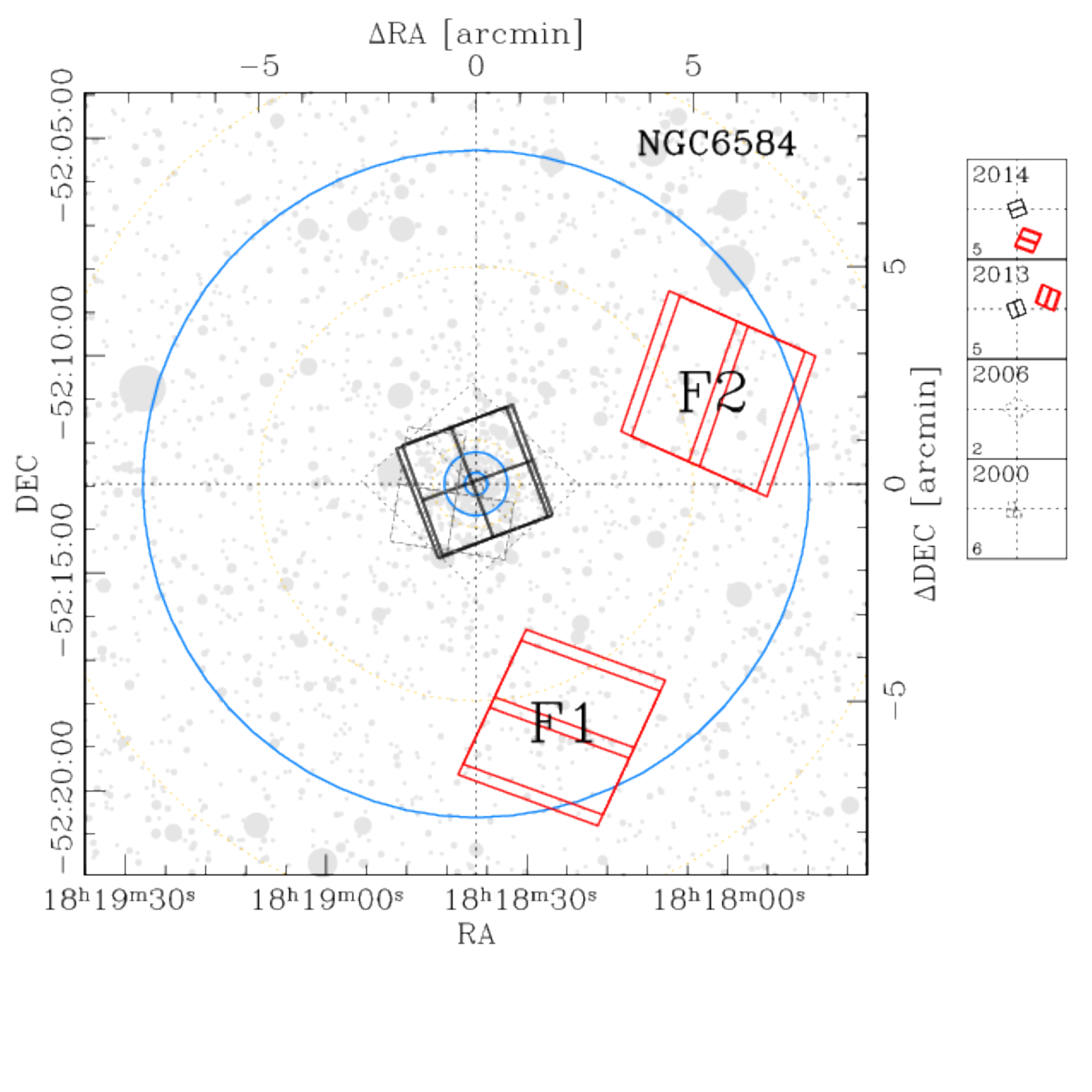}
  \includegraphics[width=7.7 cm]{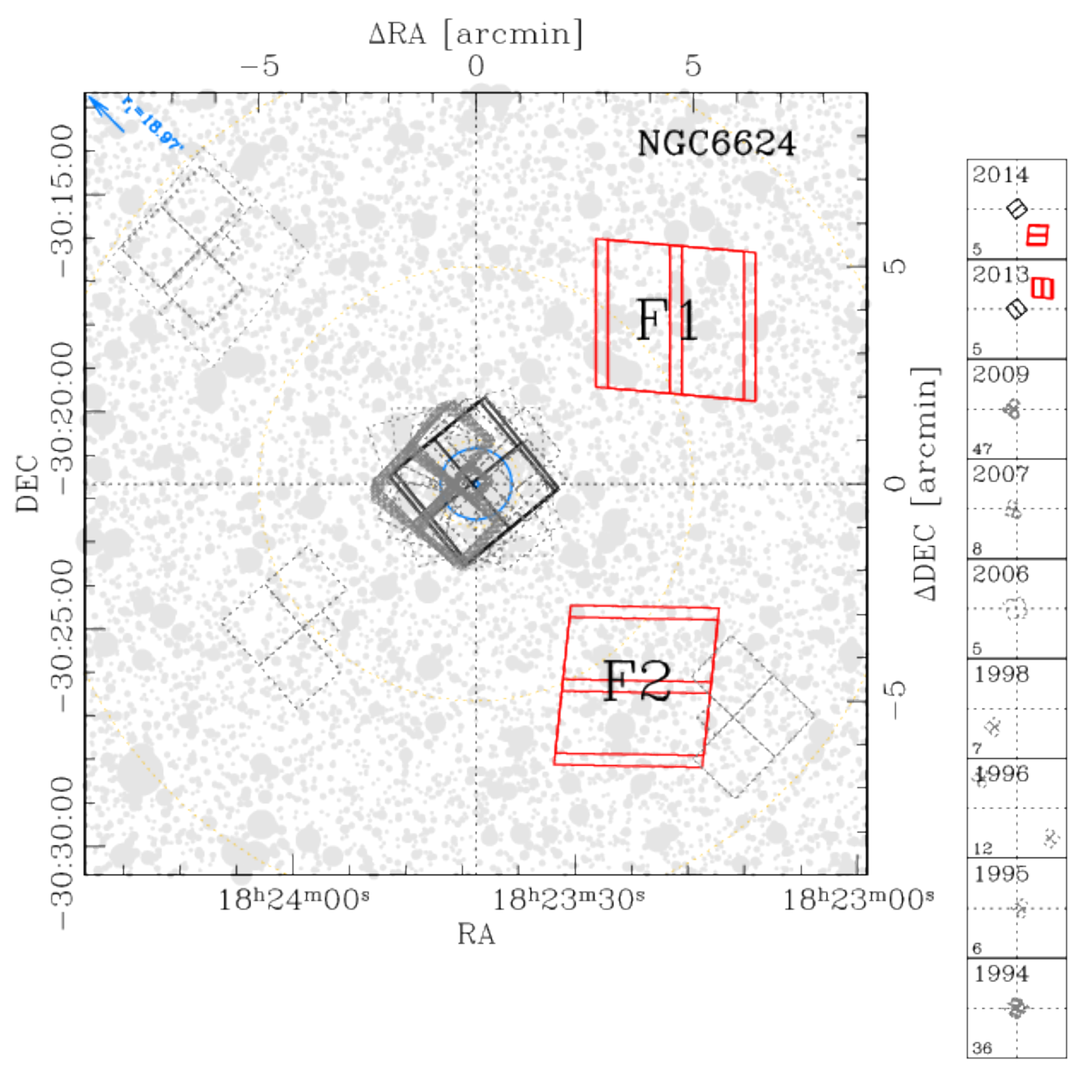}\\
  \caption{As in Figure \ref{fig:crt} but for NGC\,6441, NGC\,6496, NGC\,6535, NGC\,6541, NGC\,6584 and NGC\,6624}
  \label{fig:xcrt5}
\end{figure*}

\begin{figure*}
  \includegraphics[width=7.7 cm]{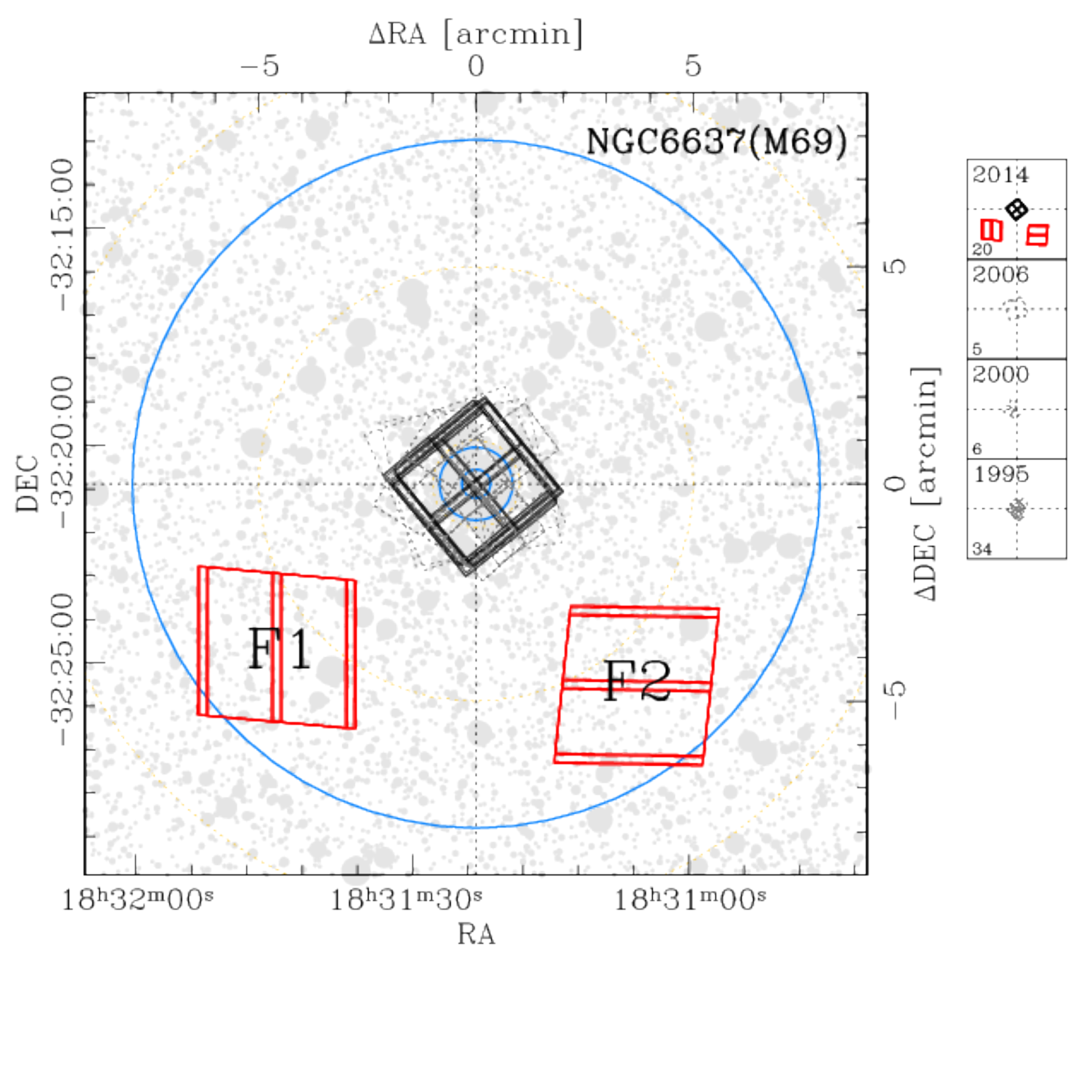}
  \includegraphics[width=7.7 cm]{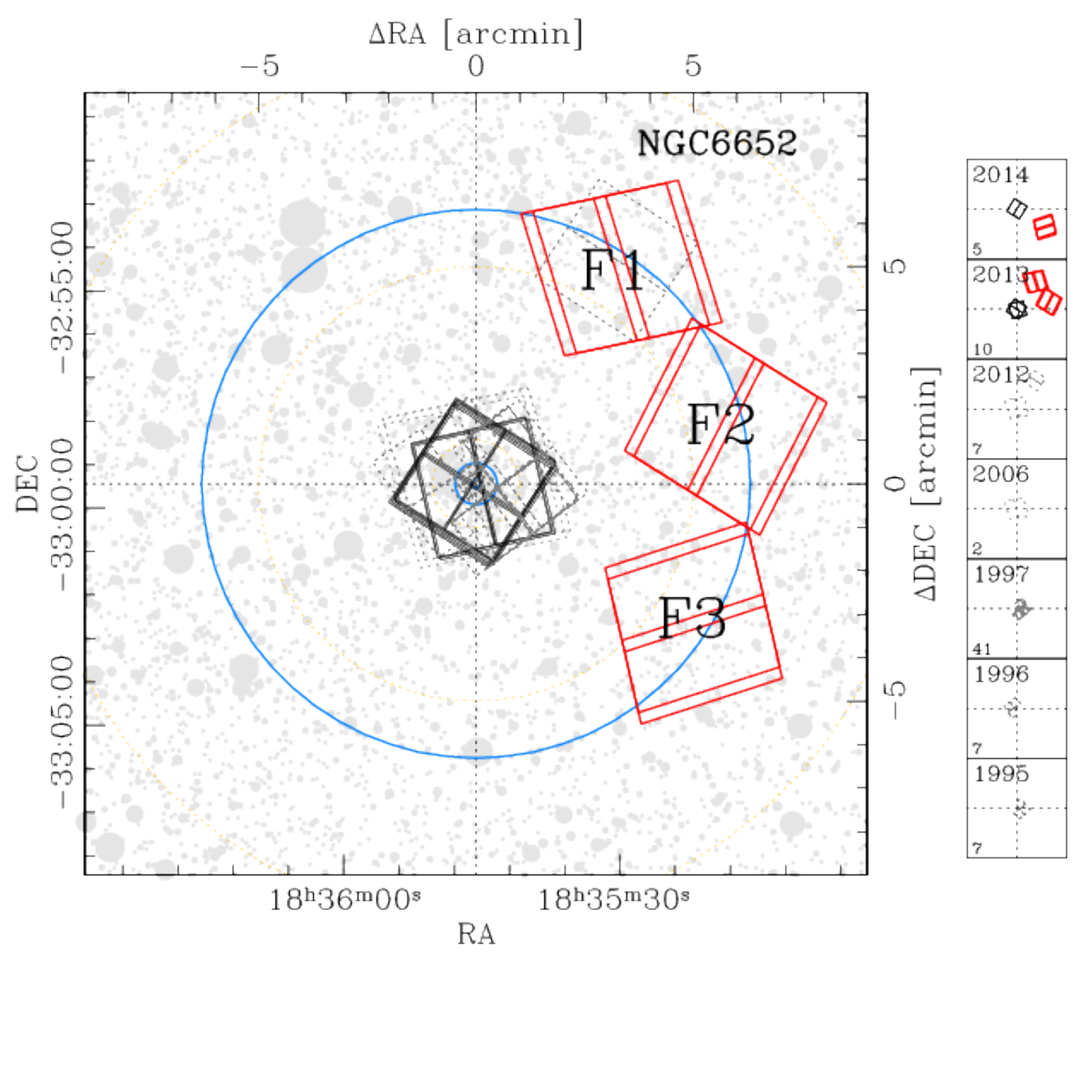}\\
  \includegraphics[width=7.7 cm]{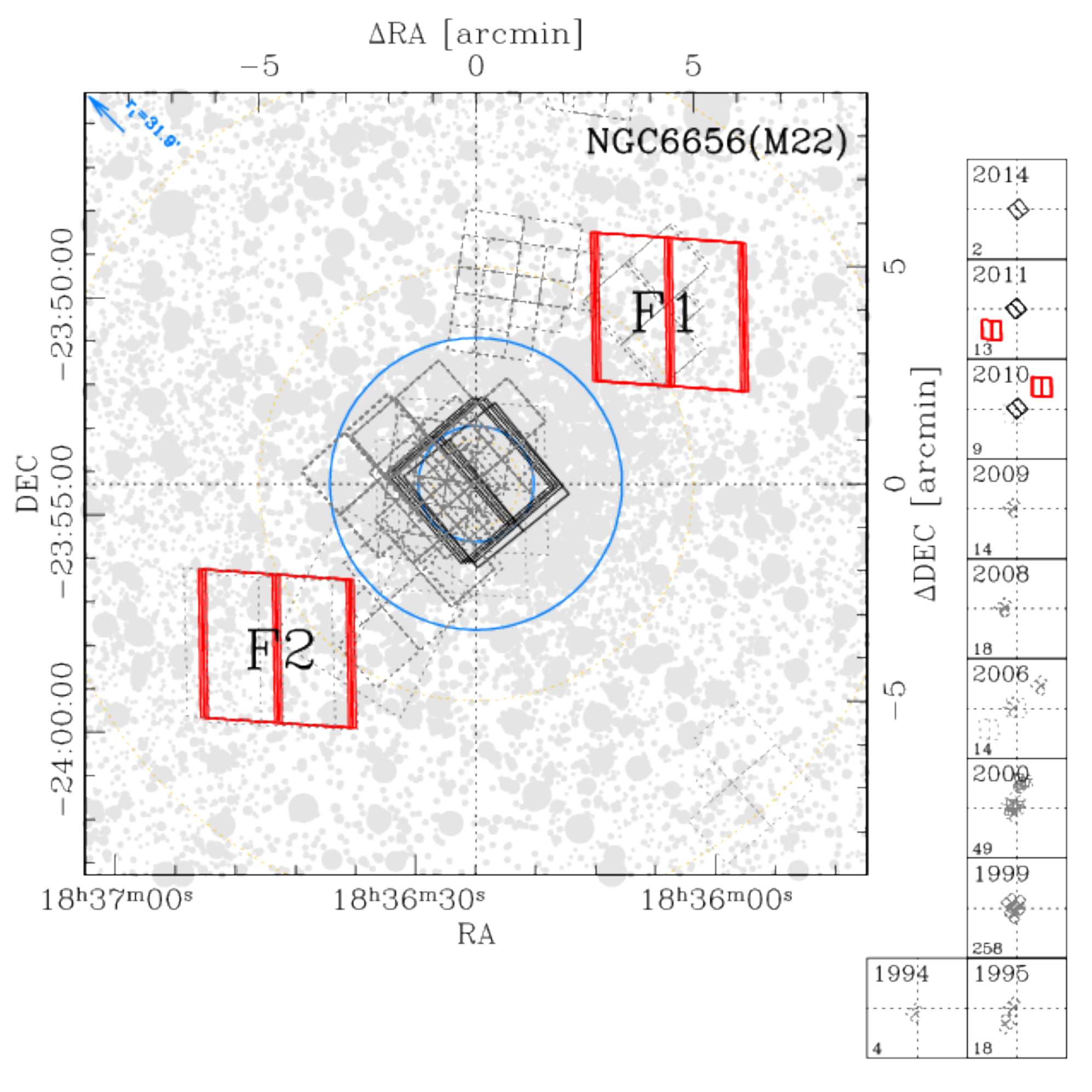}
  \includegraphics[width=7.7 cm]{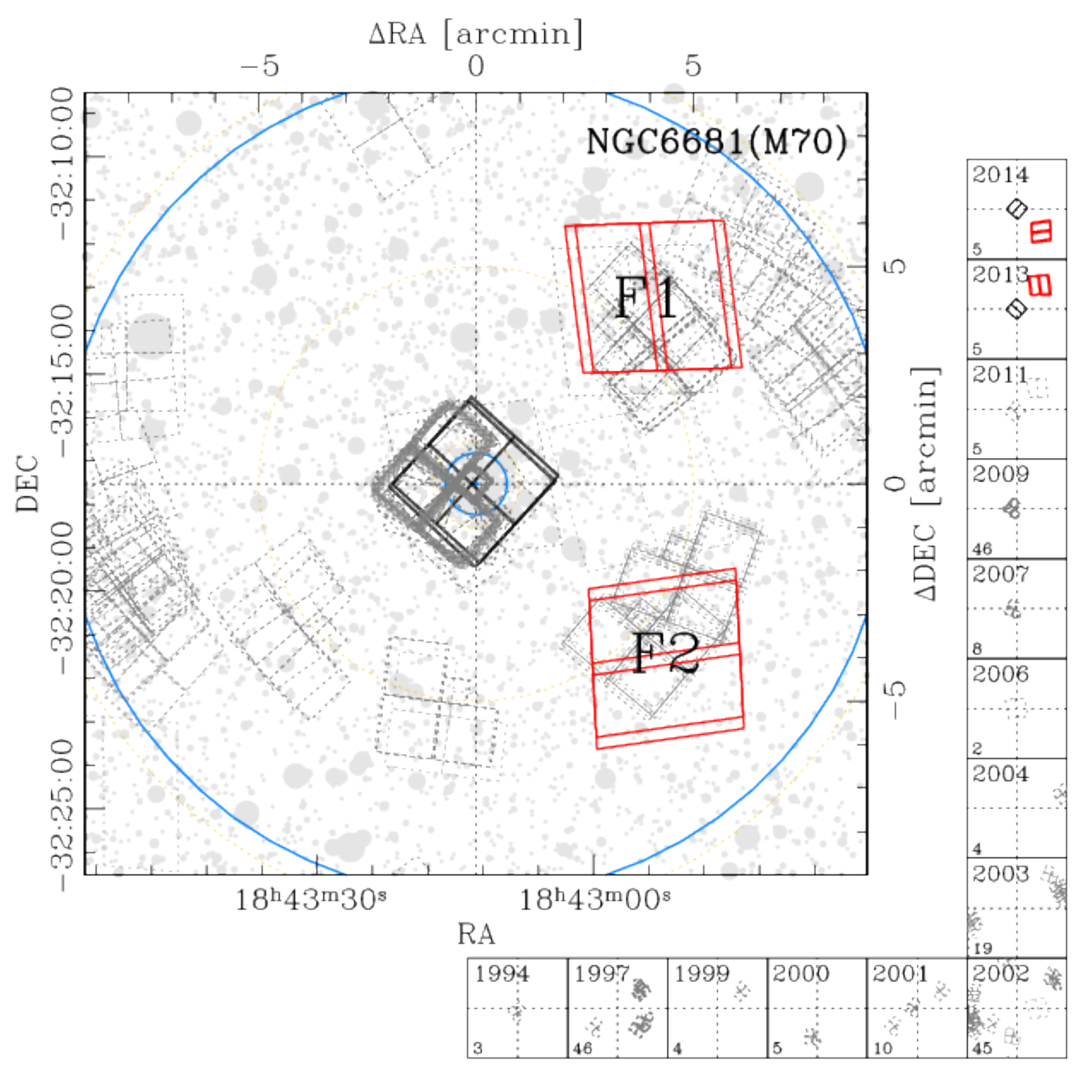}\\
  \includegraphics[width=7.7 cm]{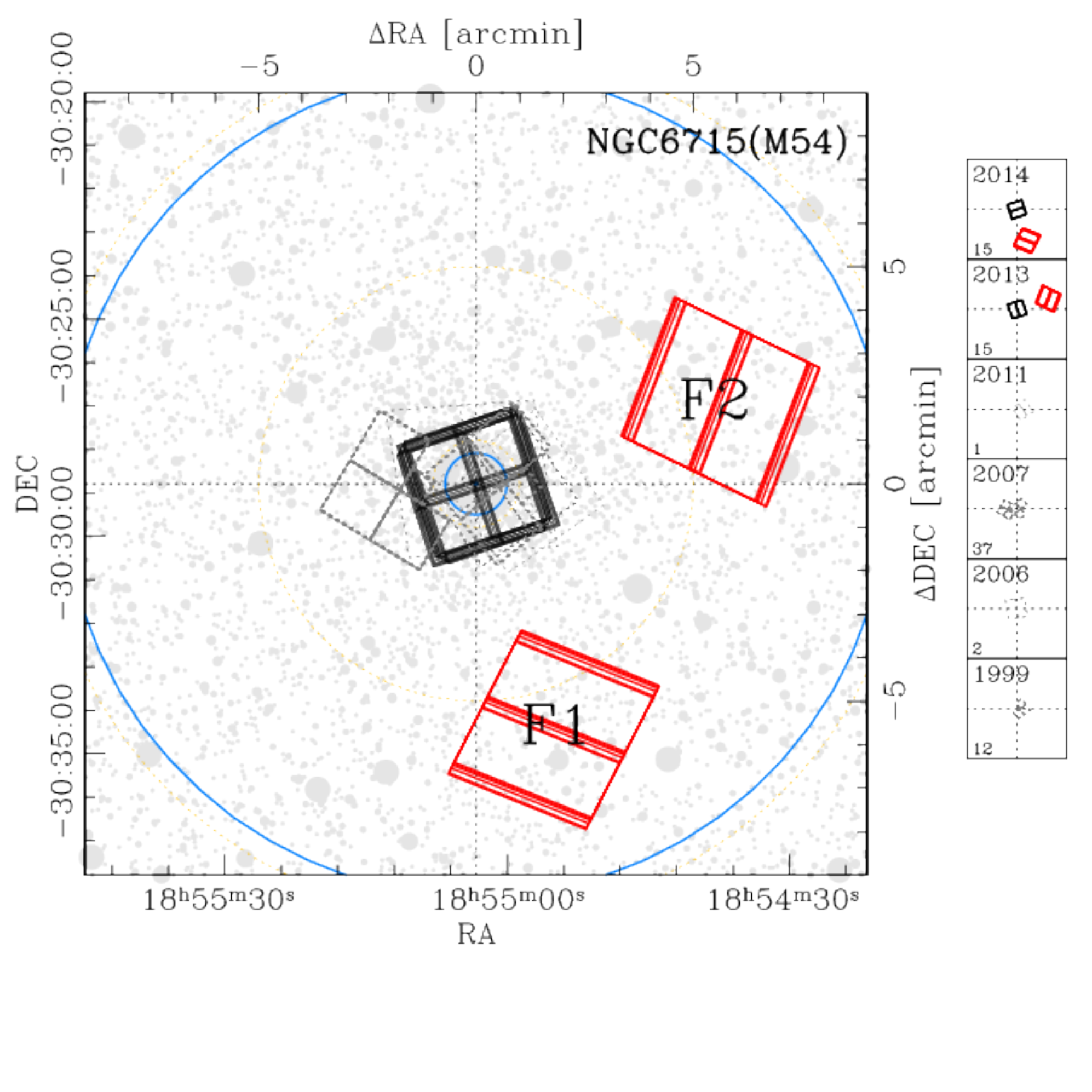}
  \includegraphics[width=7.7 cm]{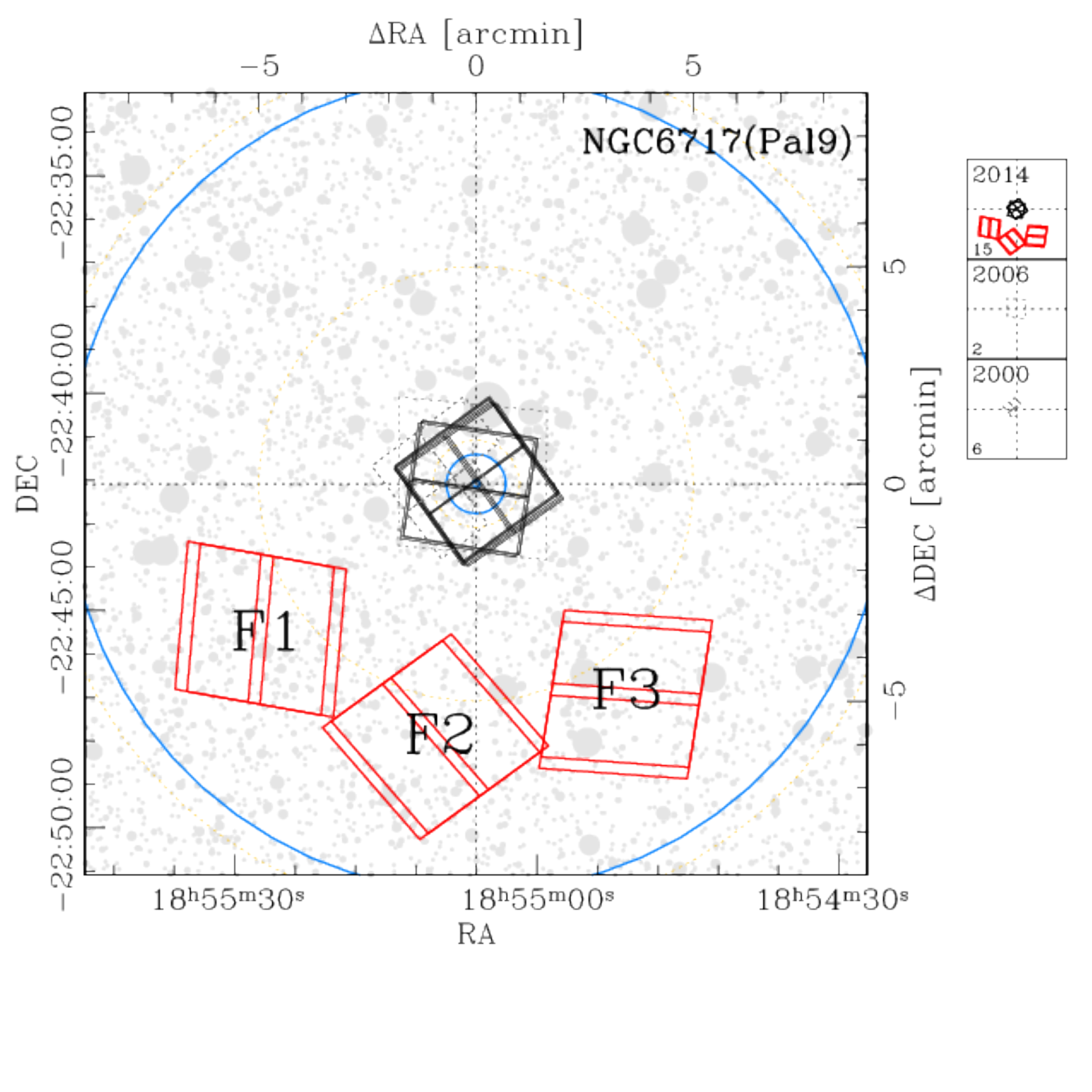}\\
  \caption{As in Figure \ref{fig:crt} but for NGC\,6637, NGC\,6652, NGC\,6656, NGC\,6681, NGC\,6715 and NGC\,6717}
  \label{fig:xcrt6}
\end{figure*}

\begin{figure*}
  \includegraphics[width=7.7 cm]{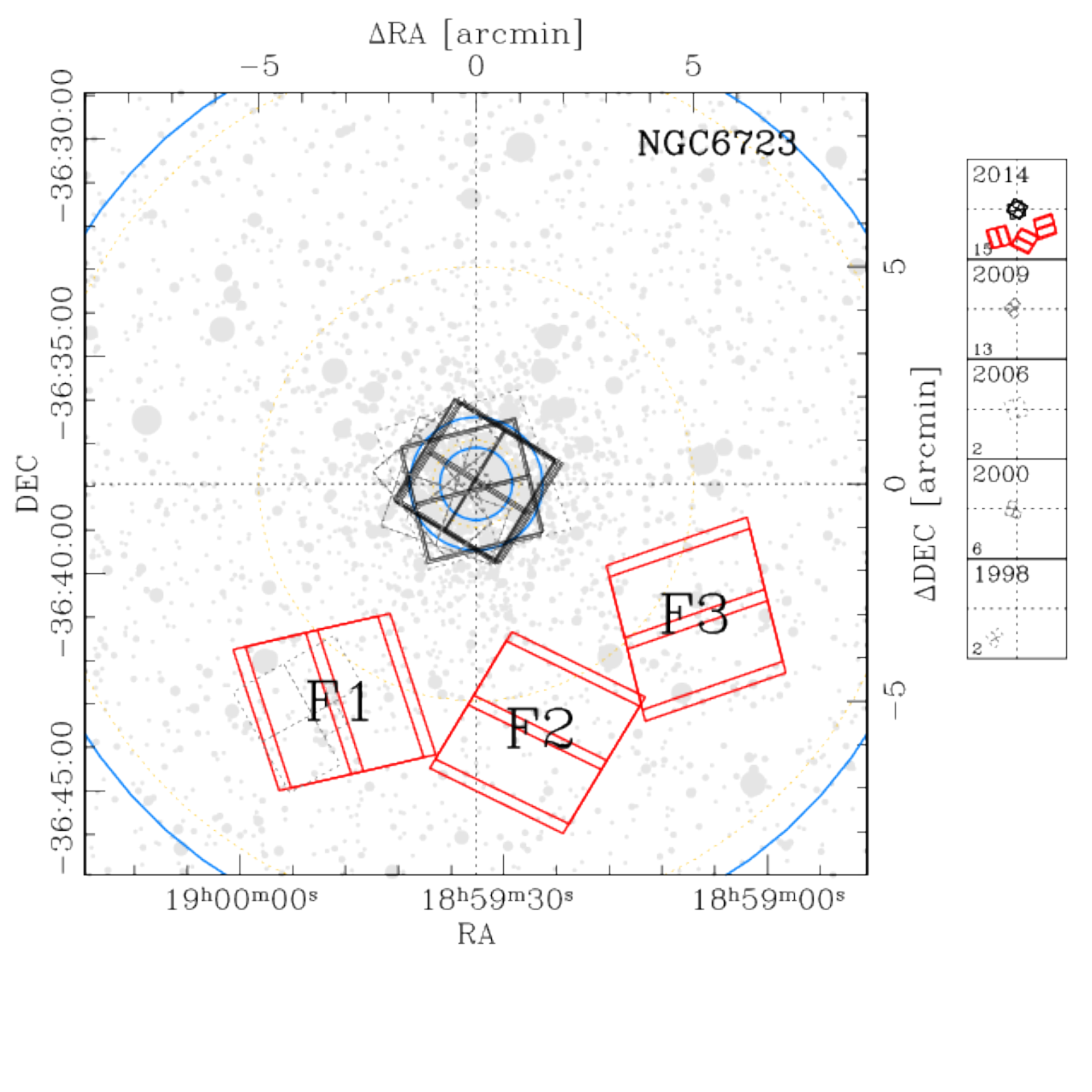}
  \includegraphics[width=7.7 cm]{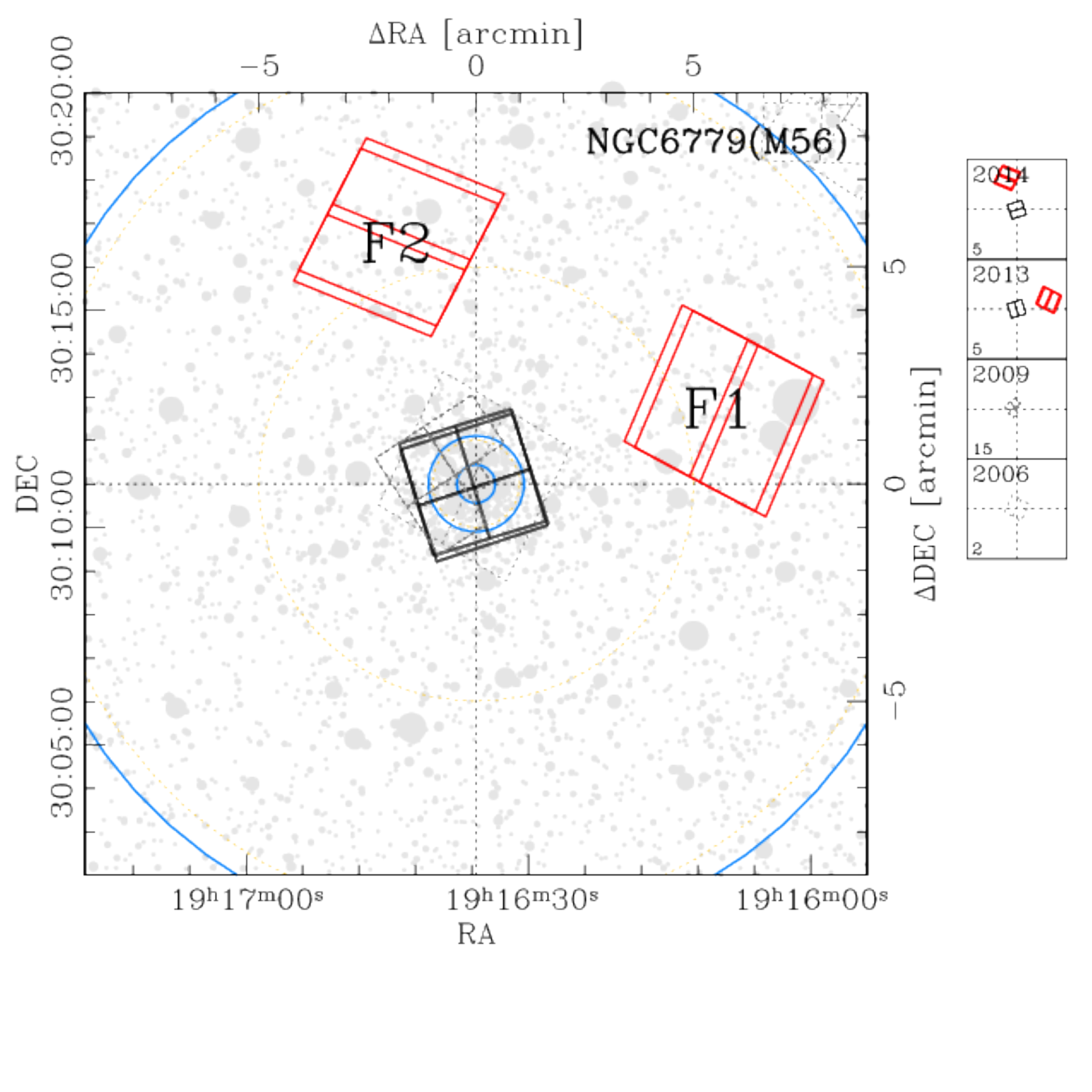}\\
  \includegraphics[width=7.7 cm]{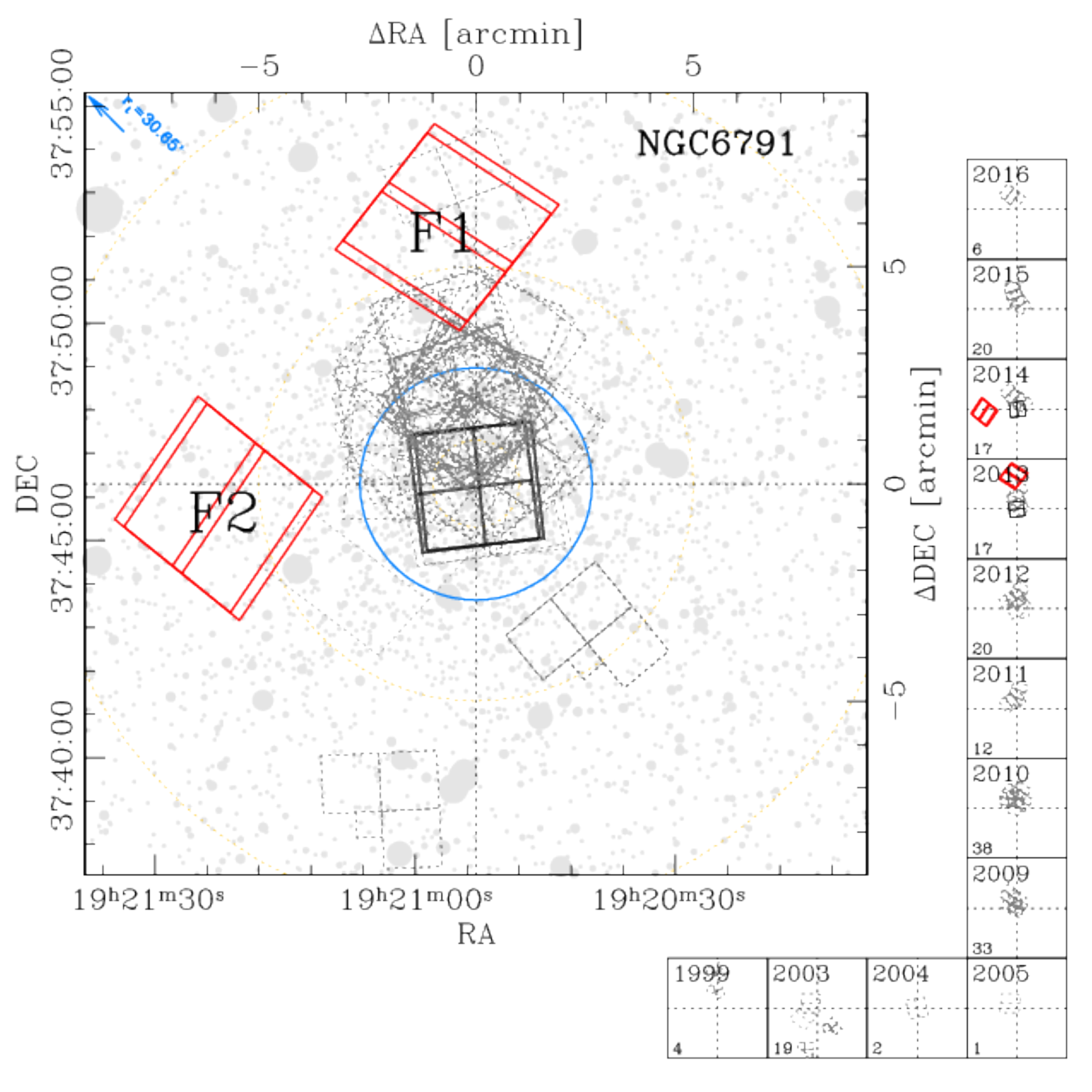}
  \includegraphics[width=7.7 cm]{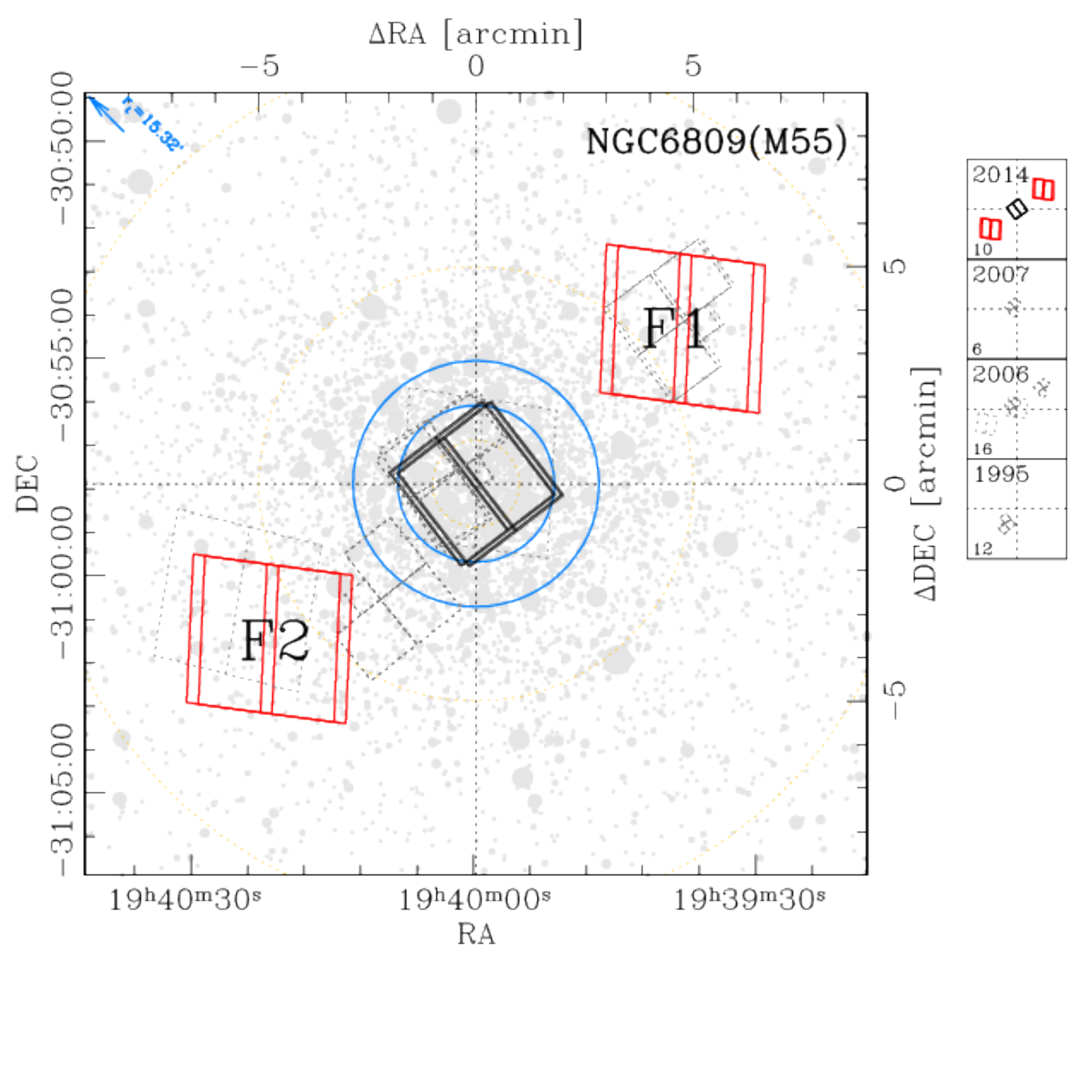}\\
  \includegraphics[width=7.7 cm]{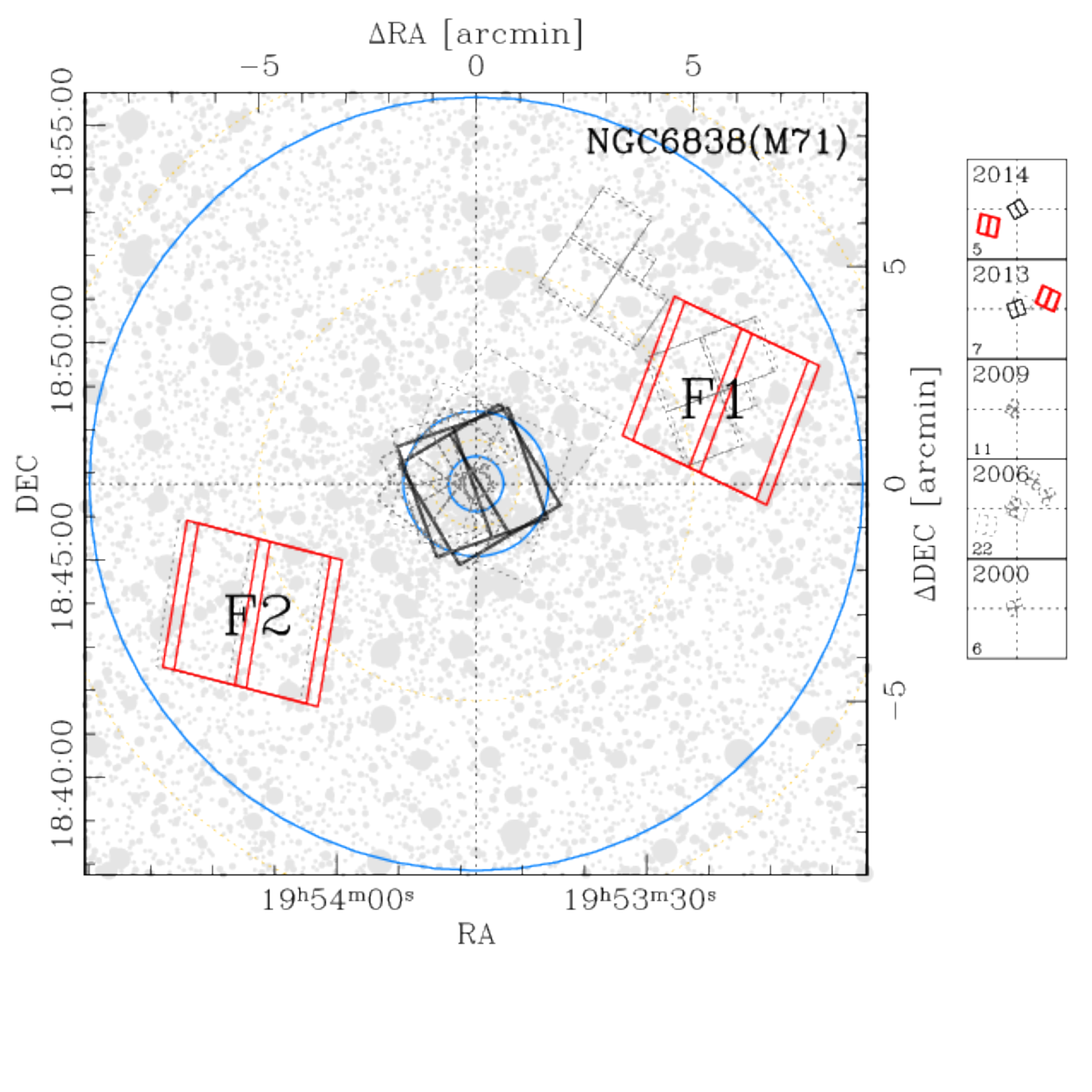}
  \includegraphics[width=7.7 cm]{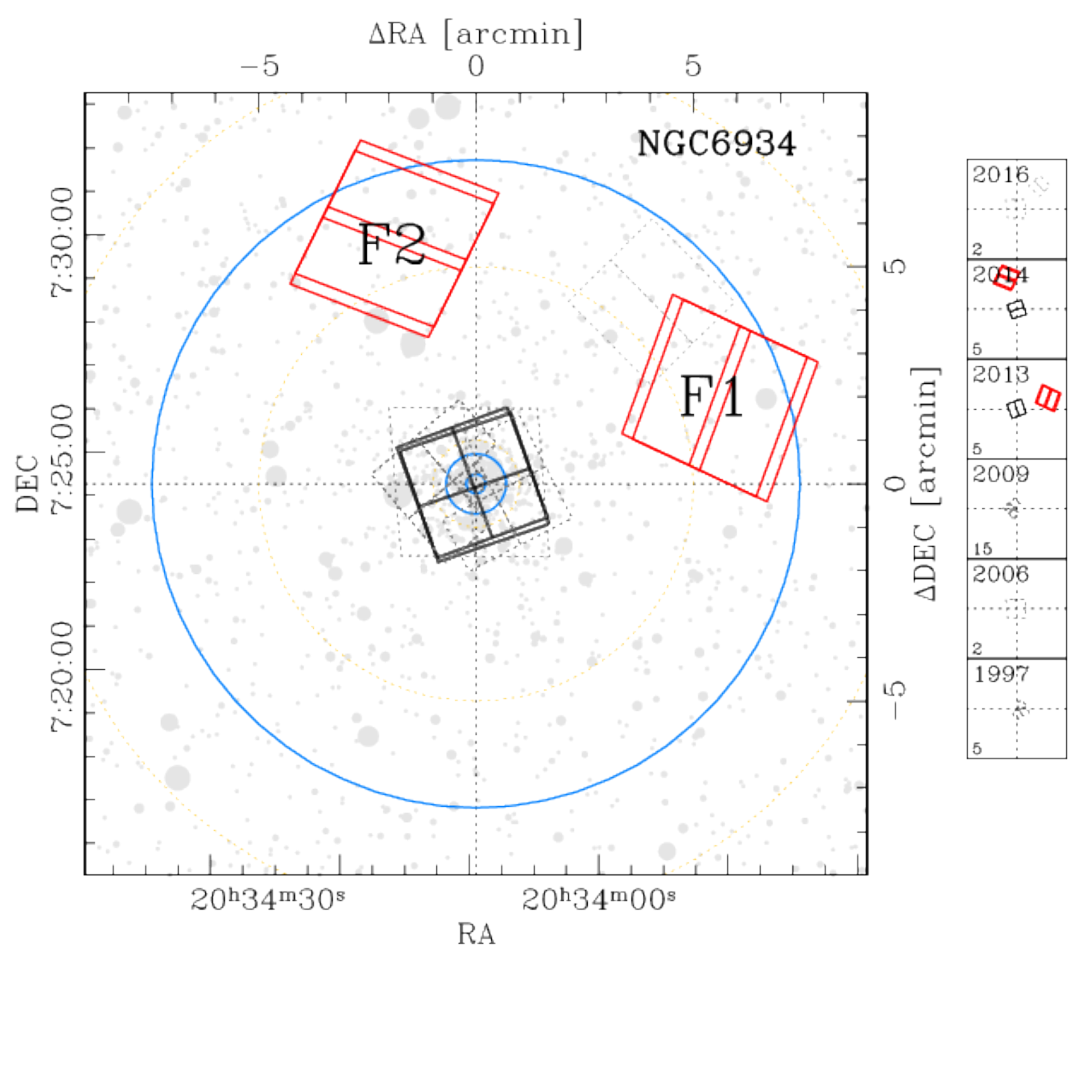}\\
  \caption{As in Figure \ref{fig:crt} but for NGC\,6723, NGC\,6779, NGC\,6791, NGC\,6809, NGC\,6838 and NGC\,6934}
  \label{fig:xcrt7}
\end{figure*}

\begin{figure*}
  \includegraphics[width=7.7 cm]{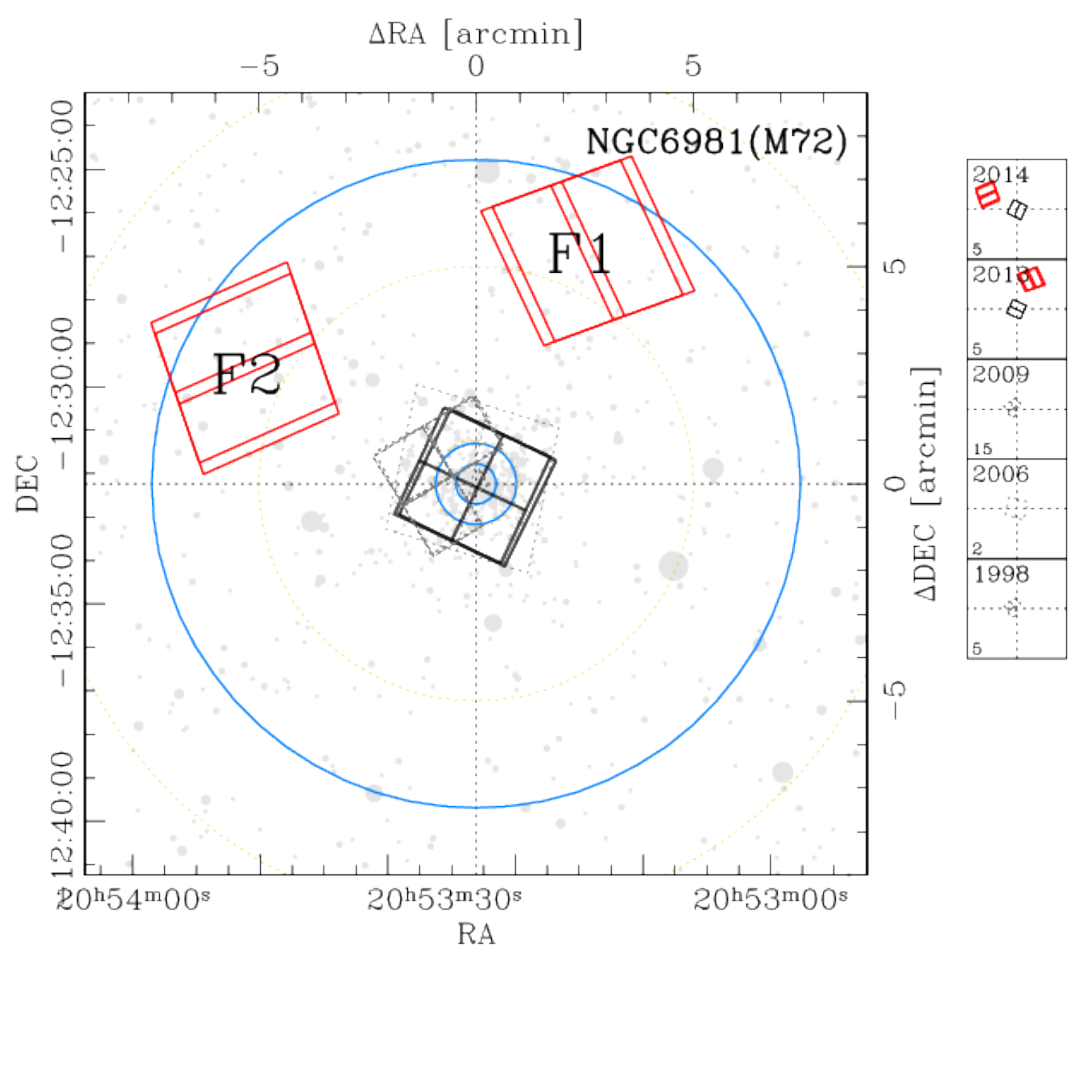}
  \includegraphics[width=7.7 cm]{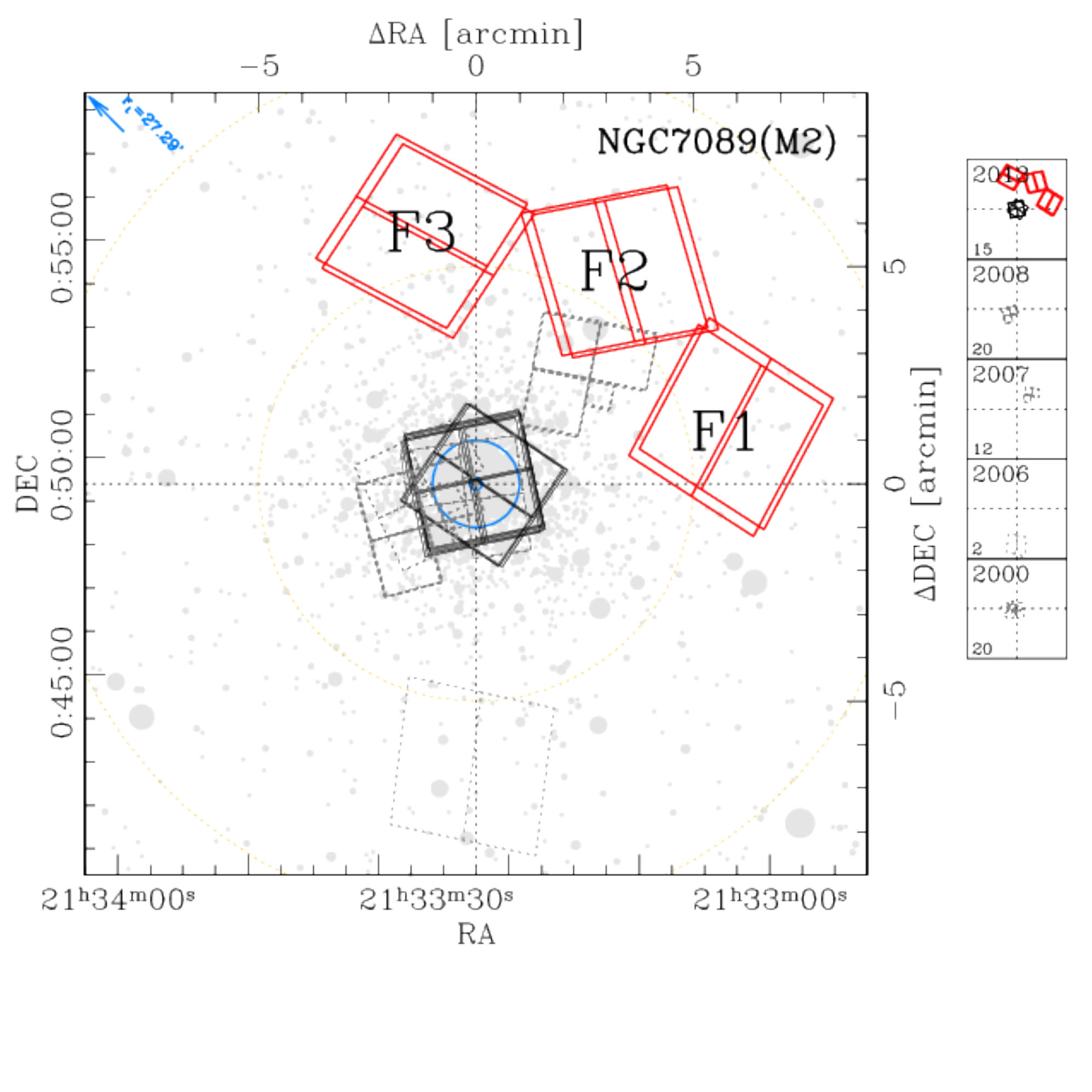}\\
  \includegraphics[width=7.7 cm]{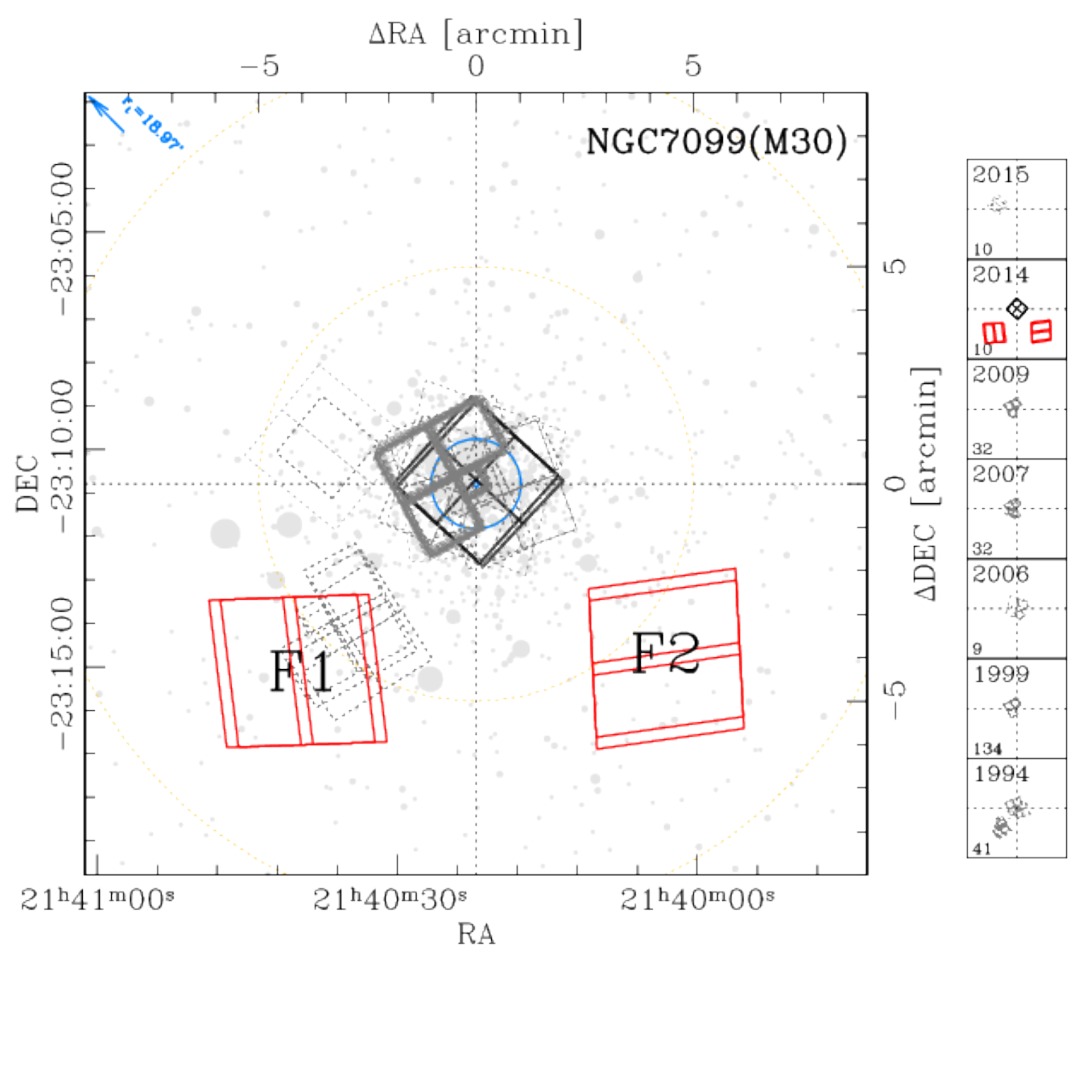}
  \caption{As in Figure \ref{fig:crt} but for NGC\,6981, NGC\,7089 and NGC\,7099}
  \label{fig:xcrt8}
\end{figure*}

\begin{figure*}
  \includegraphics[height=22 cm]{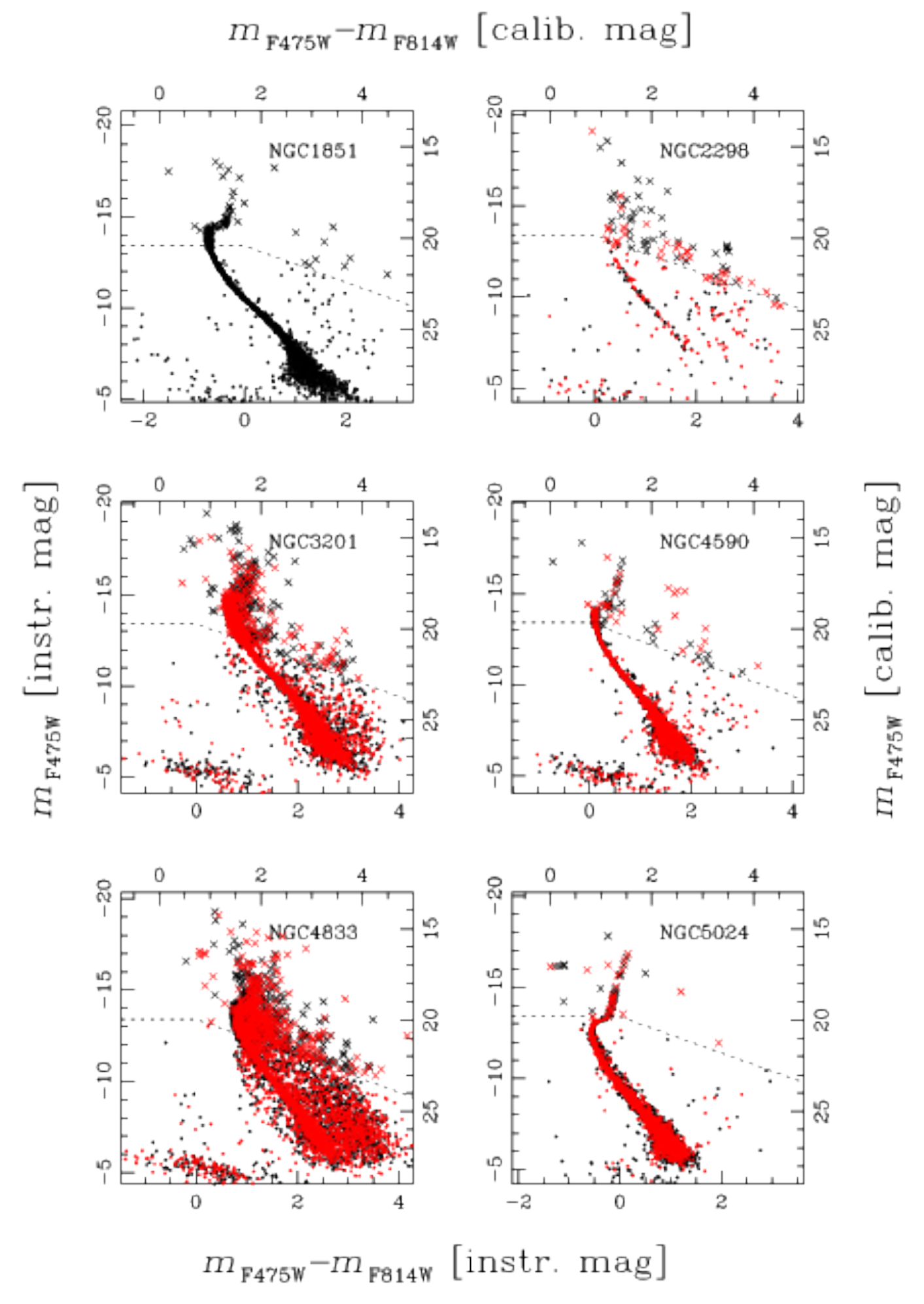}
  \caption{Obtained CMDs for NGC\,1851, NGC\,2298, NGC\,3201, NGC\,4590 and NGC\,5024. For each cluster, CMDs for all fields have been merged together. The colour-code is such that black dots represent stars measured in F1; red dots, stars of F2; green dots stars of F3; blue dots stars of F4 and orange dots represent stars measured in F5.
    Dashed lines represent the saturation level for all F1, saturated stars are represented by crosses.}
  \label{fig:xcmd1}
\end{figure*}

\begin{figure*}
  \includegraphics[height=22 cm]{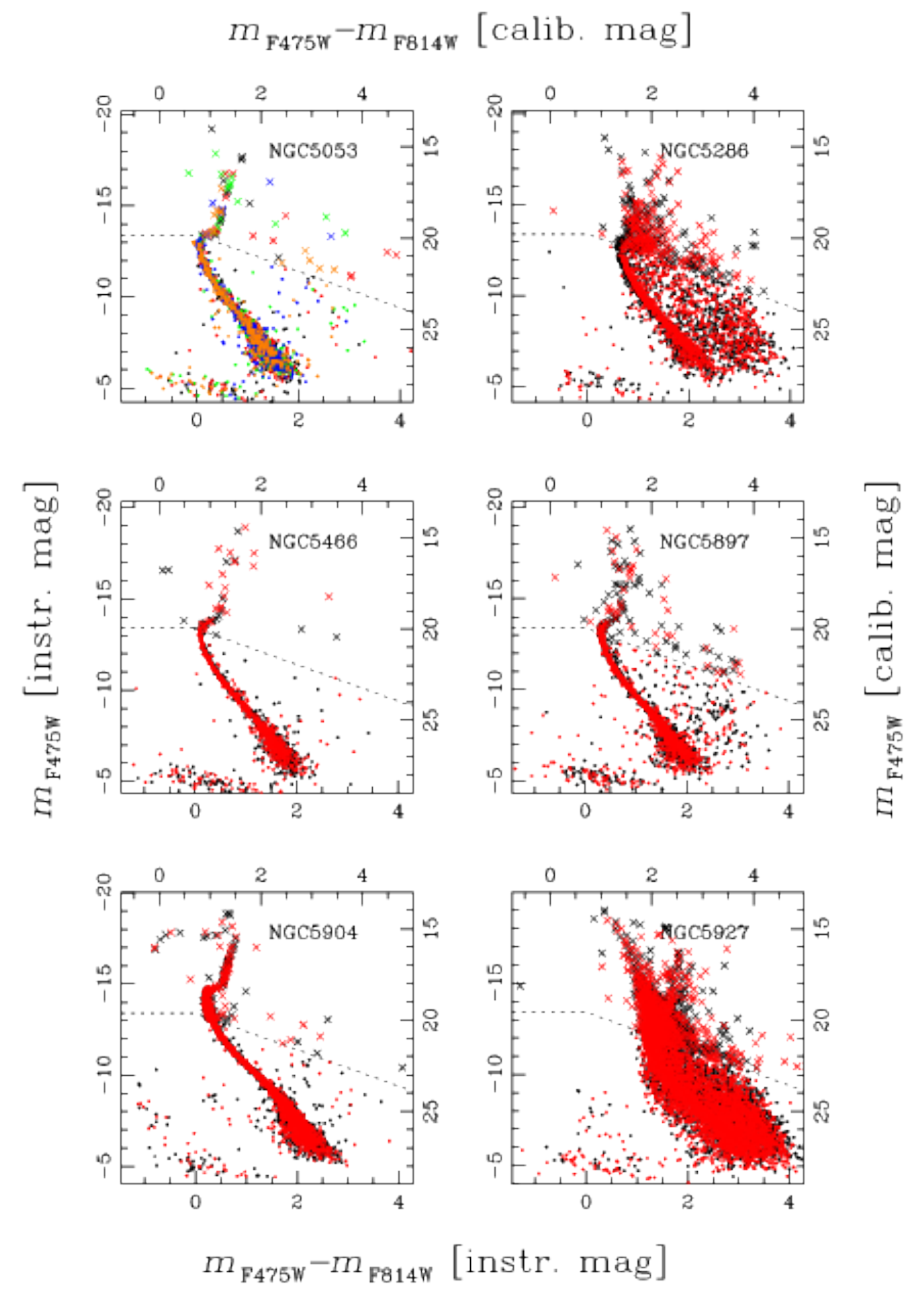}
  \caption{As in Figure \ref{fig:xcmd1} but for NGC\,5053, NGC\,5286, NGC\,5466, NGC\,5897, NGC\,5904 and NGC\,5927}
  \label{fig:xcmd2}
\end{figure*}

\begin{figure*}
  \includegraphics[height=22 cm]{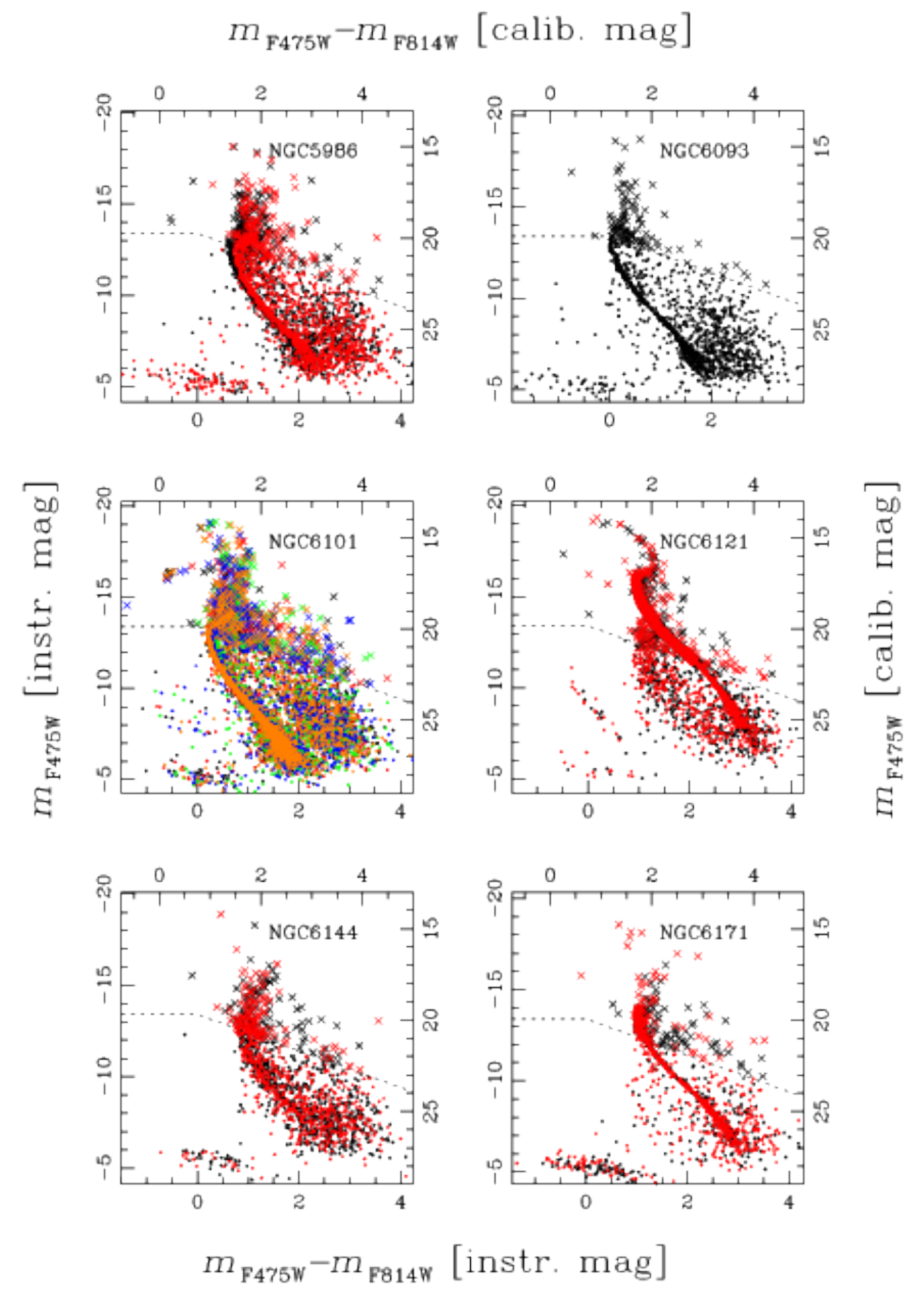}
  \caption{As in Figure \ref{fig:xcmd1} but for NGC\,5986, NGC\,6093, NGC\,6101, NGC\,6121, NGC\,6144 and NGC\,6171}
  \label{fig:xcmd3}
\end{figure*}

\begin{figure*}
  \includegraphics[height=22 cm]{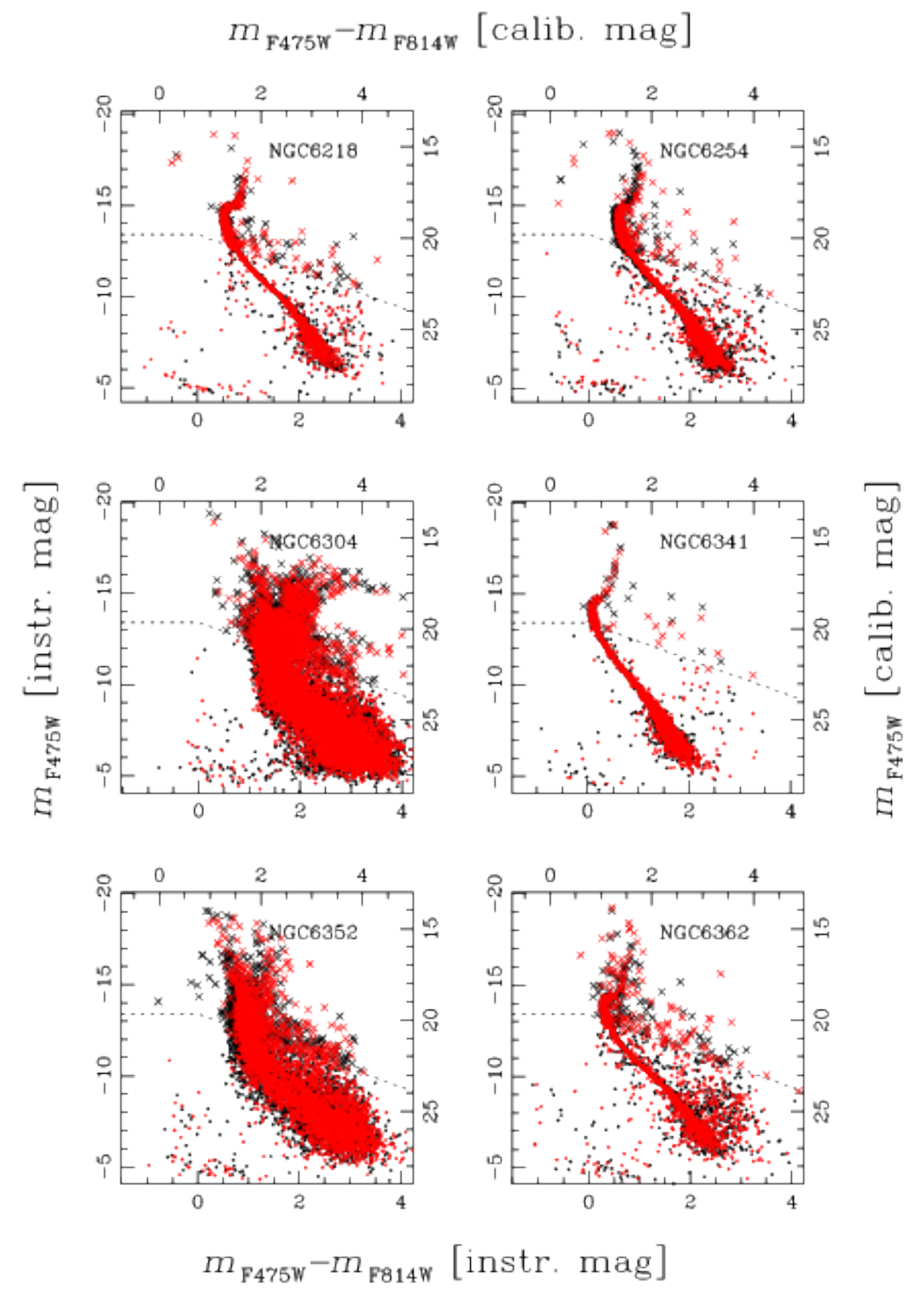}
  \caption{As in Figure \ref{fig:xcmd1} but for NGC\,6218, NGC\,6254, NGC\,6304, NGC\,6341, NGC\,6352 and NGC\,6362}
  \label{fig:xcmd4}
\end{figure*}

\begin{figure*}
  \includegraphics[height=22 cm]{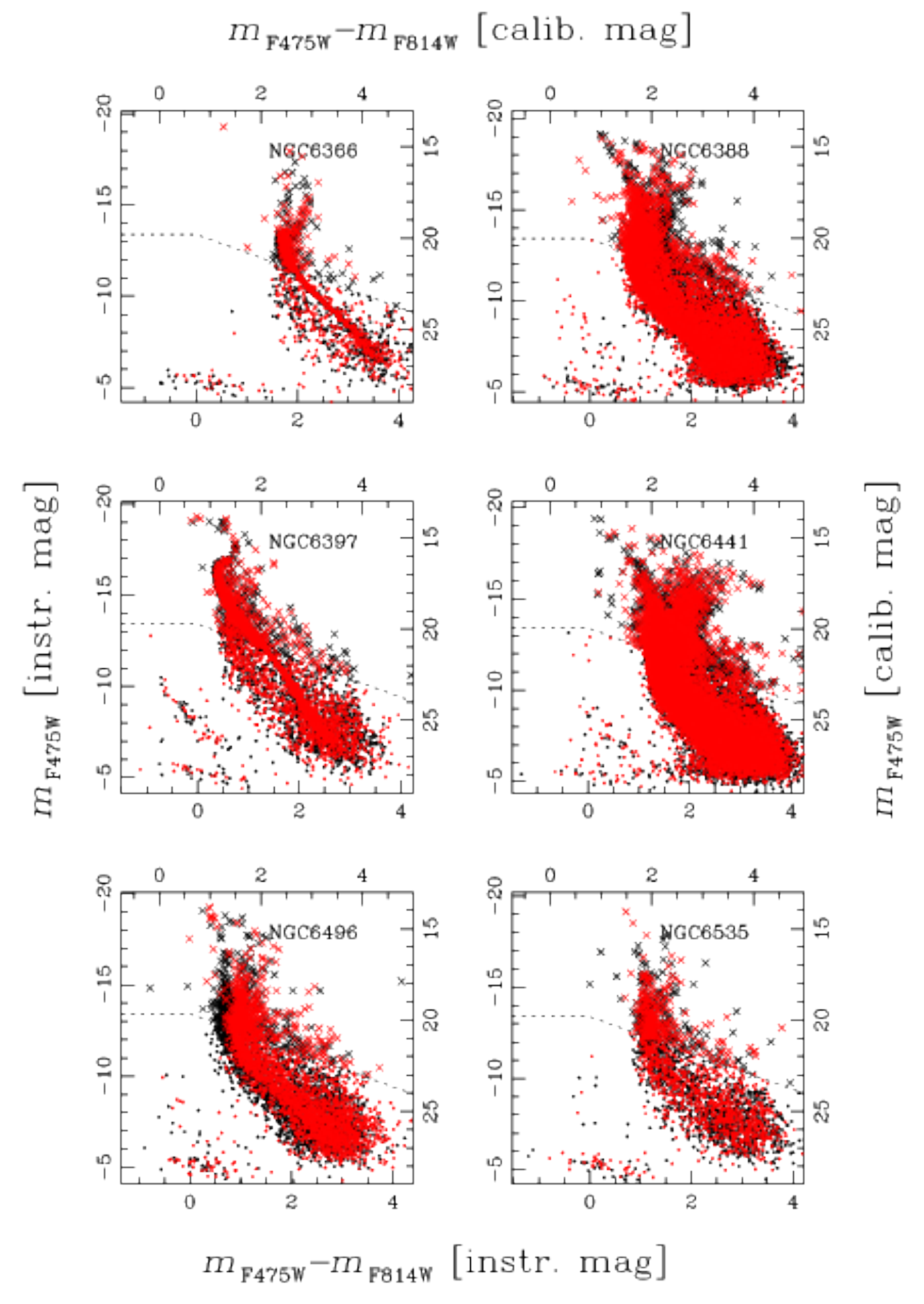}
  \caption{As in Figure \ref{fig:xcmd1} but for NGC\,6366, NGC\,6388, NGC\,6397, NGC\,6441, NGC\,6496 and NGC\,6535}
  \label{fig:xcmd5}
\end{figure*}

\begin{figure*}
  \includegraphics[height=22 cm]{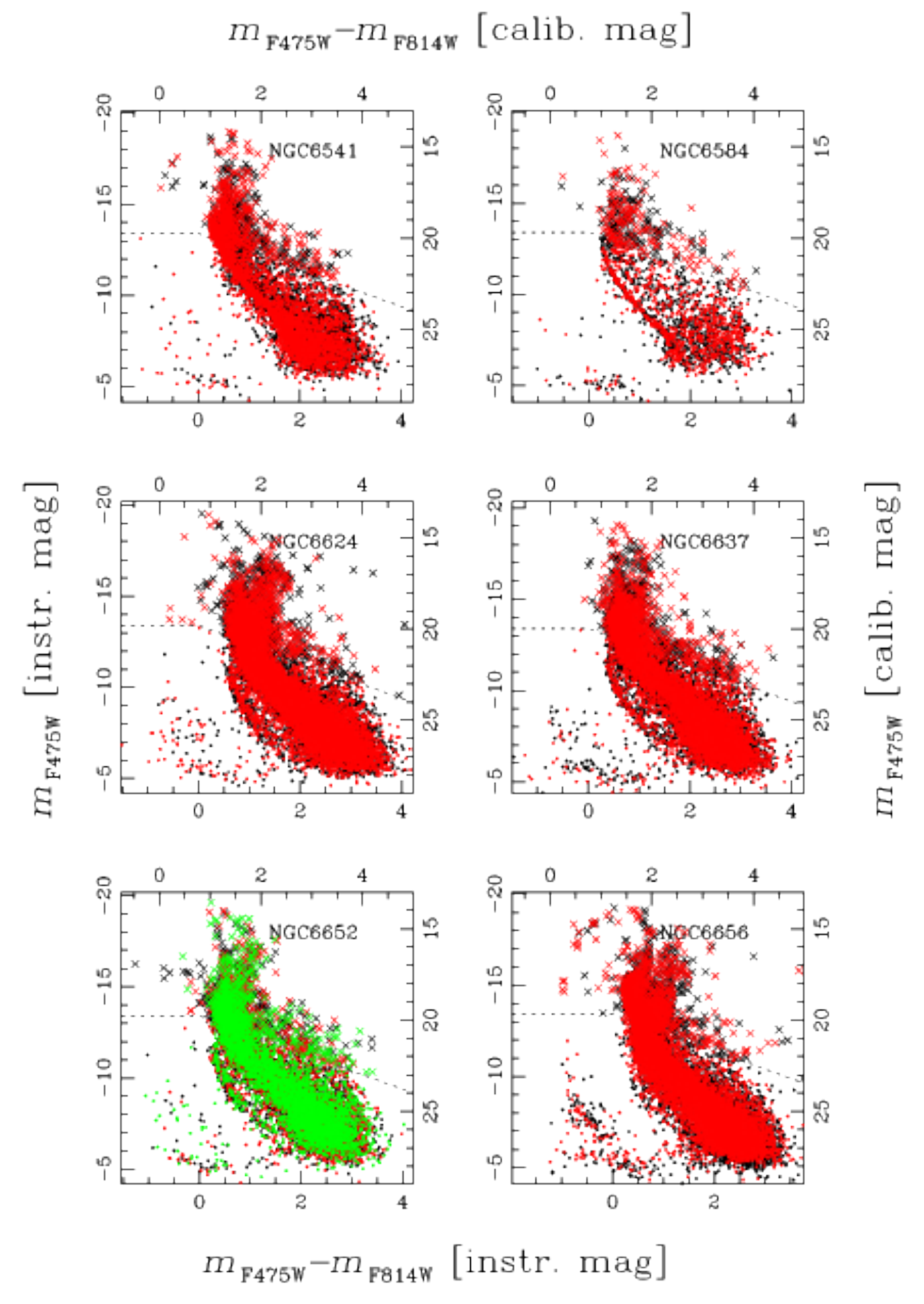}
  \caption{As in Figure \ref{fig:xcmd1} but for NGC\,6541, NGC\,6584, NGC\,6624, NGC\,6637, NGC\,6652 and NGC\,6656}
  \label{fig:xcmd6}
\end{figure*}

\begin{figure*}
  \includegraphics[height=22 cm]{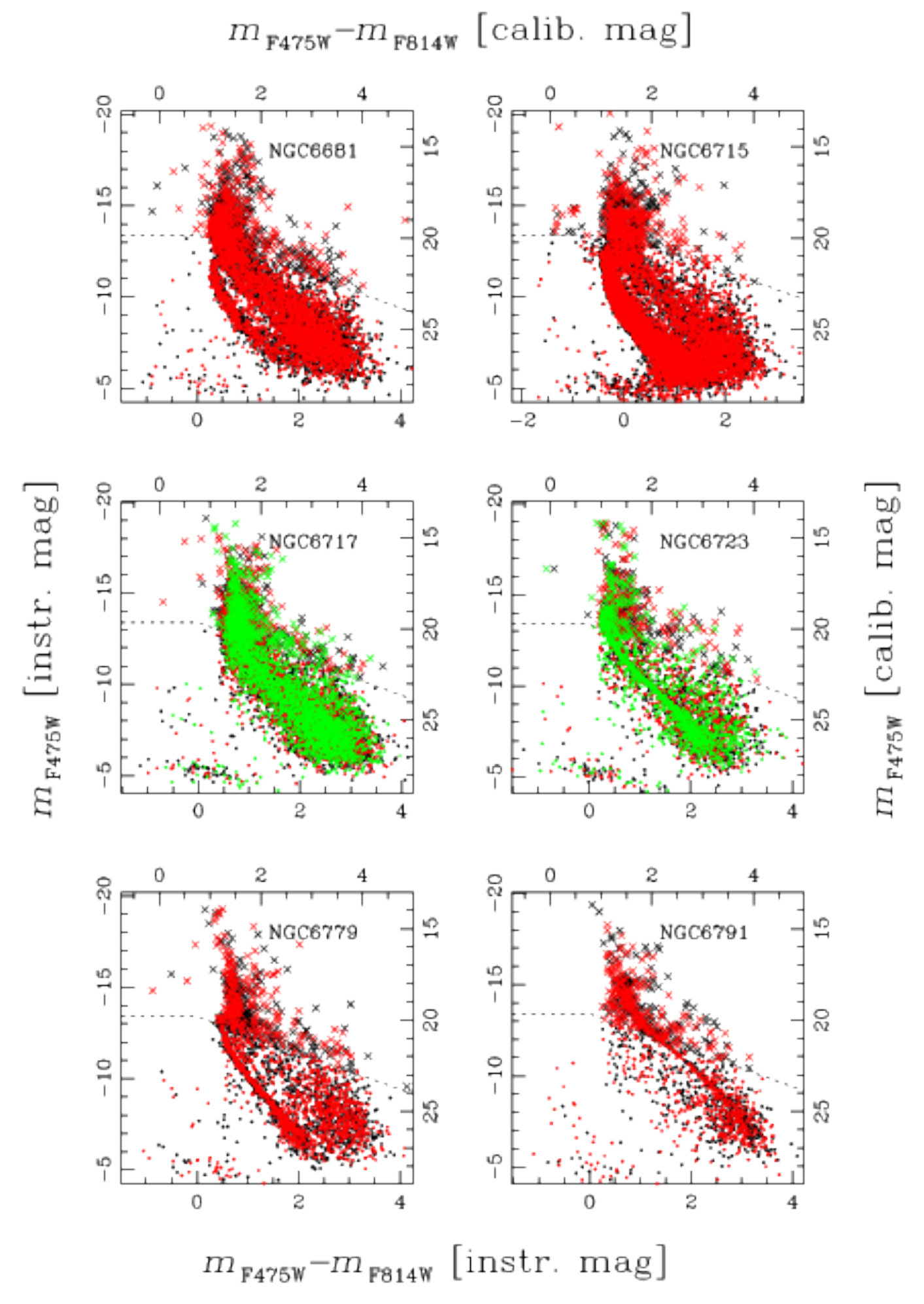}
  \caption{As in Figure \ref{fig:xcmd1} but for NGC\,6681, NGC\,6715, NGC\,6717, NGC\,6723, NGC\,6779 and NGC\,6791}
  \label{fig:xcmd7}
\end{figure*}

\begin{figure*}
  \includegraphics[height=22 cm]{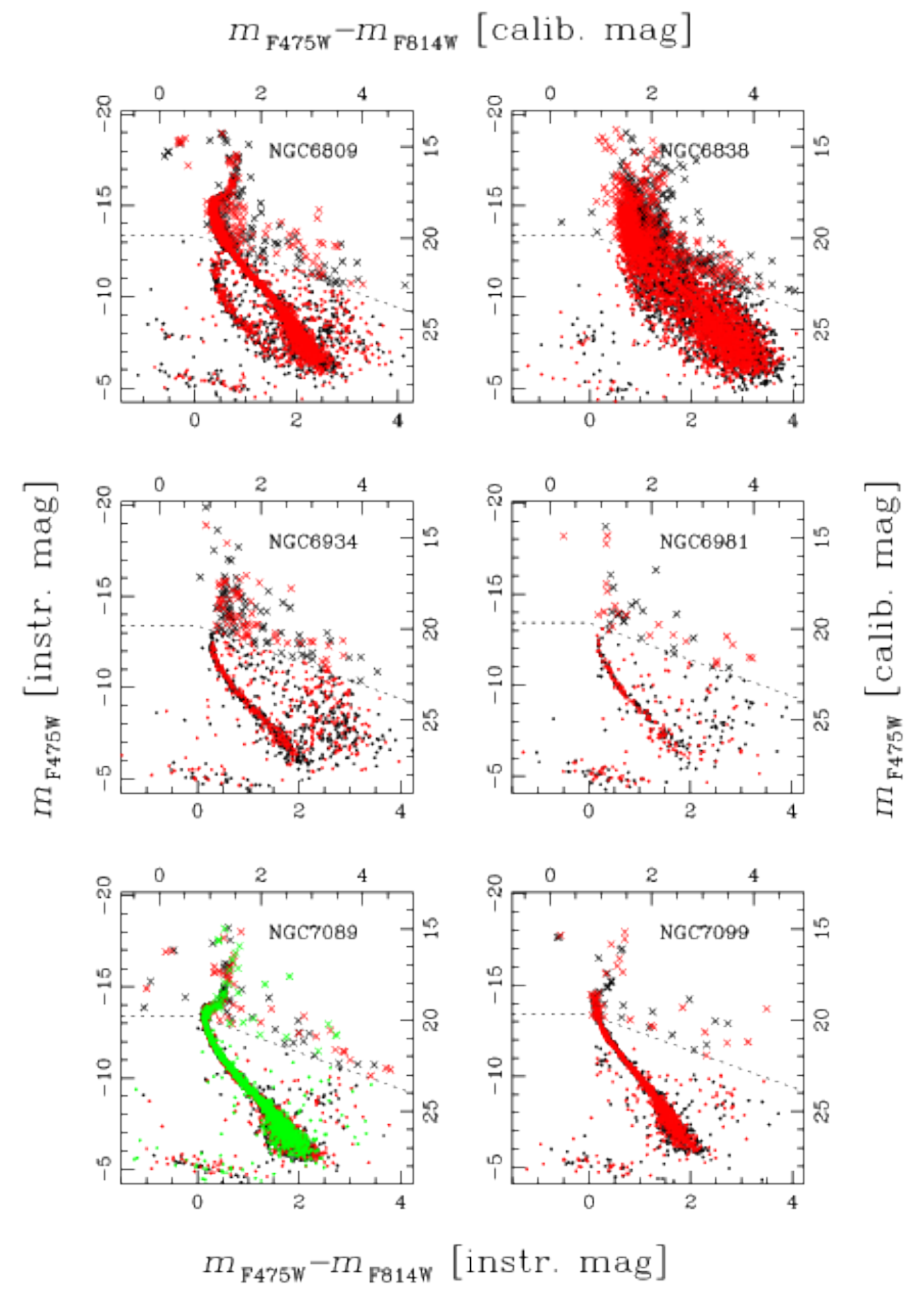}
  \caption{As in Figure \ref{fig:xcmd1} but for NGC\,6809, NGC\,6838, NGC\,6934, NGC\,6981, NGC\,7089 and NGC\,7099}
  \label{fig:xcmd8}
\end{figure*}

In this section we complement Figures \ref{fig:crt} and \ref{fig:cmd} with those referred to the rest of the sample.
In particular, Figures \ref{fig:xcrt1}, \ref{fig:xcrt2}, \ref{fig:xcrt3}, \ref{fig:xcrt4}, \ref{fig:xcrt5}, \ref{fig:xcrt6}, \ref{fig:xcrt7} and \ref{fig:xcrt8} complement the sample of finding charts presented in Figure \ref{fig:crt}.
Figures \ref{fig:xcmd1}, \ref{fig:xcmd2}, \ref{fig:xcmd3}, \ref{fig:xcmd4}, \ref{fig:xcmd5}, \ref{fig:xcmd6}, \ref{fig:xcmd7} and \ref{fig:xcmd8} are analogs to the lower-right panel of Figure \ref{fig:cmd}.


\bsp
\label{lastpage}
\end{document}